\documentclass[fleqn,usenatbib]{mnras}
\usepackage[normalem]{ulem}

\usepackage{newtxtext,newtxmath}

\usepackage[T1]{fontenc}
\usepackage{cuted}

\DeclareRobustCommand{\VAN}[3]{#2}
\let\VANthebibliography\thebibliography
\def\thebibliography{\DeclareRobustCommand{\VAN}[3]{##3}\VANthebibliography}

\usepackage{graphicx}	%
\usepackage{amsmath}	%
\usepackage{physics}
\usepackage{cuted}
\usepackage{lipsum}
\usepackage{mathtools}
\usepackage{empheq}
\usepackage{gensymb}

\newcommand{\fagn}{f_{\rm agn}}
\newcommand{\lbol}[1]{%
  L_{\rm bol} \geq 10^{#1} \, \mathrm{erg \, s^{-1}}%
}
\newcommand*\conj[1]{\widebar{#1}}

\title[AGN-channel population inference]{
Unifying spatial correlation with hierarchical population inference to constrain the AGN channel for binary black hole mergers
}

\author[L. Pouw et al.]{
Lucas Pouw$^{1}$\thanks{E-mail: pouw@strw.leidenuniv.nl},
Lorenzo Speri$^{2,1}$,
Niccol\`{o} Veronesi$^{3}$,
Elena Maria Rossi$^{1}$
\\
$^{1}$Leiden Observatory, Leiden University, PO Box 9513, 2300 RA Leiden, The Netherlands\\
$^{2}$European Space Agency (ESA), European Space Research and Technology Centre (ESTEC), Keplerlaan 1, 2201 AZ Noordwijk, the Netherlands\\
$^{3}$Department of Physics and Astronomy, Washington State University, Pullman, WA 99164, USA
}

\date{Accepted XXX. Received YYY; in original form ZZZ}

\pubyear{\the\year{}}

\begin{document}
\label{firstpage}
\pagerange{\pageref{firstpage}--\pageref{lastpage}}
\maketitle

\begin{abstract}
The origin of the stellar-mass binary black holes (BBHs) observed in gravitational wave (GW) events by the LIGO-Virgo-KAGRA (LVK) detectors remains uncertain. 
One promising formation pathway is the active galactic nucleus (AGN) channel, in which BBHs embedded in AGN accretion discs experience enhanced merger rates and may produce rapidly spinning black holes in the upper mass gap.
Previous constraints on the AGN contribution to the LVK merger population have relied either on spatial correlations between GW localisations and AGN catalogues or on modelling the intrinsic properties of GW sources. 
We develop a hierarchical Bayesian framework that unifies these approaches within a single likelihood for the fraction of AGN-hosted GW events, $\fagn$. 
This framework, combined with the latest gravitational wave transient catalogue GWTC-5.0 and the electromagnetic Quaia AGN catalogue, cannot exclude an AGN contribution to the observed merger rate.
For unobscured AGN in the optical and infrared bands with bolometric luminosities $L_{\rm bol}\geq\{10^{44.5},10^{45.5},10^{46.5}\}\,\mathrm{erg\,s^{-1}}$, we exclude $\fagn \geq \{0.52,0.41, 0.61\}$ at the 95 percent credible level under our fiducial model. 
The methodology and results presented in this work lay down the pathway for combining the full GW posterior with electromagnetic observations to infer the properties of GW populations and AGN discs.

\end{abstract}

\begin{keywords}
gravitational waves -- transients: black hole mergers -- galaxies: active -- methods: statistical
\end{keywords}

\section{Introduction}\label{sect:intro}

The second part of the fourth observing run (O4b) of the LIGO-Virgo-KAGRA (LVK) gravitational wave (GW) detectors has increased the number of observed GW events to 390 \citep{Abbott:2021gwtc2, Abbott:2023gwtc3, Abbott:2024gwtc21, ligo:2025gwtc40, lvk:2026gwtc5}.
These GW signals are produced during the mergers of compact objects, predominantly stellar-mass black holes (BHs). However, the physical processes by which BHs form binaries and produce observable GWs remain debated.

Binary black hole (BBH) formation channels can be split into two broad categories: isolated binary evolution and dynamical formation. In isolated binary evolution models, stellar evolution processes are dominant in generating the properties of BBHs.  Dynamical formation models explain BBH properties through gravitational interactions in a dynamically active environment. Each of these categories has a variety of subchannels (see \citealt{Mapelli:2021, Mandel:2022rev} for reviews). We focus on the active galactic nucleus (AGN) channel: a dynamical subchannel in which the hierarchical merger of BBHs is aided by an AGN accretion disc (see \citealt{Ford:2025rev} for a review).

The AGN channel is of interest because it naturally explains rapidly spinning BHs in the upper mass gap, such as those that produced the event GW231123 \citep{LIGO:2025gwm, Li:2025gwm, Delfavero:2025ldk}. Additionally, mergers in AGN discs are expected to produce electromagnetic (EM) counterparts to the GW signal, providing a unique opportunity for multimessenger astrophysics \citep{McKernan:2019ctp, Tagawa:2023ctp}. Moreover, the efficiency of AGN-assisted BH mergers probes unconstrained properties of AGN accretion discs, such as their gas density, size, lifetime \citep{Ford:2025rev, Tagawa:2026agn}, geometry \citep{Tomar:2026adk}, and viscosity \citep{Vaccaro:2026agn}. Other factors that may influence the merger rate of this channel include the number density of BHs in the surrounding nuclear star cluster \citep{Ford:2025rev}, the Eddington ratio, and the supermassive BH mass \citep{Yang:2019q, Vaccaro:2026agn}.

There are multiple ways in which the AGN channel explains the formation and merger of BHs and their binaries. The progenitor BHs could originate from stellar evolution in star-forming regions within the disc \citep{Stone:2017insitu, Gilbaum:2022, EpsMart:2025}, or from pre-existing (B)BHs in the nuclear star cluster that align their orbits with the disc \citep{Ostriker:1983, Syer:1991, Bartos:2017orb, Fabj:2020, Generozov:2023, Nasim:2023, Whitehead:2025}. Once in the disc, BHs form binaries through dynamical captures, aided by gas dynamical friction and accretion torques \citep{Tagawa:2020, boekholt:2023, delaurentiis:2023, Rowan:2023bbh, Rowan:2024cap, Rowan:2024gas, Whitehead:2024, Dodici:2024}. Binary formation may be especially efficient in migration traps: annular regions where the radial migration of BHs through the disc halts \citep{Paardekooper:2006, Paardekooper:2010}, increasing the number density of BHs and therefore the number of close gravitational encounters \citep{McKernan:2012tra, McKernan:2024fact, Bellovary:2016, Secunda:2019tra, Vaccaro:2026trap}. However, they may only exist in AGN within the limited bolometric luminosity range $L_{\rm bol} \in \qty[10^{43.5},10^{45.5}] \, \mathrm{erg \, s^{-1}}$ \citep{Gilbaum:2022, Grishin:2024, Gilbaum:2024}. Subsequently, disc-embedded BBHs harden and merge through gas dynamical friction \citep{Kim:2008, Vaccaro:2024}, circumbinary disc interactions (\citealt{Bartos:2017orb, Ishibashi:2024}, however, see also \citealt{Munoz:2019, Duffell:2020}), and binary-single interactions \citep{McKernan:2022spin, Rowan:2025gcb, Wang:2025bsi}.

Efforts have been made to observationally constrain the origin of GW events using three methods. First, GWs may have EM counterparts, localising the GW signal to its host galaxy. Several potential counterparts from AGN have been reported, including variability in the broad-line region \citep{Cabrera:2024em}, a gamma-ray burst \citep{Zhang:2025em}, and an optical flare \citep{Graham:2020em, Li:2025a}.

Second, even without an EM counterpart, spatial correlations between GW sky localisations and potential hosts, such as AGN, can constrain the fraction of GW events originating in AGN ($\fagn$) \citep{Bartos:2017abc, Veronesi:2022}. This approach is most sensitive to rare sources in highly complete catalogues, ruling out the most luminous AGN as dominant GW hosts \citep{Veronesi:2023, Veronesi:2025}. Similarly, \citet{Zhu:2025} report evidence for a non-zero contribution from lower-luminosity AGN. Alternatively, the clustering of sources can be used to constrain $\fagn$ \citep{Moncrieff:2025, Smith:2026, Bellomo:2026}, or the association between catalogues of EM flares and GW localisations can be tested \citep{Palmese:2021flare, Veronesi:2024, Cabrera:2026em}.

Third, intrinsic BBH parameters (e.g., chirp mass, mass ratio, or component spins) provide information about their formation channel \citep{Gayathri:2023bha, Li:2024, Bartos:2026spin}. In population studies, the observed BBH parameters are compared to model predictions for their underlying distributions (``population models'') associated with different formation channels, allowing inference of the most likely origin of the observed systems. These analyses rely on unique features in the population models to distinguish them. For example, BH masses in the upper mass gap can rule out isolated binary evolution channels, but are not sufficient to distinguish between dynamical subchannels. As such, gas-induced features in BBH population models are unique probes for the AGN channel. For instance, only sustained gas accretion can produce the highest BH spins ($\chi \gtrsim 0.9$) \citep{Kritos:2024}.
However, such signatures are expected to occur predominantly in higher-generation mergers, which involve BHs that have spent sufficient time embedded in the gaseous disc. If the AGN-channel merger rate is dominated by first-generation mergers (e.g., \citealt{McKernan:2024fact, Cook:2024}), the ability to distinguish it from other formation channels using intrinsic parameters alone is limited.

Previous works have used either the spatial or the intrinsic parameter population properties of BBHs to estimate the fraction of AGN-hosted GW events. In this work, we develop the formalism to combine both spatial and intrinsic information to place more robust constraints on $\fagn$, leveraging the full GW posterior while informed by an EM catalogue. 

To this end, we build on dark-siren methods for cosmological inference with GWs, in which the GW source redshift is constrained statistically using galaxy catalogues and population models \citep{Schutz:1986, Palmese:2025}. The \textit{galaxy catalogue} method assumes only galaxies are GW sources, such that redshift catalogues provide a prior on the GW source redshift \citep{Pozzo:2012, Fishbach:2019, Gray:2020, Gray:2022}. The population (or \textit{spectral siren}) method exploits redshifting features in the BH mass population to infer the GW redshift \citep{Chernoff:1993, Taylor:2012, Farr:2019, Ezquiaga:2022}. Crucially, these methods are expressions of the same underlying hierarchical likelihood, marginalised over different GW parameters. This enables, and requires, simultaneous inference of population models for all GW parameters \citep{Mastrogiovanni:2023icaro, Gray:2023}.

We extend this framework to multiple populations of merging BHs: one originating in AGN and another representing all alternative formation channels. We demonstrate our method in a simplified case in which the intrinsic parameters of both populations follow a single fixed distribution, an aspect ignored in previous works \citep{Veronesi:2023, Veronesi:2025, Zhu:2025}. Therefore, we distinguish the AGN-origin and alternative-origin GW populations based on their different spatial distributions, informing the likelihood with an AGN catalogue. This unifies the spatial-correlation method with population studies such as \citet{Gayathri:2023bha} and \citet{Bartos:2026spin}, although we defer the inference of intrinsic population hyperparameters to future work. We infer the fraction of GW events originating from several subpopulations of unobscured AGN, defined by bolometric luminosity thresholds. We improve on previous spatial-correlation analyses \citep{Veronesi:2023, Veronesi:2025, Zhu:2025} by expanding the likelihood to include GW selection effects, redshift-evolving merger rates, full AGN redshift posteriors, a spatially varying EM selection function, and by improved modelling of the AGN-hosted GWs that have no host detection. Moreover, in Appendix \ref{apx:V23} we show that we recover the \citet{Veronesi:2023} likelihood exactly when removing these additions, thereby proving EM-informed population inference has implicitly been applied in previous spatial-correlation analyses.

This paper is organised as follows. We first characterise the GW and AGN observations at our disposal (Sect. \ref{sect:data}), and then we derive the likelihood that we apply to this data, specifying our modelling choices (Sect. \ref{sect:method}). In Sect. \ref{sect:mockdata}, we use mock data to showcase the potential of the EM-informed population inference method for constraining $\fagn$. The results of the application to real data are shown in Sect. \ref{sect:results} and discussed in Sect. \ref{sect:discussion}. We conclude and summarise our work in Sect. \ref{sect:conclusions}. 

Throughout this work, we adopt the \texttt{Planck15} cosmology \citep{Planck15} from the \texttt{astropy} library \citep{Astropy:2022} with $H_{0} = 67.9 \, \mathrm{km \, s^{-1} \, Mpc^{-1}}$ and $\Omega_{\rm m} = 0.3065$, consistent with the parameter estimation of the LVK Collaboration \citep{Abbott:2021gwtc2, Abbott:2024gwtc21, Abbott:2023gwtc3, ligo:2025gwtc40, lvk:2026gwtc5}.

\section{Data}\label{sect:data}

We combine GW and EM data in our analysis. In this section, we characterise the GW data (Sect. \ref{subsect:GWdata}) and the AGN catalogue (Sect. \ref{subsect:EMdata}) that we use to calibrate our mock data analysis (Sect. \ref{sect:mockdata}) and with which we infer $\fagn$ using real data (Sect. \ref{sect:results}).

\subsection{Gravitational wave data}\label{subsect:GWdata}

We analyse GW events from BBH mergers in the cumulative GW catalogue GWTC-5.0 \citep{lvk:2026gwtc5}, released by the Gravitational Wave Open Science Center \citep{GWOSC:2026}. We select our sample of 256 GW events by requiring a network signal-to-noise ratio (SNR) greater than 10 for events from O1 and O2, and a false alarm rate (FAR) below $1/\mathrm{year}$ for events from O3 and O4, resulting in a homogeneous sample as shown by \citet{Essick:2023sel}. We use posterior samples obtained with the \texttt{IMRPhenomXPHM} waveform model \citep{Pratten:2021} for events up to O3, and its updated version \texttt{IMRPhenomXPHM-SpinTaylor} \citep{Colleoni:2025xphm} for O4a and O4b events. All GW data are downloaded from the official Zenodo repositories of the LVK collaboration\footnote{
\hyperlink{https://zenodo.org/records/6513631}{GWTC-2.1},
\hyperlink{https://zenodo.org/records/8177023}{GWTC-3.0},
\hyperlink{https://zenodo.org/records/17014085}{GWTC-4.0},
\hyperlink{https://zenodo.org/records/20348005}{GWTC-5.0 (Part 1)},
\hyperlink{https://zenodo.org/records/20348006}{GWTC-5.0 (Part 2)},
\hyperlink{https://zenodo.org/records/20292639}{hyperparameter posteriors},
\hyperlink{https://zenodo.org/records/19500052}{Search-sensitivity estimates}
}.

The published posterior samples were obtained using joint parameter-estimation (PE) priors on luminosity distance and component masses that are uniform in volume for O1-O2 events and uniform in source frame for O3-O4 events. To ensure consistency with our mock-data analysis (Sect. \ref{sect:mockdata}), we reweight all posterior samples to a PE prior that is uniform in comoving volume using importance resampling. In addition, we reweight them to our reference population model for intrinsic GW parameters (see Sect. \ref{sect:method}). We plot the redshift errors of the reweighted GW events as a function of redshift and luminosity distance in Fig. \ref{fig:data}, along with the AGN data that we present in the next section.

Across our final GW sample, the median of the single-event median luminosity distances is $2.3_{-1.8}^{+6.0}$ Gpc, which is $0.41_{-0.30}^{+0.77}$ in redshift. The subscript and superscript indicate the 5th and 95th percentiles in our sample, respectively. Similarly, the median 1$\sigma$ relative luminosity distance error is $\Delta d_{L} / d_{L} = 0.28_{-0.11}^{+0.09}$, and the median 1$\sigma$ relative redshift error is $\Delta z / z = 0.23_{-0.08}^{+0.08}$. The median 90 percent credible level (CL) localisation volume is $\log_{10}(V_{90\%} \, [\mathrm{Mpc^{3}}]) = 9.1_{-2.6}^{+1.4}$, and the median 90 percent CL localisation area is $\Delta \Omega_{90\%} = 1142_{-1108}^{+10458} \, \mathrm{deg^2}$.

The reweighted posterior samples are subsequently converted into 3D multi-order sky maps using the \texttt{ligo-skymap-from-samples} function in the \texttt{ligo.skymap} Python library \citep{Singer:2016gtd, Singer:2016slm}. Multi-order sky maps partition the sky into pixels, with a finer resolution in regions of higher probability. Each pixel encodes the posterior distribution of the GW luminosity distance along its line of sight (LOS). The LOS distance posteriors are parametrised as a uniform-in-volume prior multiplied by a Gaussian likelihood\footnote{Although this parametrisation is inconsistent with our adopted luminosity-distance prior, it does not introduce a significant bias (see Fig. \ref{fig:ppplot}).}.

We account for GW selection effects in our likelihood using the cumulative search-sensitivity estimates released with GWTC-5.0 \citep{Essick:2021sel, Essick:2025sel}. This is an injection campaign that provides semianalytic SNR estimates for O1 and O2, and real injections for O3 and O4. 

\subsection{Electromagnetic data}\label{subsect:EMdata}

\subsubsection{The AGN catalogue: Quaia}

We inform the GW likelihood with the AGN catalogue Quaia \citep{Quaia:2024}. Quaia is the cross-section between the catalogue of quasar candidates in Gaia DR3 \citep{GaiaDR3} and the unWISE infrared AGN catalogue \citep{Lang:2014wise, Meisner:2019wise}, with additional selections in colour and proper motion to increase the purity of the quasar sample. The redshifts of the resulting quasar sample have been estimated using a $k$-Nearest Neighbours model. The training dataset consists of the Gaia photometry and SDSS spectroscopic redshifts for the AGN that are both in Quaia and in SDSS DR16Q \citep{Lyke:2020}. The resulting catalogue contains spectro-photometric redshift estimates for 1,295,502 quasar candidates with a Gaia G-band magnitude below 20.5.

Quaia is ideal for GW studies because of its all-sky footprint and high redshift completeness (see Sect. \ref{sect:completeness}), which increases catalogue support in all GW localisation volumes. Additionally, its sky-uniformity allows for a simplified analysis where the redshift-dependent AGN selection function is approximated as constant across the entire sky, with the Galactic plane as the only exception.

\begin{figure}
    \centering
    \includegraphics[width=\linewidth]{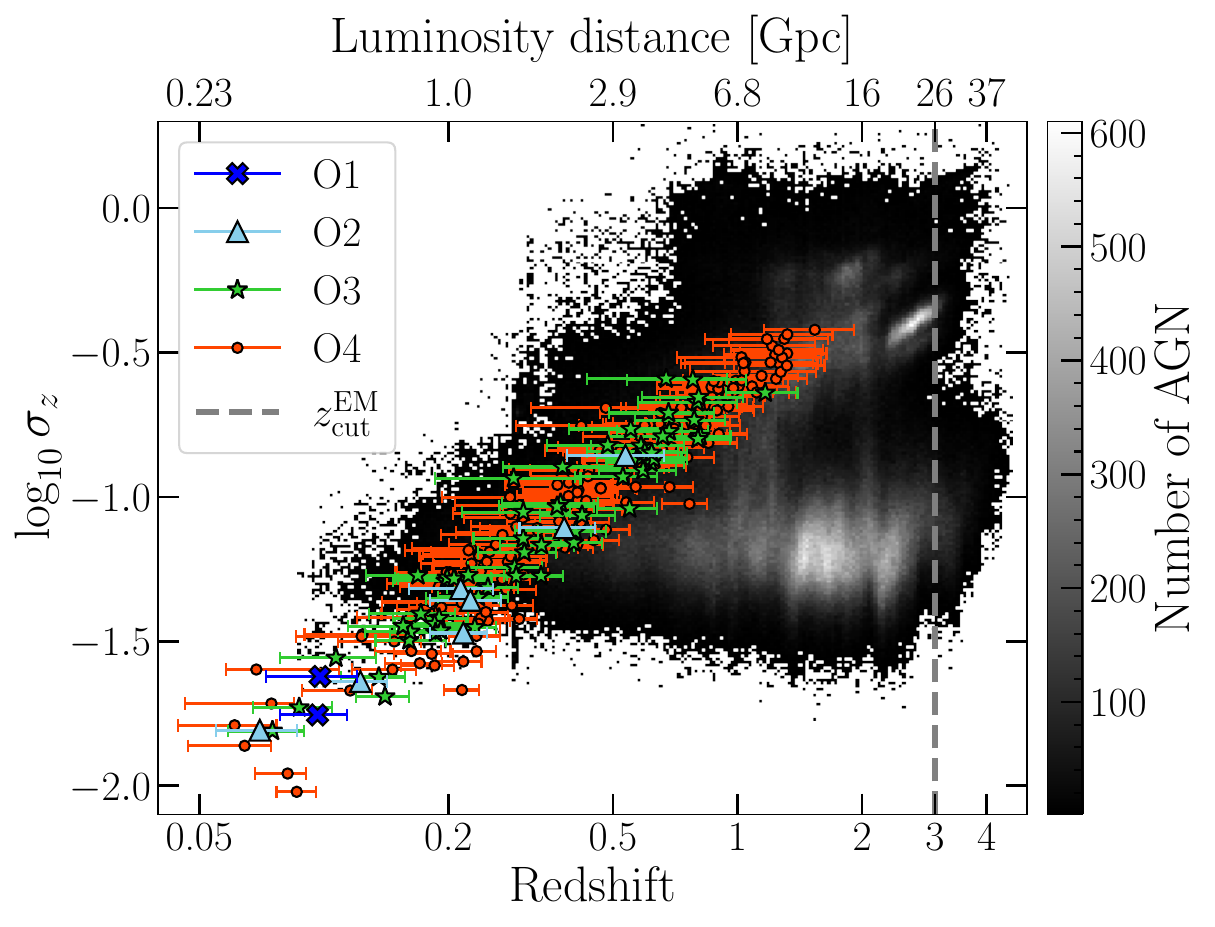}
    \caption{The redshift error $\sigma_z$ as a function of redshift $z$ of gravitational wave (GW) and AGN observations on a log-log scale. The AGN catalogue, Quaia, is shown as a 2D histogram (grey scale). In our analysis, we cut away all AGN above $z_{\rm cut}^{\rm EM} = 3$. We show the redshift measurements from the different GW observing runs (O1-O4) as color coded error bars. The horizontal error bars show the 16th and 84th posterior percentiles, while the vertical axis shows the half-difference of these.
    }
    \label{fig:data}
\end{figure}

We show the distribution of Quaia redshift estimates and of their uncertainties as a 2D histogram in Fig. \ref{fig:data}. For the majority of AGN, the redshift uncertainty remains approximately constant with distance, at $\sigma_{z,\rm agn} \approx 0.06$. The median redshift uncertainty across the full catalogue is $\sigma_{z,\rm agn} \approx 0.09$. Above $z \gtrsim 0.5$, the GW redshift uncertainties are larger than those of most AGN within their localisation volumes. 
At lower redshifts, the uncertainties become comparable, with the AGN uncertainties even exceeding those of the GWs below $z \lesssim 0.2$. Therefore, AGN are not perfectly localised in redshift, and cannot be treated as point sources as assumed in \citet{Veronesi:2025}.
In Appendix \ref{subsubsect:neglect_err} we show that neglecting AGN redshift uncertainties in EM-informed population inference biases $\fagn$ downwards.

\subsubsection{Selecting AGN subpopulations in Quaia}

Our goal is to infer the fraction of GW events originating from AGN. Since all AGN catalogues are incomplete due to survey limits and dust obscuration, we focus on AGN subpopulations for which the catalogue completeness is high enough to be informative. For example, optical and infrared surveys are expected to detect most luminous, unobscured AGN. In this case, $\fagn$ denotes the fraction of GW events originating from that AGN subpopulation.

Here, we consider the subpopulation of AGN that is unobscured in the optical and infrared, such that they could have been observed by both Gaia and WISE. We divide this selection of AGN into smaller subpopulations, characterised by a minimum bolometric luminosity. The thresholds we choose are $\log_{10} (L_{\rm bol}^{\rm thresh} [\mathrm{erg \, s^{-1}}]) \in \{44.5, 45.5, 46.5\}$, in line with previous spatial-correlation studies \citep{Veronesi:2023, Veronesi:2025}. We make this selection based on the mean bolometric luminosity of each AGN. This classification is imperfect due to the uncertainty in the bolometric luminosity estimation. The consequence is that some AGN will have a true bolometric luminosity that is lower than the threshold we set, contaminating the subcatalogue. We do not expect this to influence our results, since it mainly affects a minority of AGN near the threshold luminosity. AGN that have a true luminosity above the threshold but do not enter the subcatalogue result in a lower completeness of the subcatalogue without introducing a bias. 

The method for calculating AGN bolometric luminosities from Gaia $G_{\rm RP}$ magnitudes is laid out in Sect. 2.3.1. of \citet{Veronesi:2025}, which we modify by adopting a refitted slope of the mean AGN spectral energy distribution up to $z=3$ of 0.623. We compare these bolometric luminosity estimates with those calculated by \citet{WuShen:2022sdss} for AGN in SDSS DR16Q. On average, the Quaia bolometric luminosities are overestimated by 0.1 dex relative to the SDSS estimates, with the discrepancy depending on both redshift and bolometric luminosity. To correct for this offset, we fit a plane to the difference between the logarithm of the bolometric luminosity estimates from Quaia and SDSS as a function of the Quaia bolometric luminosity and redshift. We obtain $\log_{10} L_{\rm bol}^{\rm SDSS} - \log_{10} L_{\rm bol}^{\rm Quaia} = - 0.0005 \log_{10} L_{\rm bol}^{\rm Quaia} - 0.0427 z^{\rm Quaia} - 0.0205$. We therefore calculate the corrected luminosities as $\log_{10} L_{\rm bol}^{\rm corr} = 0.9995 \log_{10} L_{\rm bol}^{\rm Quaia} - 0.0427 z^{\rm Quaia} - 0.0205$. The difference distribution between the corrected luminosities and those of the \citet{WuShen:2022sdss} sample has a mean of zero and a standard deviation of 0.202 dex.

Only AGN that spatially coincide with GW localisations inform our likelihood, as we will show explicitly in Sect. \ref{sect:mockdata}. Since no GW event in our sample has posterior support above $z=3$, we select the 1,259,043 Quaia sources with a mean redshift of $z_{\rm cut}^{\rm EM} \leq 3$. For simplicity in modelling the catalogue completeness, we remove the sources with Galactic latitudes $|b| \leq 10 \degree$, approximately corresponding to the Galactic plane. This leaves 1,243,157 objects. Of those, we select subpopulations based on the bolometric luminosity. We obtain 1,243,143 AGN brighter than $\lbol{44.5}$, of which 1,084,513 exceed $\lbol{45.5}$ and 198,402 have $\lbol{46.5}$.

\subsubsection{Completeness estimation}\label{sect:completeness}

\begin{figure}
    \centering
    \includegraphics[width=\linewidth]{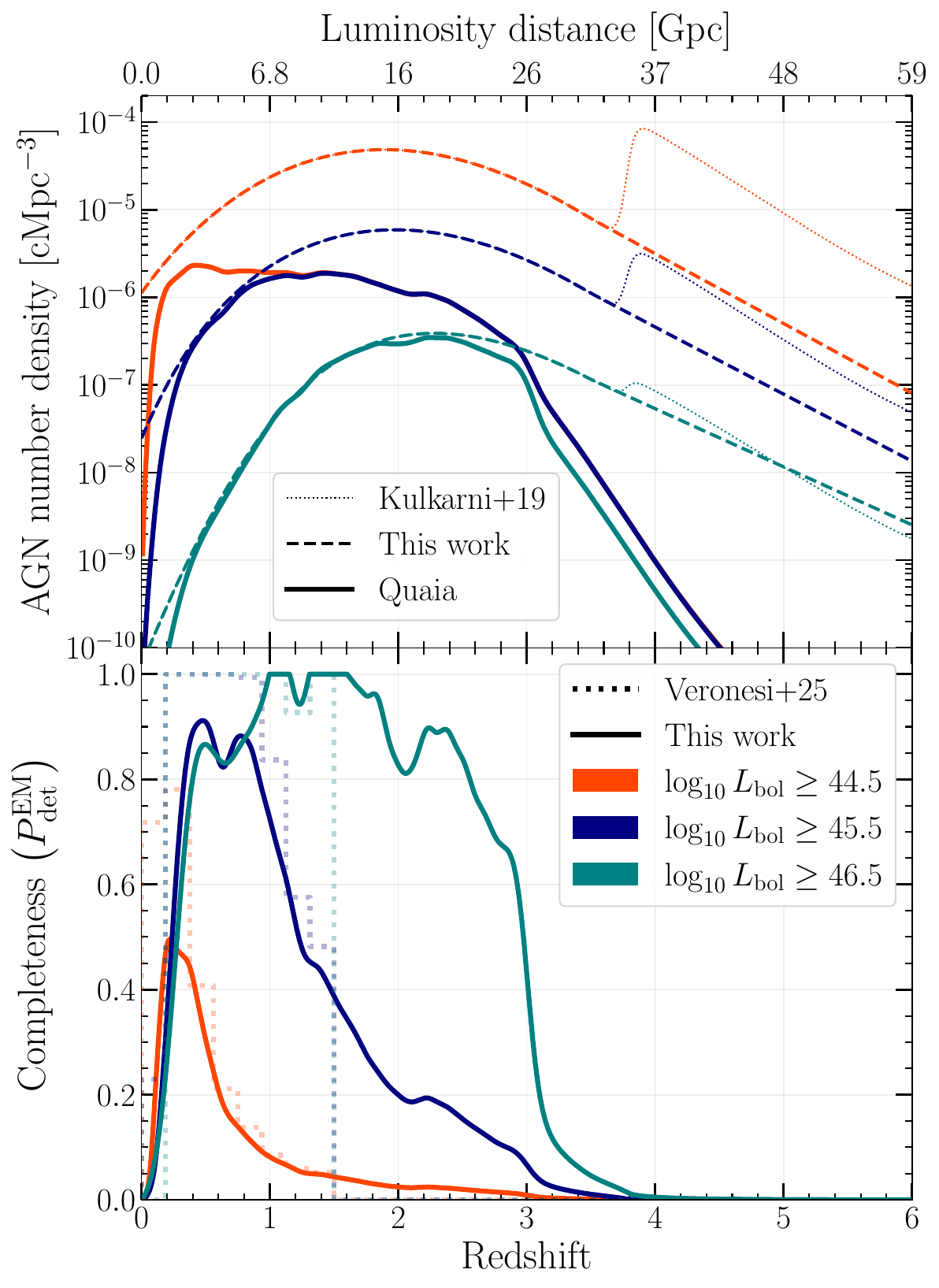}
    \caption{
    \textit{Upper:} Comoving AGN number density as a function of redshift. Colors correspond to different subcatalogues of Quaia, selected using the indicated bolometric luminosity threshold in $\mathrm{erg \, s^{-1}}$. We show the expected number density evolution derived from the quasar luminosity function (QLF) of \citet{kulkarni:2019} (dotted lines) and from the adjusted QLF used in this work (dashed lines). Solid lines show the observed number densities of AGN in the Quaia catalogue that have a mean observed redshift below $z=3$. 
    \textit{Lower}: Completeness of each subcatalogue as a function of redshift, given by the ratio of the observed and expected AGN number densities. Dotted lines show the completeness obtained by \citet{Veronesi:2025}. Completeness values are capped at 1.
    }
    \label{fig:completeness}
\end{figure}

The completeness, or detection probability $P_{\rm det}^{\rm EM}(z, \Omega)$, is the probability that a potential GW host at redshift $z$ and sky position $\Omega$ is in the EM catalogue, marginalised over the luminosity of the host. We consider all observed AGN potential GW hosts. 

We assume AGN to be uniformly distributed over the sky, and that the completeness of Quaia is constant everywhere but in the Galactic plane, where we set $P_{\rm det}^{\rm EM}(z, \Omega) = 0$. Outside the Galactic plane, we approximate $P_{\rm det}^{\rm EM}(z, \Omega)$ as the ratio of the observed AGN number density ($n_{\rm obs}$) to the expected AGN number density ($n_{\rm exp}$):
\begin{equation}\label{eq:fc_principle}
    P_{\rm det}^{\rm EM}(z, \Omega) = \frac{n_{\rm obs}(z)}{n_{\rm exp}(z)} \, .
\end{equation}

The expected AGN number density is given by a quasar luminosity function (QLF). In particular, we use the Model 1 QLF as a function of redshift and absolute magnitude at $1450\text{\AA}$, $\Phi(z, M_{1450})$, from \citet{kulkarni:2019}\footnote{They assume a flat cosmology with ($\Omega_{\rm m}$, $\Omega_{\Lambda}$) = (0.3, 0.7) and $H_{0} = 70 \, \mathrm{km \, s^{-1} \, Mpc^{-1}}$. We do not expect this difference with our cosmology to significantly impact our results.}. 
We integrate the QLF over the absolute magnitude range that defines the AGN subpopulation to obtain the number density: 
\begin{equation}\label{eq:n_exp_z}
    n_{\rm exp}(z) = \int_{M_{1450} \leq M_{1450}^{\rm thresh}} \dd M_{1450} \, \Phi(z, M_{1450}) \, ,
\end{equation}
where $M_{1450}^{\rm thresh}$ is the absolute magnitude corresponding to $L_{\rm bol}^{\rm thresh}$, calculated using the bolometric corrections of \citet{Runnoe:2012}. 

This QLF is suited to the AGN in Quaia, as it describes the population of optically unobscured AGN. However, this QLF exhibits a known unphysical peak at $z \approx 4$ and for $z > 7.5$ the model diverges. Therefore, we replace the magnitude-integrated QLF above a redshift $z = 3.5$ by an exponential decay of the form $y_{3.5} \exp[a (z - 3.5)]$, fitted in the range $3 \leq z \leq 3.5$. Here, $y_{3.5}$ is the PDF value at $z=3.5$, ensuring continuity at this point, and $a$ is a fit parameter. The AGN number densities before and after this adjustment are shown in the top panel of Fig. \ref{fig:completeness}.

We construct the observed AGN number density from the full AGN redshift posteriors. The redshift posterior of AGN $k$ is calculated as
\begin{align}\label{eq:agn_zpost}
    p_{\rm agn}(z | \hat{z}_{k}, \hat{L}_{\mathrm{bol}, k}) = \frac{p_{\rm agn}(\hat{z}_{k} | z) \pi_{\rm agn}(z | \hat{L}_{\mathrm{bol}, k})}{\int_{z_{\rm min}}^{z_{\rm max}} \dd z \, p_{\rm agn}(\hat{z}_{k} | z) \pi_{\rm agn}(z | \hat{L}_{\mathrm{bol}, k})} \, ,
\end{align}
where measured quantities are indicated with a hat and we neglect the uncertainty in the measured bolometric luminosity. The normalisation is between $z_{\rm min} = 10^{-6}$ and $z_{\rm max} = 10$. As mentioned in Sect. \ref{subsect:EMdata}, the AGN redshift errors in Quaia do not increase with redshift (see Fig. \ref{fig:data}). We therefore model the redshift likelihood $p_{\rm agn}(\hat{z}_{k} | z)$ as a Gaussian centred on the observed redshift $\hat{z}_{k}$, with a redshift-independent uncertainty $\sigma^{k}_{z, \mathrm{agn}}$ for each AGN $k$. The prior is calculated by multiplying the expected number density of AGN that have a bolometric luminosity of $\hat{L}_{\mathrm{bol}, k}$ by the comoving volume element,
\begin{equation}\label{eq:agn_zprior}
    \pi_{\rm agn}(z | \hat{L}_{\mathrm{bol}, k}) \propto \Phi\left(z, M_{1450}(\hat{L}_{\mathrm{bol}, k}) \right)\frac{\dd V_{\rm c}}{\dd z \dd \Omega} \, .
\end{equation}
The redshift prior on all AGN ($\pi_{\rm agn}(z)$) is obtained by integrating this equation over absolute magnitude, analogous to Eq. \ref{eq:n_exp_z}. The total number of observed AGN between redshifts $z$ and $z + \dd z$ is then given by $\dd z \sum_{k}^{N_{\rm agn}} p_{\rm agn}(z | \hat{z}_{k}, \hat{L}_{\mathrm{bol}, k})$. We convert this to a comoving number density by dividing by the comoving volume element and the sky area covered by the survey. The coverage fraction of Quaia is $f_{\rm cover} = 1 - \sin(10\degree) \approx 0.826$. In summary, we calculate
\begin{equation}
    n_{\rm obs}(z) = 
    \qty(4\pi f_{\rm cover} \dv{V_{\rm c}}{z \dd \Omega})^{-1} \sum_{k}^{N_{\rm agn}} p_{\rm agn}(z | \hat{z}_{k}, \hat{L}_{\mathrm{bol}, k}) \,
    .
\end{equation}

The number densities and completenesses we obtain are shown in Fig. \ref{fig:completeness}. The top panel shows the expected and observed comoving AGN number density for each subcatalogue. The ratio of these quantities is the completeness, which is plotted in the bottom panel. We cap the completeness at 1. For comparison, we also show the completeness calculated by \citet{Veronesi:2025}, who applied a catalogue cut at $z_{\rm cut}^{\rm EM} = 1.5$. We improve on their analysis by raising this cut to $z_{\rm cut}^{\rm EM} = 3$, and by incorporating the full AGN redshift posteriors in the completeness estimation. Our completeness matches \citet{Veronesi:2025} well up to $z=1.5$. By summing the full AGN redshift posteriors, we obtain non-zero completeness above $z=3$. This is because an AGN observed at, for example, $z=2.9$ with an error of $\sigma_{z, \rm agn} = 0.2$ has a significant chance of truly being located above $z=3$.

Our completeness estimation is most reliable in regions where the AGN number density is high. In low-density regions, Poissonian fluctuations may cause the completeness to be strongly over- or underestimated. This can be seen at redshifts $z < 0.2$ in Fig. \ref{fig:completeness}, where the completeness drops to zero due to the absence of AGN, even though the detection probability should approach unity towards $z = 0$ in a flux-limited catalogue. However, as explained in more detail in Appendix \ref{subsubsect:agn_selfunc}, the absolute error introduced in our likelihood (see Sect. \ref{sect:method}) is small, as the selection function is suppressed by $\pi_{\rm agn}(z)$ in such low-density regions.

\section{Method}\label{sect:method}

To infer the fraction of AGN-hosted GW events we adopt the population inference framework based on hierarchical Bayesian models widely used in the LVK collaboration \citep{Loredo:2004nn,Thrane:2019,Mandel:2018mve, Vitale:2020aaz}.
The goal of population inference is to estimate the hyperparameters $\Lambda$ describing population properties of GW events of astrophysical origin. To infer these hyperparameters from a GW dataset $D = \{d_i\}$ containing $N_{\rm GW}$ detected events we use the likelihood \citep{Mandel:2018mve, Vitale:2020aaz}:
\begin{align}\label{eq:hierarchical_llh}
    p(D|\Lambda) \propto \prod_{i=1}^{N_{\rm GW}} \frac{1}{\alpha(\Lambda)} \int \dd \theta \, p_{\rm GW}(d_i|\theta) \, p_{\rm pop}(\theta | \Lambda) \, ,
\end{align}
where $p_{\rm GW}(d|\theta)$ is the single-event GW likelihood describing the relation between the data and the single-event parameters $\theta$ (e.g., BBH component masses and luminosity distance), and $p_{\rm pop}(\theta | \Lambda)$ is the population prior from which single-event parameters are drawn. The detection efficiency $\alpha(\Lambda)$ takes the observational limitations of GW detectors into account to avoid Malmquist bias. 

The detection efficiency is defined as the expected fraction of detectable GW events generated from a population characterised by population parameters $\Lambda$ \citep{Mandel:2018mve}:
\begin{equation}\label{eq:detection_efficiency}
    \alpha(\Lambda) \equiv \int \dd \theta \, P_{\rm det}^{\rm GW}(\theta) \, p_{\rm pop}(\theta | \Lambda) \, ,
\end{equation}
where $P_{\rm det}^{\rm GW}(\theta)$ is the fraction of detectable data realisations that can result from the true event parameters $\theta$, which depends on the choice of a detectability threshold. As stated in Sect. \ref{subsect:GWdata}, we apply SNR and FAR cuts in our analysis of real GW data (Sect. \ref{sect:results}), and we estimate the detection efficiency using the official LVK injection campaign (see Appendix C1 of \citealt{Essick:2025sel} for details). However, for our mock data analysis (Sect. \ref{sect:mockdata}) we instead adopt a redshift threshold to define GW detectability, and we compute Eq. \ref{eq:detection_efficiency} numerically with our simulated $P_{\rm det}^{\rm GW}(z)$. 
This allows us to efficiently compute the detection efficiency for a large number of mock datasets without detailed SNR calculations, while still capturing the main impact of GW selection effects on our inference.

Since a GW event is either generated through the AGN channel ($A$) or through an alternative channel ($\conj{A}$), our population prior is a mixture model of these AGN-origin and alternative-origin GW populations. The fraction of GW events coming from the AGN channel is the population parameter $\fagn$, which we keep separate from $\Lambda$ for clarity:
\begin{align}\label{eq:ppop_full}
    p_{\rm pop}(\theta|\Lambda, \fagn) 
    &= \fagn p_{\rm pop}(\theta | \Lambda, A) + (1 - \fagn) p_{\rm pop}(\theta | \Lambda, \conj{A}) \, .
\end{align}

We split $\theta$ into spatial ($z, \Omega$) and intrinsic parameters ($\theta'$), such that $\theta = \{z, \Omega, \theta'\}$. We assume that the population of $\theta'$ is independent of redshift and sky position, and that this population is given by an identical reference distribution for both the AGN-origin and alternative-origin GWs, denoted $p_{\rm pop}^{\rm ref}(\theta'|\Lambda)$ \citep{Bartos:2026spin}. The population prior for GWs generated by channel $a \in \{A, \conj{A}\}$ is then given by
\begin{align}
p_{\rm pop}(\theta|\Lambda,a)
= p_{\rm pop}^{\rm ref}(\theta'|\Lambda)
\,p_{\rm pop}(z,\Omega|\Lambda,a) \, .
\end{align}
If the reference distribution is not specified, it is implicitly assumed to be identical to the LVK PE priors \citep{lvk:2026gwtc5}. This is the case for previous spatial-correlation analyses that ignored intrinsic GW parameters \citep{Veronesi:2023, Veronesi:2025, Zhu:2025}. However, since the LVK PE priors are not informed by the astrophysical population, adopting them may bias GW analyses \citep{Mandel:2026pe, Mould:2026pe}. Since the source-frame BBH component masses are strongly degenerate with the GW redshift, they are the most important intrinsic parameters in our analysis. We adopt a \texttt{Broken Powerlaw + 2 Peaks} parametrisation for the BBH mass distribution, evaluated at the maximum-a-posteriori (MAP) hyperparameter values inferred by LVK under their \texttt{Default BBH} model with a \texttt{Madau-Dickinson} rate evolution with redshift \citep{ligo:gwtc5pop}. All remaining parameter models are taken to be identical to the LVK PE priors. We discuss these modelling choices in Sect. \ref{subsect:AGNchoices}.

The spatial population priors are derived in detail in Appendix \ref{subsect:zpops}. Here, we review the final expressions. The population prior on the redshift and sky position of AGN-origin GWs is given by
\begin{align}\label{eq:pagn}
p_{\rm pop}(z,\Omega|&\Lambda,A)
\propto \frac{1}{1+z} 
\Bigg[ 
\bigl(1-P_{\rm det}^{\rm EM}(z,\Omega)\bigr) \frac{\pi_{\rm agn}(z)}{4\pi} \\
&+ \left \langle P_{\rm det}^{\rm EM} \right \rangle \frac{1}{N_{\rm agn}} \sum_{k=1}^{N_{\rm agn}} p_{\rm agn}(z|\hat z_k,\hat L_{{\rm bol},k}) \delta(\Omega-\hat\Omega_k) \nonumber
\Bigg] \, .
\end{align}
There are two terms inside the brackets, representing the spatial GW priors when the host AGN is, or is not observed. The probability of detecting a host AGN, or equivalently, the completeness of the AGN catalogue enters as the function $P_{\rm det}^{\rm EM}(z, \Omega)$ (see Sect. \ref{sect:completeness}), and its spatial average $\left \langle P_{\rm det}^{\rm EM} \right \rangle$. If we have observed the host AGN, the AGN positions in an EM catalogue constitute a prior from which the GW origin is drawn. We construct this prior by summing up the spatial posteriors of all $N_{\rm agn}$ sources, assuming all AGN to have an equal probability of hosting a GW event in their source frame. This assumption is discussed in Sect. \ref{subsect:AGNchoices}. The AGN redshift posteriors are given by Eq. \ref{eq:agn_zpost}, and we assume that the AGN sky-position error is negligible compared to the GW localisation areas. If the host AGN is not in our catalogue, it will be a random draw from the spatial prior on all AGN. We calculate this prior from the observed QLF (see Sect. \ref{sect:completeness}), neglecting clustering. The $1/(1+z)$ prefactor accounts for the time dilation between the source and detector frames, and it is the reason we numerically normalise the population prior in redshift.

The alternative-origin GW population prior is given by
\begin{align}\label{eq:pnonagn}
p_{\rm pop}(z,\Omega|\Lambda,\conj{A})
\propto \frac{1}{4\pi} \frac{\psi(z|\Lambda)}{1+z} \frac{\dd V_{\rm c}}{\dd z} \, ,
\end{align}
representing GW hosts that are distributed uniformly in comoving volume ($V_{\rm c}$), producing GWs at a rate that evolves in redshift according to the Madau-Dickinson cosmic star-formation rate (SFR) parametrisation \citep{Madau:2014},
\begin{equation}
    \psi(z | \Lambda) = \frac{(1 + z)^{\gamma}}{1 + \left(\frac{1 + z}{1 + z_{\rm peak}}\right)^{\kappa}} \, ,
\end{equation}
where $\Lambda = \{\gamma, z_{\rm peak}, \kappa\}$ are hyperparameters. Our fiducial model has $\{\gamma, 1 + z_{\rm peak}, \kappa\} = \{3.3, 2.55, 6.1\}$, given by the fit of \citet{Strolger:2020}. Since BBH formation is more efficient at lower metallicities \citep{vanSon:2025Z} and there is a time delay between BBH formation and merger \citep{Fishbach:2021tdel, Padhyegurjar:2026dtd}, a common alternative choice for the rate evolution $\psi(z | \Lambda)$ is the low-metallicity SFR convolved with a time-delay distribution (e.g., \citealt{Fishbach:2018lowZ, Yang:2020zevo}). However, our simplistic model is consistent with the measured BBH merger rate in GWTC-5.0 \citep{ligo:gwtc5pop}. Still, an even simpler model of a single power law, $\propto (1 + z)^{\gamma}$ with $\gamma = 2.5_{-0.7}^{+0.7}$, is favoured with a log-Bayes factor of 0.15. Therefore, the disfavoured complexity we add by including the hyperparameters $\fagn$, $z_{\rm peak}$, and $\kappa$ is reflected in our results.

Filling Eq. \ref{eq:ppop_full} into Eq. \ref{eq:hierarchical_llh}, and defining the evidence for channel $a$ generating the GW event $i$ as
\begin{equation}\label{eq:hypothesis_evidence}
    Z_{a}(d_i|\Lambda) \equiv \int \dd \theta \, p_{\rm GW}(d_i|\theta) p_{\rm pop}(\theta | \Lambda, a) \, ,
\end{equation}
we obtain the hyperlikelihood:
\begin{align}\label{eq:llh_full}
    p(D | \Lambda, \fagn) \propto \prod_{i=1}^{N_{\rm GW}} \frac{\fagn Z_{A}(d_i | \Lambda) + \qty(1-\fagn) Z_{\conj{A}}(d_i | \Lambda)}{\alpha(\Lambda, \fagn)} 
     \, .
\end{align}
Here we recognise the functional form of the likelihood presented in \citet{Veronesi:2023}, now including GW selection effects through $\alpha$. We further improve on their method by incorporating redshift-dependent merger rates, full AGN redshift posteriors instead of neglecting AGN redshift errors, an EM selection function that varies within the GW localisation volume, and a consistent modelling of the AGN distribution ($\pi_{\rm agn}$). In Appendix \ref{apx:V23} we show that we recover the \citet{Veronesi:2023} likelihood when removing these additions, thereby proving EM-informed population inference has implicitly been applied in previous spatial-correlation analyses. By adopting the same reference population for the non-spatial parameters of all GWs, we have eliminated the dependence of $\fagn$ on these parameters. We emphasise that this does not make the likelihood agnostic of the intrinsic parameter population, as the dependence on $\fagn$ is only lost under specific model assumptions on both the AGN- and alternative-origin GW parameters. By making these choices explicit, we unify the spatial-correlation method with population studies such as \citet{Gayathri:2023bha} and \citet{Bartos:2026spin}. However, we distinguish the GW populations only based on their different spatial distributions, and we defer the inference of intrinsic population hyperparameters to future work.

\begin{figure}
    \centering
    \includegraphics[width=\linewidth]{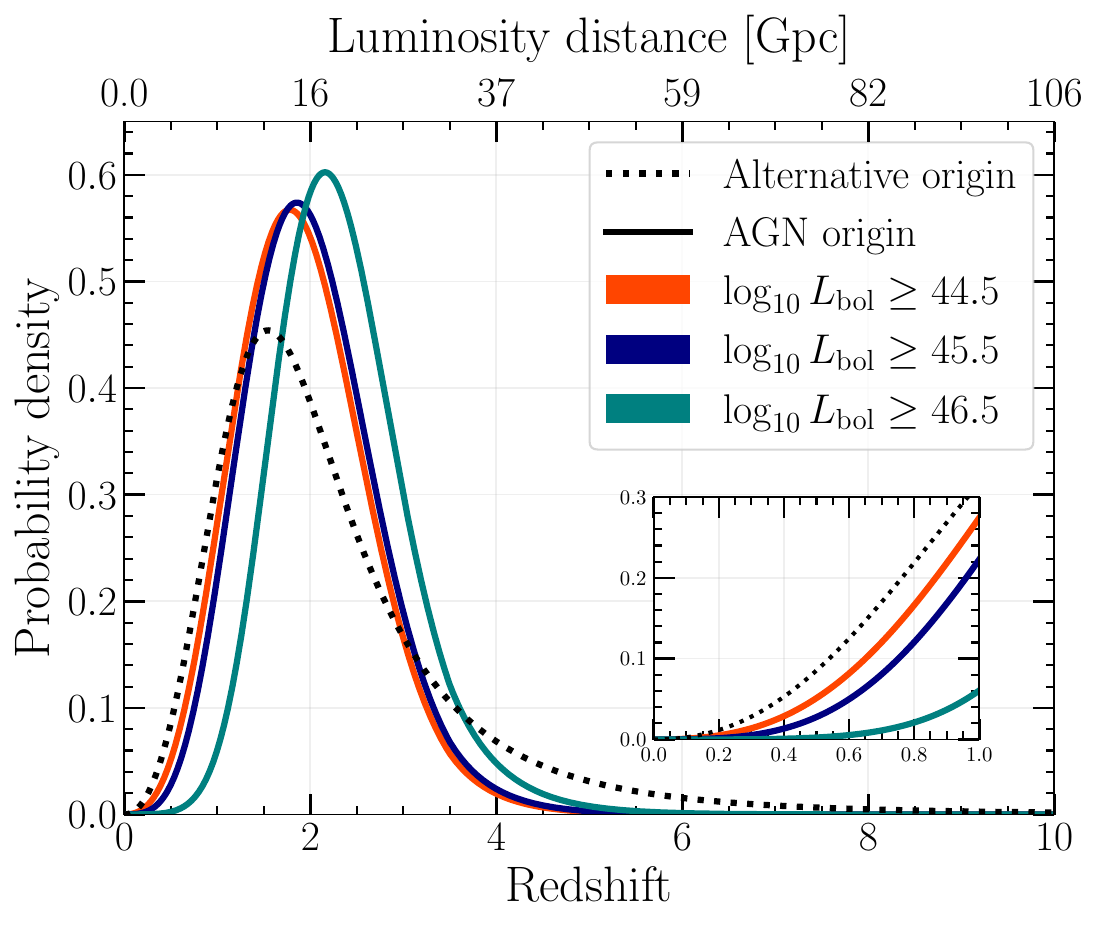}
    \caption{Redshift distributions of the two GW populations we consider: AGN-origin (solid lines, \ref{eq:pagn}) and alternative-origin GWs (dotted line, Eq. \ref{eq:pnonagn}). Different colors correspond to GWs generated by different subpopulations of unobscured AGN, characterised by bolometric luminosity thresholds in $\mathrm{erg \, s^{-1}}$. The inset shows a zoom on the redshift range $0 \leq z \leq 1$, where most GWs in our observed sample lie.
    }
    \label{fig:all_zpops}
\end{figure}

In Fig. \ref{fig:all_zpops}, we show our fiducial redshift model for the alternative-origin GWs (Eq. \ref{eq:pnonagn}) alongside the redshift distributions of AGN-channel mergers without observed hosts ($\propto \pi_{\rm agn}(z) / (1 + z)$, Eq. \ref{eq:pagn}) for several bolometric luminosity thresholds. The inset shows a zoom on the range that contains most GW observations (see Fig. \ref{fig:data}), and is therefore most informative in our analysis. Since the redshift distribution of AGN-hosted GWs is different from the alternative-origin GWs, it is possible to constrain $\fagn$ even without an AGN catalogue. In particular, more dissimilar populations are easier to distinguish. This can be seen from Eq. \ref{eq:llh_full}, where $\fagn$ is proportional to the difference in evidences.

Conversely, if the AGN-origin and alternative-origin GW populations have identical redshift distributions, information on $\fagn$ can only come from the spatial prior constructed from the AGN catalogue. Even for a perfect, complete catalogue with exact redshifts, however, its constraining power is fundamentally limited by the rarity of the AGN subpopulation. A denser catalogue more closely resembles the average redshift distribution, reducing its ability to distinguish the AGN-hosted GW population from the alternative.

Note that $\fagn$ depends on the maximum redshift at which GWs can be emitted ($z_{\rm max}$). Namely, variations in $z_{\rm max}$ affect the normalisation of our two GW populations differently, changing the alternative-origin and AGN-origin redshift models in Fig. \ref{fig:all_zpops} relative to each other. We set $z_{\rm max} = 10$ in the rest of this work, but we verified that this choice does not significantly impact our results for values in the range $5 \leq z_{\rm max} \leq 10$.

\section{Mock data analysis}\label{sect:mockdata}

We validate the likelihood derived in Sect. \ref{sect:method} by applying it to mock data. We show that our framework consistently recovers injected values of $\fagn$ (Sect. \ref{subsect:setup}). We then quantify the information gained from using AGN catalogues with varying completeness and redshift uncertainties (Sect. \ref{subsect:infogain}). In Appendix \ref{subsect:bias}, we use this mock data to quantify the biases that could result from common ways of inconsistently treating the EM data: neglecting AGN redshift errors and estimating the EM selection function from measured data.

\subsection{Validation and computational details}\label{subsect:setup}

We show that $\fagn$ is consistently recovered using the likelihood derived in Sect. \ref{sect:method}. To this end, we simulate 200 data realisations with an injected value of $\fagn$ that is randomly sampled from a uniform distribution between 0 and 1. A single data realisation consists of a set of GW sky maps (see Sect. \ref{subsubsect:mockgws}) and an AGN catalogue (see Sect. \ref{subsubsect:mockagn}). For computational efficiency, we only simulate the subpopulation of AGN brighter than $\lbol{46.5}$, which are the rarest.

Our mock data analysis only differs from our inference on real data (see Sect. \ref{sect:results}) in the calculation of the GW detection efficiency. Instead of adopting cuts in SNR and FAR, we set a threshold in redshift for GW detection. This allows us to capture the main impact of GW selection effects on our inference, without detailed modelling of intrinsic parameters, such as BBH masses, that are essential to calculate SNR estimates. Our mock GW selection function results in a sharper threshold in redshift than would result from an SNR cut. This is because high-redshifts GW events could still result in a detectable SNR if the BBH masses are large enough, blurring the detector horizon. This does not impact our conclusions.

\subsubsection{Mock GW observations}\label{subsubsect:mockgws}

Here we explain the generative process for a single realisation of a mock GW dataset. First, we sample a value of $\fagn$ uniformly between 0 and 1, and we choose the total number of GWs injected in the entire mock universe ($N^{\rm tot}_{\rm inj}$, Tab. \ref{tab:mockgw}). The realised number of GWs from AGN is sampled from a binomial distribution with $n = N^{\rm tot}_{\rm inj}$ and $p=\fagn$. All other GWs belong to the alternative-origin population.

The true coordinates of GWs from each population are sampled from their respective population prior. All sky positions are sampled uniformly on the sky. The true redshifts of alternative-origin GWs are sampled from Eq. \ref{eq:pnonagn} or the black dotted line in Fig. \ref{fig:all_zpops}. The true redshifts of AGN-origin GWs are sampled proportional to $\pi_{\rm agn}(z) / (1 + z)$. This is the first term in Eq. \ref{eq:pagn} and the teal solid line in Fig. \ref{fig:all_zpops}.

We give the observed GW coordinates a random offset from the true GW coordinates. The magnitude of this offset is set by the size of the localisation volume. We match the localisation volume distribution to real GW observations by sampling $V_{90\%}$ from the GWTC-4.0 distribution, which is statistically indistinguishable from that of GWTC-5.0. However, for simplicity, we do not make $V_{90\%}$ redshift dependent. We adopt a spherical Gaussian likelihood in comoving coordinates. The observed GW coordinates are sampled from a Maxwell-Boltzmann distribution centred on the true GW coordinates. The scale parameter ($\sigma$) of this distribution is calculated from its relation to the cumulative distribution function at the 90th percentile, which encloses the sampled volume $V_{90\%}$. This results in $\sigma = \left( \tfrac{3V_{90\%}}{ 4\pi} \right)^{1/3} \big / \sqrt{2 P^{-1}(3/2, \, 0.9)}$, where $P^{-1}$ denotes the inverse to the regularised lower incomplete gamma function.

To simulate GW selection effects, we consider a GW detected if the data realisation lies below the redshift detection threshold $z^{\rm GW}_{\rm thresh}$. Our mock data matches the GWTC-4.0 distribution of $V_{90\%}$ as a function of luminosity distance best when using $z^{\rm GW}_{\rm thresh} \approx 0.3$ with $N^{\rm tot}_{\rm inj} = 10^5$, which results in 160 detected GWs on average.

Finally, we generate posteriors on the true GW coordinates for the detected GW events. Posterior samples are drawn directly from the likelihood, thereby assuming a prior that is uniform in comoving volume. Posterior samples are converted into GW sky maps using the \texttt{ligo-skymap-from-samples} function in the \texttt{ligo.skymap} Python library \citep{Singer:2016gtd, Singer:2016slm}.

\subsubsection{Mock AGN observations}\label{subsubsect:mockagn}

We simulate observed catalogues of AGN with $\lbol{46.5}$. Each source is characterised by two sky coordinates and a redshift. We do not sample AGN luminosities, and we adopt a selection function in redshift and sky position only.

We start by generating a complete catalogue with true AGN coordinates in two steps. First, we add the true positions of the simulated (both detected and undetected) GWs from AGN to the catalogue. We then add AGN positions that are uncorrelated to GWs, which are treated as noise in our analysis. Sky positions are uniformly sampled on the celestial sphere, and redshifts are sampled in such a way that the total redshift distribution of AGN is $\pi_{\rm agn}(z)$. A complete catalogue contains $3.5 \cdot 10^5$ sources, corresponding to the volume-integrated QLF up to $z=10$ (see Sect. \ref{sect:completeness}).

We then sample observed AGN coordinates. The observed sky coordinates are set equal to their true values, reflecting our assumption of negligible sky-position errors. Observed redshifts are drawn from Gaussian likelihoods truncated to $10^{-6} \leq z \leq 10$, centred on the true redshifts, and with widths $\sigma_{z,\rm agn}$. Our simulated redshift errors are independent of the true redshift, consistent with most Quaia sources (Fig. \ref{fig:data}). We randomly sample $\sigma_{z,\rm agn}$ from the Quaia redshift-error distribution. The AGN redshift posteriors are then calculated with Eq. \ref{eq:agn_zpost}.

We match the completeness of our simulated catalogue to that of Quaia. We remove all AGN in the Galactic plane ($|b| \leq 10^\circ$) and those with observed redshifts above $z_{\rm cut}^{\rm EM}=1.5$. For computational efficiency, we do not choose $z_{\rm cut}^{\rm EM} = 3$ as done in our real-data analysis. We show in Sect. \ref{subsubsect:real_em_data} that this choice does not influence our results, since $z_{\rm cut}^{\rm EM} > z_{\rm thresh}^{\rm GW}$, where $z_{\rm thresh}^{\rm GW} = 0.3$. We divide the remaining sample into eight linear redshift bins over $0<z<1.5$ and randomly retain, in each bin, a fraction of AGN equal to the estimated completeness of \citet{Veronesi:2025}. In our analysis, we approximate the AGN selection function using the method described in Sect. \ref{sect:completeness}.

\subsubsection{Evaluating the likelihood}\label{subsubsect:eval_llh}

With the mock data generated, we can evaluate the hierarchical likelihood (Eq. \ref{eq:hierarchical_llh}). Using Bayes' theorem, we write the single-event GW likelihood as
\begin{equation}
    p_{\rm GW}(d | z, \Omega) \propto \frac{p_{\rm GW}(z, \Omega | d)}{\pi_{\rm PE}(z, \Omega)} \, ,
\end{equation}
where the PE prior $\pi_{\rm PE}(z, \Omega)$ is uniform in comoving volume and the posterior $p_{\rm GW}(z, \Omega | d)$ is provided by the sky maps.

We then compute the redshift population prior. Computation time from evaluating and summing many AGN redshift posteriors can be saved in multiple ways in this step of the process (see also \citealt{Gray:2022, Gray:2023}). For example, we pre-calculate the AGN redshift posteriors. Moreover, we marginalise over the LOSs in only the 99.9 percent CL sky area of each GW to compute the evidence for the AGN-origin hypothesis. We expect this to introduce a bias towards lower $\fagn$ that is only noticeable when using more than $\sim 10^3$ GWs. Note that the redshift prior still has to be normalised by dividing by the total number of AGN on the whole sky. We integrate over redshift on a linear grid between $z_{\rm min} = 10^{-6}$ and $z_{\rm max} = 10$, using the minimum number of grid points such that the difference between adjacent grid values is smaller than the smallest $1\sigma$ AGN or GW redshift error ($\approx 0.01$, see Fig. \ref{fig:data}), thus capturing all redshift information.

Finally, we correct for selection effects by evaluating Eq. \ref{eq:detection_efficiency}, where $P_{\rm det}^{\rm GW}(z)$ is determined from Monte Carlo simulations of the GW generation process described in Sect. \ref{subsubsect:mockgws}.

\begin{figure}
    \centering
    \includegraphics[width=\linewidth]{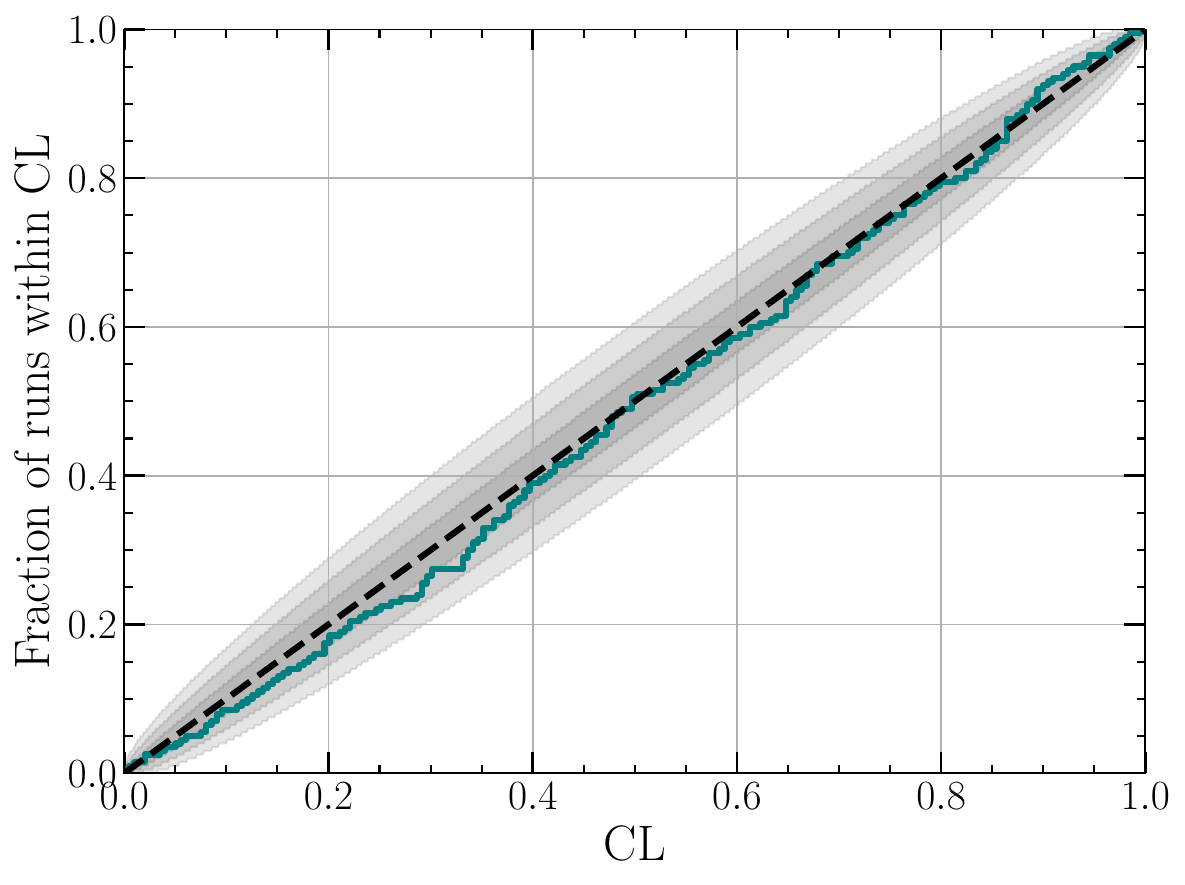}
    \caption{Probability-probability plot for the inference of $\fagn$, showing the fraction of times the true parameter value falls within a given posterior credible level (CL), plotted against that CL. This plot is made with 200 sets of simulated data. The mock GW data is calibrated on GWTC-4.0. The mock AGN catalogue mimics the distribution of Quaia sources brighter than $\lbol{46.5}$. The grey shaded regions show the size of 1$\sigma$, 2$\sigma$ and 3$\sigma$ fluctuations expected from random data realisations. Injected values of $\fagn$ are recovered without significant bias.}
    \label{fig:ppplot}
\end{figure}

After inferring $\fagn$ from each dataset, we construct the probability-probability plot \citep{Sidery:2014PP, RomeroShaw:2020PP} seen in Fig. \ref{fig:ppplot}. This plot shows the fraction of times the true parameter value falls within a given posterior credible level, plotted against that level. Inference is consistent if the graph follows the diagonal, which is the case in Fig. \ref{fig:ppplot}.

\subsection{How to obtain optimal constraints}\label{subsect:infogain}

We explore several properties of GW and EM data that influence the constraining power of our framework. In Sect. \ref{subsubsect:limiting_cases}, we demonstrate the potential improvement in the precision of $\fagn$ measurements attainable with a perfect AGN catalogue for different GW detection horizons. We then examine the redshift precision and completeness required of realistic AGN catalogues to approach these optimal constraints (Sect. \ref{subsubsect:real_em_data}).

\subsubsection{The empty-catalogue and perfect-catalogue limits}\label{subsubsect:limiting_cases}

The amount of information provided by an AGN catalogue depends on its completeness and the precision of the redshift measurements\footnote{In addition, the shape of the AGN redshift distribution and their rarity also matter. However, investigating this in detail is outside the scope of this work.}. The most information will be added by a catalogue that is 100 percent complete with perfectly measured redshifts. We call this the perfect-catalogue limit. Conversely, the weakest possible constraint will be put on $\fagn$ in the empty-catalogue limit, where no EM catalogue is used.

\begin{figure}
    \centering
    \includegraphics[width=\linewidth]{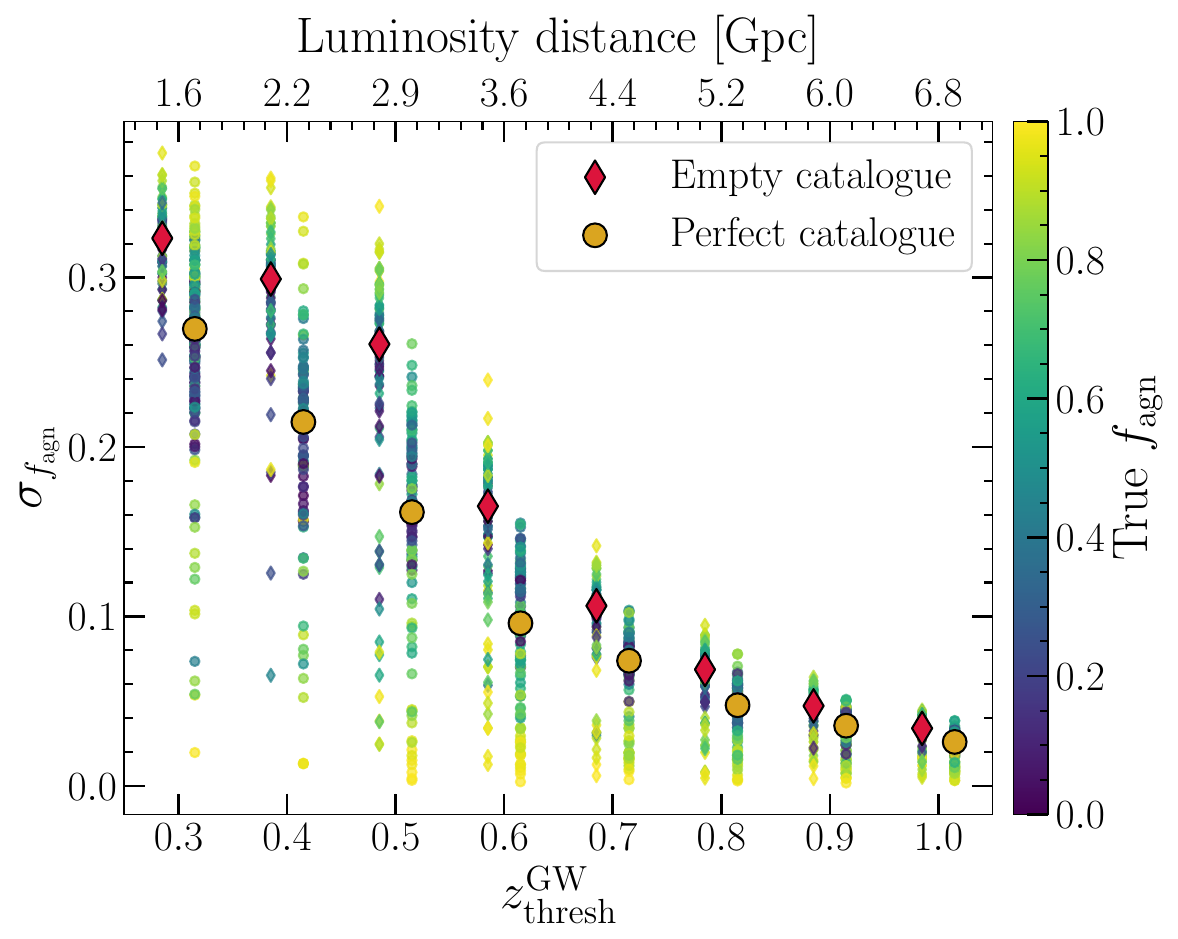}
    \caption{The 1$\sigma$ uncertainty on $\fagn$ ($\sigma_{\fagn}$) as a function of the GW redshift detection threshold ($z_{\rm thresh}^{\rm GW}$) in two limits: the perfect-catalogue limit (circles) and the empty-catalogue limit (diamonds). For each detection threshold and limit, we produce 100 mock data realisations for which we infer $\fagn$. The constraints from individual realisations are shown in the background, coloured according to the true injected value of $\fagn$. The medians are indicated with larger markers.  All $\sigma_{\fagn}$ are calculated assuming a total of $10^5$ injected GW events (see Tab. \ref{tab:mockgw}). We consider an AGN population with bolometric luminosities $\lbol{46.5}$.}
    \label{fig:sig_v_thresh}
\end{figure}

In Fig. \ref{fig:sig_v_thresh} we show the 1$\sigma$ error on $\fagn$ ($\sigma_{\fagn}$) in these two limits as a function of $z_{\rm thresh}^{\rm GW}$. For each detection threshold and limit, we produce 100 mock data realisations for which we infer $\fagn$. We plot the individual constraints and the median $\sigma_{\fagn}$ for each dataset. The colour bar indicates the injected value of $\fagn$. At fixed $z_{\rm thresh}^{\rm GW}$, we assume the error on $\fagn$ scales with the total number of injected GWs as $\sigma_{\fagn} \propto (N_{\rm tot}^{\rm inj})^{-1/2}$. To save computation time in generating Fig. \ref{fig:sig_v_thresh} we estimate $\fagn$ on smaller mock data samples, extrapolating to $N_{\rm tot}^{\rm inj} = 10^5$. Our choices for the injected number of GWs are listed in Tab. \ref{tab:mockgw}.

\begin{table}\label{tab:mockgw}
    \centering
    \caption{Our choices for the GW redshift detection threshold ($z^{\rm GW}_{\rm thresh}$) and the total number of injected GWs ($N^{\rm tot}_{\rm inj}$) for making Fig. \ref{fig:sig_v_thresh}, and the corresponding expected number of detected GWs ($N^{\rm exp}_{\rm det}$). We scale the errors in this figure to the expected constraints that would have resulted from using $10^5$ injections, or $N^{\rm exp}_{\rm det}\big|_{N^{\rm tot}_{\rm inj}=10^{5}}$ detections. We indicate the 16th and 84th percentile on the expected number of detections that results from sampling $\fagn$ uniformly between 0 and 1.}
    \begin{tabular}{c|c|c|c}
        \hline\hline
        $z^{\rm GW}_{\rm thresh}$ & $N^{\rm tot}_{\rm inj}$ & $N^{\rm exp}_{\rm det}$ & $N^{\rm exp}_{\rm det}\big|_{N^{\rm tot}_{\rm inj}=10^{5}}$ \\
        \hline
        0.3 & $1 \cdot 10^5$ & $160^{+101}_{-107}$ & $160^{+101}_{-107}$ \\[2pt]
        0.4 & $1 \cdot 10^5$ & $365^{+231}_{-244}$ & $365^{+231}_{-244}$ \\[2pt]
        0.5 & $1 \cdot 10^5$ & $687^{+473}_{-430}$ & $687^{+473}_{-430}$ \\[2pt]
        0.6 & $5 \cdot 10^4$ & $600^{+379}_{-359}$ & $1199^{+757}_{-717}$ \\[2pt]
        0.7 & $2 \cdot 10^4$ & $397^{+214}_{-231}$ & $1985^{+1070}_{-1155}$ \\[2pt]
        0.8 & $1 \cdot 10^4$ & $286^{+176}_{-164}$ & $2865^{+1755}_{-1645}$ \\[2pt]
        0.9 & $5 \cdot 10^3$ & $209^{+107}_{-116}$ & $4180^{+2140}_{-2320}$ \\[2pt]
        1.0 & $3 \cdot 10^3$ & $164^{+90}_{-80}$ & $5450^{+2983}_{-2683}$ \\
        \hline
    \end{tabular}
\end{table}

The relative reduction in $\sigma_{\fagn}$ between the perfect- and empty-catalogue limits stays roughly constant when varying $z_{\rm thresh}^{\rm GW}$, at $30 \pm 8$ percent. The smallest relative gain is for $z_{\rm thresh}^{\rm GW} = 0.3$, with a 17 percent reduction in $\sigma_{\fagn}$. This is because the observed GWs are located at lower redshifts than the nearest AGN with $\lbol{46.5}$ (see Fig. \ref{fig:all_zpops}), limiting the spatial overlap between AGN and GWs that is essential for our method. The highest relative gain is for $z_{\rm thresh}^{\rm GW} = 0.6$, with a $42$ percent reduction in $\sigma_{\fagn}$.

Fig. \ref{fig:sig_v_thresh} shows that a perfect AGN catalogue can provide as much information as up to hundreds of GWs. For example, the expected constraints in the perfect-catalogue limit at $z_{\rm thresh}^{\rm GW} = 0.5$ with $\approx 687$ detected GW events are comparable to those in the empty-catalogue limit at $z_{\rm thresh}^{\rm GW} = 0.6$ with $\approx 1199$ events (see Tab. \ref{tab:mockgw}). The same comparison for $z_{\rm thresh}^{\rm GW} = 0.8$ and $z_{\rm thresh}^{\rm GW} = 0.9$ shows that a perfect AGN catalogue may substitute $\gtrsim10^3$ GWs, although the absolute reduction in $\sigma_{\fagn}$ becomes limited as the number of observed GWs grows. Note that $\sigma_{\fagn}$ at higher redshift thresholds may be inaccurate, as the localisation volumes of our mock GWs are always sampled from the same, redshift-independent distribution. In reality, localisation volumes increase with redshift, decreasing their constraining power. At the same time, a further detector horizon will also be accompanied by an increased number of GW events that are better localised than currently modelled, increasing precision.

Fluctuations in $\sigma_{\fagn}$ are predominantly caused by statistical variations in the true value of $\fagn$. Especially when $\fagn$ is near unity, there is a large scatter on $\sigma_{\fagn}$ around its median value. This is due to the GW detection threshold, which governs the number of detected GWs. Since our redshift population prior for AGN-origin GWs lies at higher redshifts than the alternative, higher values of $\fagn$ result in less detected GWs on average, resulting in flat posteriors when the number of detected GWs is too low to constrain the prior. However, when a data realisation occurs with more observed GWs at high redshift than expected, we obtain a very narrow posterior towards $\fagn=1$, as this is unlikely to occur for low values of $\fagn$.

\subsubsection{Realistic EM data: redshift errors and selection effects}\label{subsubsect:real_em_data}
With the best- and worst-case scenarios laid out, we can investigate the properties of realistic catalogues necessary for obtaining optimal constraints on $\fagn$. To this end, we vary the redshift errors of the AGN and the redshift completeness of the catalogue. We limit the completeness of our mock catalogues by removing all AGN above a redshift $z_{\rm cut}^{\rm EM}$, and those within the Galactic plane ($|b| \leq 10\degree$). All AGN are given the same redshift error, instead of sampling $\sigma_{z, \rm agn}$ from Quaia, as was done for Fig. \ref{fig:ppplot}. We fix $z_{\rm thresh}^{\rm GW} = 1.0$.

\begin{figure}
    \centering
    \includegraphics[width=\linewidth]{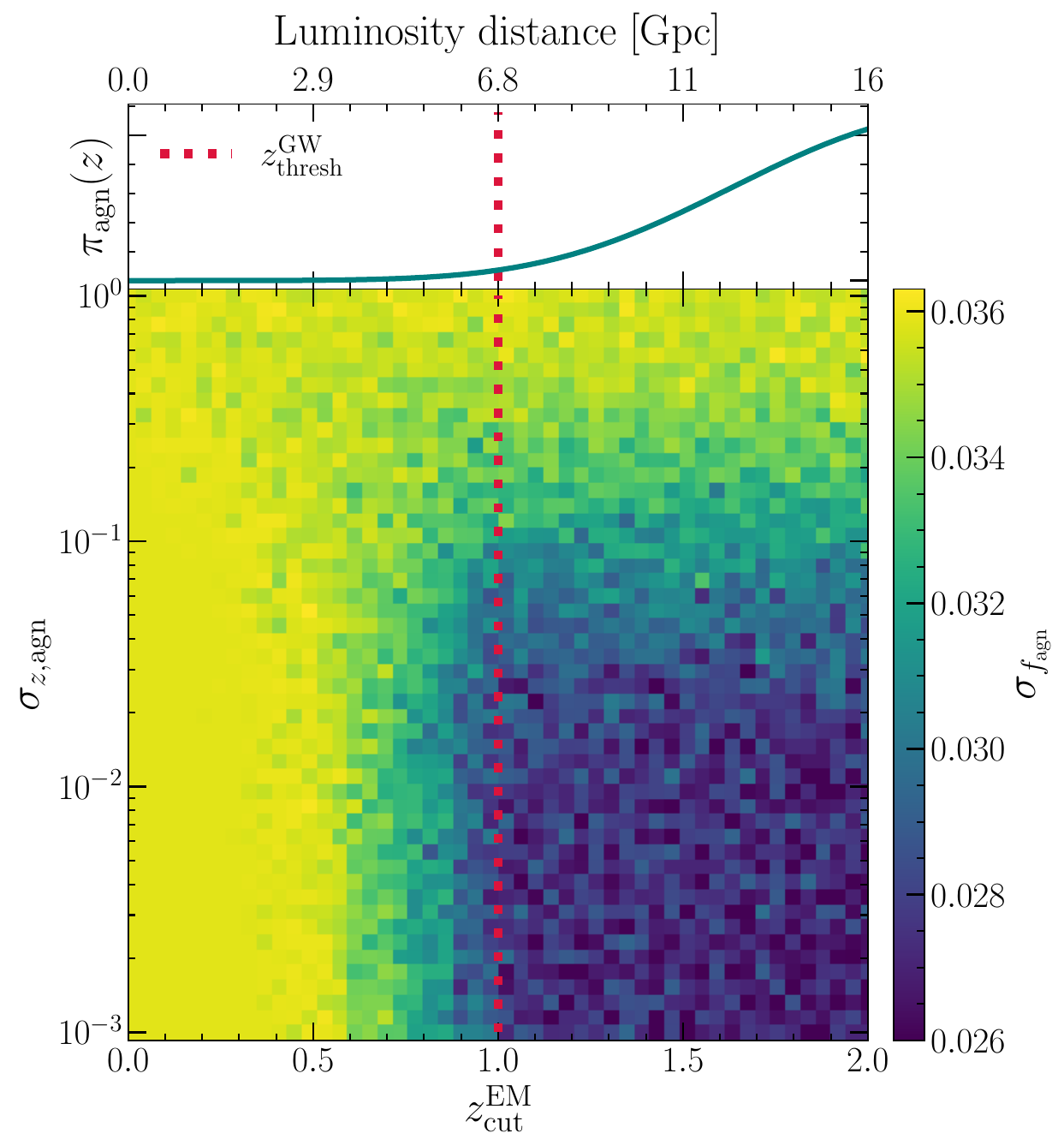}
    \caption{The median 1$\sigma$ uncertainty on $\fagn$ ($\sigma_{\fagn}$) is plotted as a function of the AGN redshift error ($\sigma_{z, \rm agn}$) and the redshift cut ($z_{\rm cut}^{\rm EM}$) of the AGN catalogue. The AGN catalogue is complete in redshift up to $z_{\rm cut}^{\rm EM}$, and empty inside the Galactic plane (latitudes $|b| \leq 10\degree$). We simulate an AGN population with bolometric luminosities $\lbol{46.5}$, the redshift distribution of which is seen in the upper panel (solid line). The GW redshift detection threshold ($z_{\rm thresh}^{\rm GW}$) is indicated with a dotted line.}
    \label{fig:cat_quality_effect}
\end{figure}

The result is shown in Fig. \ref{fig:cat_quality_effect}, with which we distinguish three regimes in $z_{\rm cut}^{\rm EM}$. Constraints on $\fagn$ only start to improve once $z_{\rm thresh}^{\rm EM} \gtrsim 0.6$, and they do not improve further for $z_{\rm thresh}^{\rm EM} > z_{\rm thresh}^{\rm GW}$. This shows the necessity of spatial overlap between GWs and AGN for EM-informed population inference: there are no GWs above $z_{\rm thresh}^{\rm GW}$ and there are no AGN below $z \approx 0.6$, as can be seen from $\pi_{\rm agn}(z)$ in the top panel. We conclude that the AGN in the redshift range $0.6 \lesssim z < z_{\rm thresh}^{\rm GW}$ are the ones informing the likelihood.

Reducing $\sigma_{z,\rm agn}$ yields increasingly large improvements in precision on $\fagn$ as $z_{\rm thresh}^{\rm EM}$ approaches $z_{\rm thresh}^{\rm GW}$. However, once $\sigma_{z,\rm agn} \lesssim 0.01$, further reductions provide little additional benefit because the AGN redshift uncertainties become negligible compared to those of the GWs. Since the median 1$\sigma$ redshift error for detected mock GW events is $\sigma_{z, \rm GW} \approx 0.1$, we conclude that the typical AGN redshift uncertainty should be a factor 10 smaller than that of the GW sample for near-optimal constraints.

Note that a complete catalogue with poorly-measured AGN redshifts does not give an improved constraint with respect to an empty catalogue. One might expect the well-measured sky positions of the AGN to still provide information on $\fagn$, even though their redshifts are uncertain. However, the less certain the AGN redshifts are, the higher the number of AGN that are compatible with the GW redshift posterior. In the limit of completely unknown redshifts, all AGN along the entire LOSs in the GW sky localisation area are compatible hosts, at which point our method reduces to a 2D sky-projected analysis. The information supplied by such a catalogue is limited by the rarity of the AGN subpopulation.

\section{Results}\label{sect:results}

In Sect. \ref{sect:mockdata} we have shown the validity and potential of our framework for estimating the AGN-hosted contribution $\fagn$ to the BBH population. We now apply it to the detected GWTC-5.0 BBH population, informed by the Quaia AGN catalogue (see Sect. \ref{sect:data}), assuming our fiducial population model (see Sect. \ref{sect:method}).

\begin{figure}
    \centering
    \includegraphics[width=\linewidth]{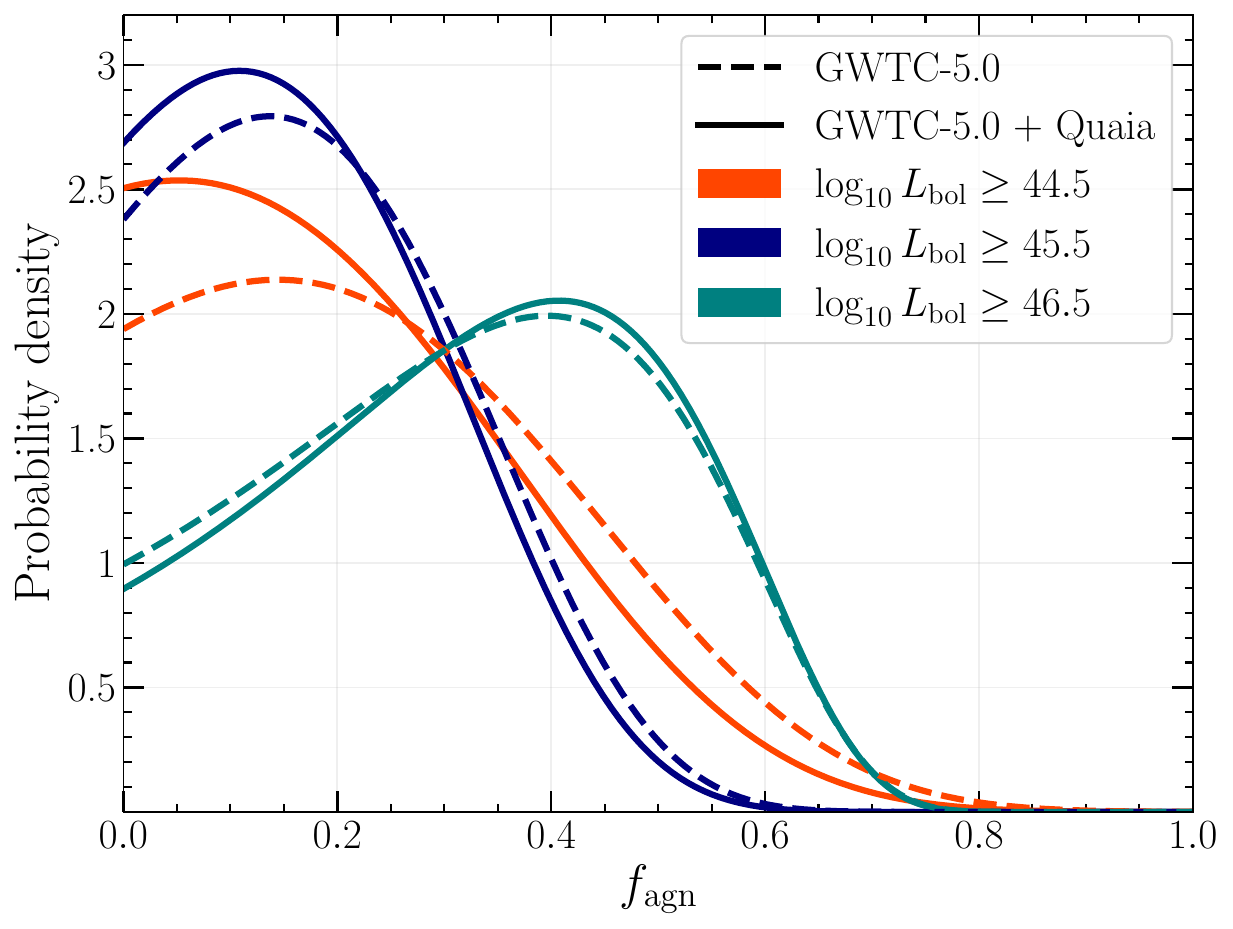}
    \caption{Posterior distributions on the fraction of gravitational wave (GW) events in the Universe hosted by subpopulations of AGN ($\fagn$). AGN subpopulations are characterised by a bolometric luminosity threshold in $\mathrm{erg \, s^{-1}}$, indicated by different colours. All AGN subpopulations considered are unobscured in the optical and infrared. Dashed lines show the posterior using only GW information. Solid lines show posteriors that are also informed by the AGN catalogue Quaia.}
    \label{fig:posteriors}
\end{figure}

Fig. \ref{fig:posteriors} shows the posterior distributions on the fraction of GW events hosted by each AGN subpopulation we consider. Dashed posteriors are solely informed by the GW data. These GW-only posteriors represent the empty-catalogue limit described in Sect. \ref{subsubsect:limiting_cases}, where the constraints are driven only by the difference in redshift distributions between AGN-hosted and alternative-origin GWs (see Fig. \ref{fig:all_zpops}). Solid posteriors are informed by both GW data and the AGN catalogue. The spatial prior constructed from observations of AGN provides more information than just their average redshift distribution, thus narrowing the posteriors compared to the empty-catalogue case (see Fig. \ref{fig:sig_v_thresh}). We find 95 percent CL upper limits on $\fagn$ of $\fagn \leq \{0.52, 0.41, 0.61\}$ for unobscured AGN with bolometric luminosities $L_{\rm bol} \geq \{10^{44.5},10^{45.5},10^{46.5}\} \, \mathrm{erg \, s^{-1}}$ when including Quaia in the analysis. Without Quaia, these upper limits are $\fagn \leq \{0.58, 0.43, 0.61\}$. We cannot exclude $\fagn = 0$ at this level of significance, even though all our posteriors peak at $\fagn > 0$.

The extent to which the posterior changes with respect to the empty-catalogue case by including Quaia is dependent on the completeness and redshift errors of the catalogue (see Fig. \ref{fig:cat_quality_effect}) and on the number of members of the AGN subpopulation. This change is small for all three AGN subpopulations, and there is no visible trend with completeness and rarity, both of which are more favourable for the most luminous AGN. This indicates that the information added by Quaia is limited by the precision of its redshift measurements, which is similar for all AGN.

\begin{figure}
    \centering
    \includegraphics[width=\linewidth]{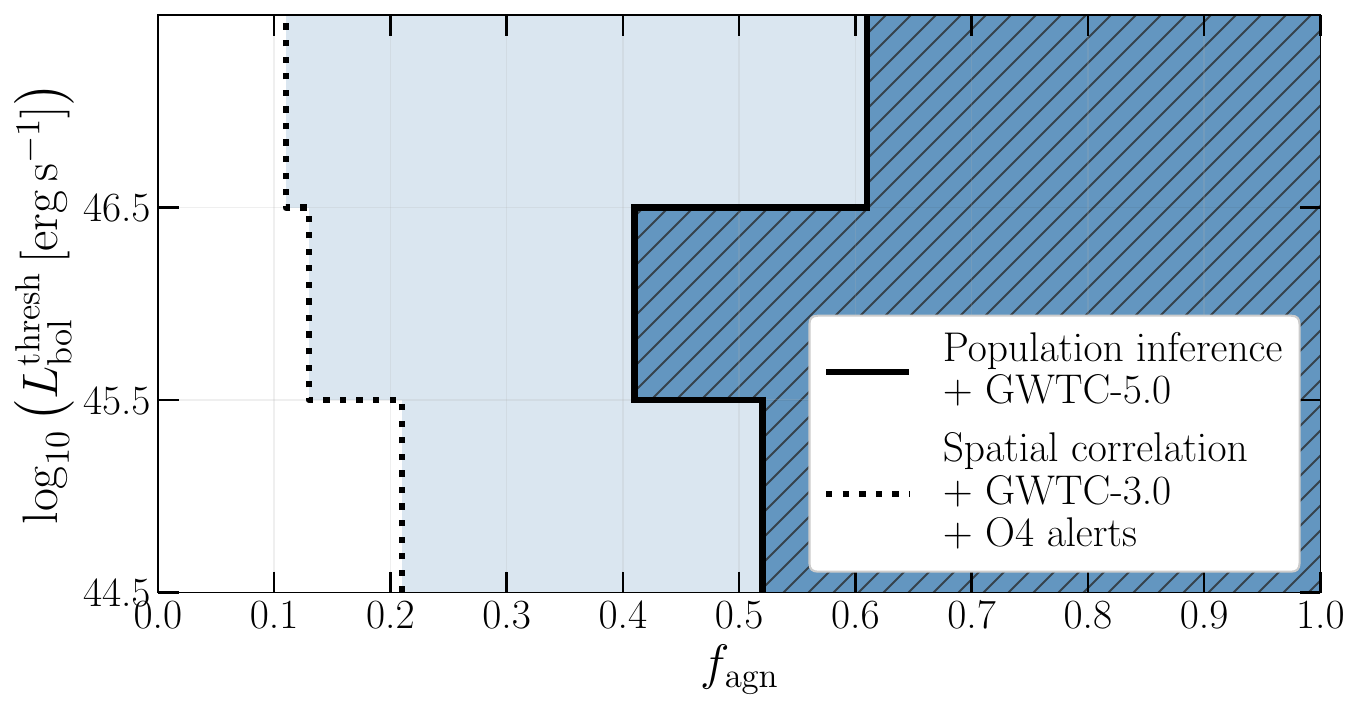}
    \caption{The 95 percent credible upper limits on the fraction GW events hosted by unobscured AGN brighter than the bolometric luminosity threshold $L_{\rm bol}^{\rm thresh}$. The shaded region enclosed by the dotted line shows the rejected values of $\fagn$ obtained using the spatial correlation of 159 GW events, including preliminary O4 events, with AGN observations up to redshift $z=1.5$ by \citet{Veronesi:2025}. The constraints we obtain applying our fiducial population model to 256 GWTC-5.0 GW events informed by AGN up to redshift $z=3$ are indicated by the hashed region bounded by the solid line. Both works use the AGN catalogue Quaia to inform the likelihood.}
    \label{fig:constraints}
\end{figure}

In Fig. \ref{fig:constraints} we compare our EM-informed population inference constraints on $\fagn$ with those from the spatial-correlation analysis of \citet{Veronesi:2025}. We have obtained our constraints using 256 GW events from BBH mergers in GWTC-5.0, informed by Quaia sources up to $z=3$, whereas \citet{Veronesi:2025} applied their analysis to a sample of 159 GW events, consisting of GWTC-3.0 and preliminary O4 alerts, combined with Quaia sources below $z=1.5$. They obtain $\fagn \leq \{0.21,0.13,0.11\}$ at 95 percent credibility. Therefore, our improved modelling of both GW and AGN data relaxes the constraints previously put on $\fagn$, even though we have analysed a larger dataset of both AGN and GWs. This shows that we remove the downward bias that resulted from neglecting AGN redshift errors (see Appendix \ref{subsect:bias}) and neglecting the difference in GW detection efficiency between the AGN-origin and alternative-origin GW populations. Indeed, if we set the detection efficiency equal to unity, we obtain biased 95 percent CL upper limits of $\fagn \leq \{0.029, 0.017, 0.012\}$. Incorporating these effects increases the uncertainty in $\fagn$, resulting in looser constraints, but also more robust ones.

It is insightful to see to what extent the individual GW events contribute to our population-level posteriors and how this correlates with their properties. Adapting Eq. 2.11 of \citet{Gupte:2025pecc}, we calculate the probability that GW event $i$ came from an AGN as
\begin{align}
    p_i(\mathrm{AGN}) = \left \langle \frac{\fagn Z_{A}(d_i)}{\fagn Z_{A}(d_i) + (1 - \fagn) Z_{\conj{A}}(d_i)} \right \rangle_{p(\fagn | d_i)} \, ,
\end{align}
where the average is over the posterior on $\fagn$ and the evidences ($Z$) are given by Eq. \ref{eq:hypothesis_evidence}.

\begin{figure}
    \centering
    \includegraphics[width=\linewidth]{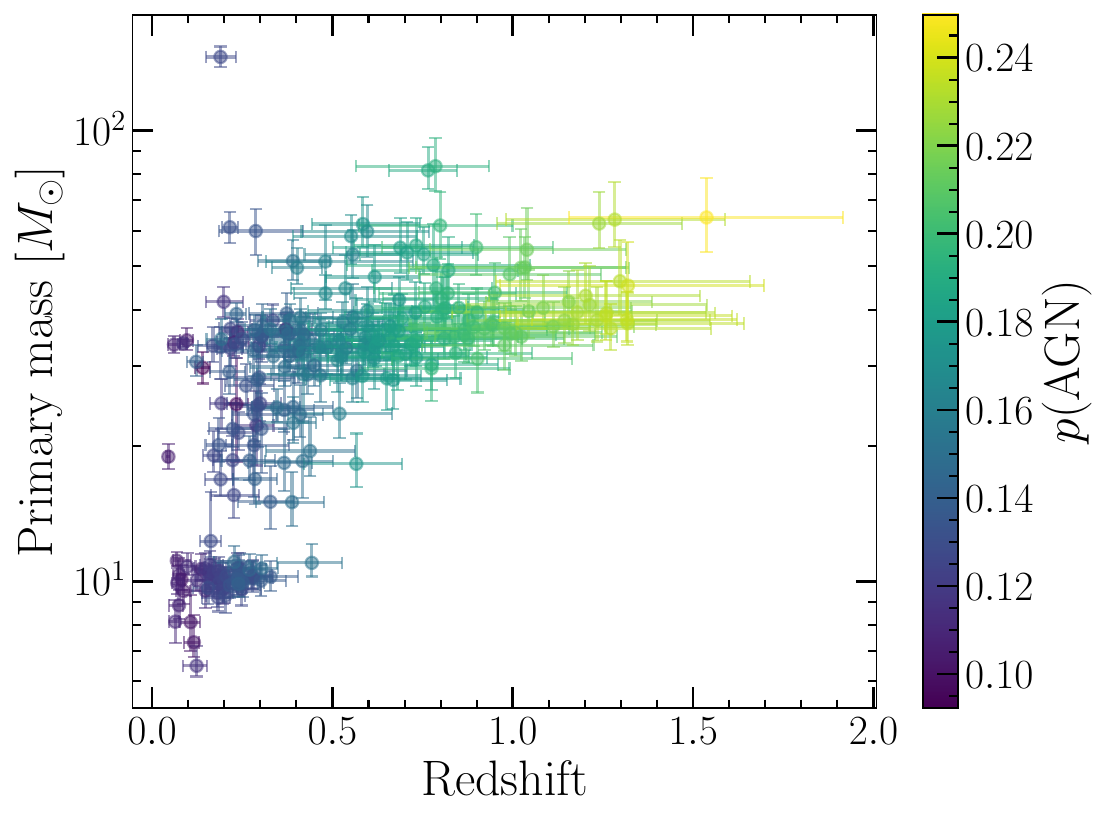}
    \caption{The posterior probability of 256 GW events from BBH mergers in GWTC-5.0 originating from unobscured AGN brighter than $\lbol{44.5}$, plotted as a function of primary mass and redshift. Error bars indicate the 16th and 84th percentile of the GW posterior. The event GW230704\_212616 has the highest AGN-origin probability, with $p(\mathrm{AGN}) \approx 0.250$. Combining the probabilities across all events, there is a 98.1 percent chance of at least one GW event coming from an AGN under our fiducial model.
    }
    \label{fig:p_AGN}
\end{figure}

In Fig. \ref{fig:p_AGN}, we show the AGN-origin probability as a function of primary BH mass and redshift for the AGN brighter than $\lbol{44.5}$, which is the broadest subpopulation we consider in this work. However, we find no qualitative difference when considering our other AGN subpopulations. We find a strong correlation between $p(\rm AGN)$ and redshift, which is a direct consequence of our modelling: we assume that the number of AGN-hosted GWs increases with redshift in the current range of observed GW data (see Fig. \ref{fig:all_zpops}). Events with the highest AGN-origin probabilities tend to have higher primary masses, consistent with predictions for hierarchical mergers. 
However, we cannot conclude from this analysis that this is a population property, as Malmquist bias also causes an absence of GW detections in the region of low primary mass and high redshift.

We find GW230704\_212616 to have the highest AGN-origin probability, with $p(\mathrm{AGN}) \approx 0.250$. Therefore, no individual GW event is more likely to have an AGN origin than an alternative origin. Nevertheless, when combining the probabilities across all events, we find a 98.1 percent probability that at least one GW in the observed dataset originated in an AGN. This should be regarded as an upper limit, since future analyses may place stronger constraints on $\fagn$, thereby reducing the AGN-origin probability of individual GW events. For comparison, if our posterior on $\fagn$ were flat, each event would have a 50 percent probability of originating in an AGN, yielding nearly a 100 percent probability of at least one AGN-origin event in the dataset. Table \ref{tab:agn_probabilities} ranks all GW events by their AGN-origin probability.

\section{Discussion}\label{sect:discussion}

We have presented the population inference method for inferring the AGN-channel contribution to the BBH population, informing the GW population with an EM source catalogue. This represents a generalisation of previous spatial-correlation analyses \citep{Bartos:2017abc, Veronesi:2023}, unifying them with the population inference analyses that leverage intrinsic GW parameters instead of spatial ones \citep{Gayathri:2023bha, Bartos:2026spin}. Our analysis is currently limited to unobscured AGN, because only for these sources do we have a sufficiently complete catalogue. Consequently, our constraints on $\fagn$ apply only to this subset of the AGN population, and do not directly constrain the contribution from obscured sources.

We find that the contribution of individual GW events to our population-level posterior is driven by the GW redshift, with higher-redshift events being more likely to have an AGN origin. This is inconsistent with the results from \citet{Bartos:2026spin}, who differentiated AGN-hosted GW events from alternative GW populations based on BBH component spins. For instance, they impose that AGN-channel mergers involve high-spin primary BHs. As a consequence, their AGN-origin probability is strongly correlated with the primary BH spin, just as our ranking traces the GW redshift, but not both primary BH spin and redshift at the same time. This inconsistency demonstrates the necessity of jointly inferring population distributions of intrinsic and spatial GW parameters to more robustly characterise candidate AGN-channel GW events.

Because we adopt a strongly parametrised model, our results may be sensitive to features of the GW population that are not captured by our model. For instance, a high-redshift GW population unrelated to the AGN channel may exist, such as one originating from dense star clusters \citep{Farah:2026}. If such a population more closely traces the spatial distribution of AGN-hosted GWs than the alternative population model we consider, its presence could bias our inference towards higher values of $\fagn$. However, this is mitigated by the inclusion of an AGN catalogue enforcing the GW-AGN association directly, thereby making our inference more robust. 

\subsection{Redshift evolution of the AGN-channel merger rate}\label{subsect:rates}

\begin{figure}
    \centering
    \includegraphics[width=\linewidth]{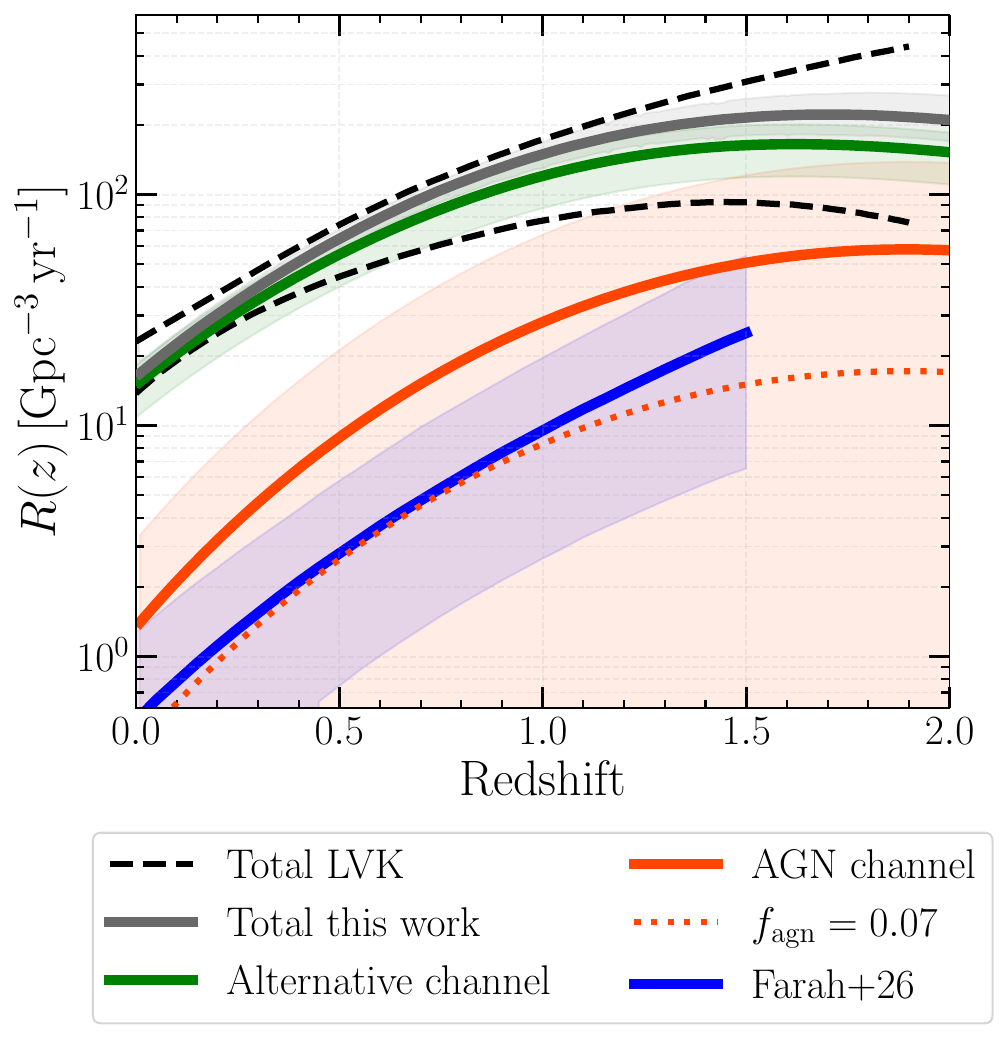}
    \caption{The BBH merger rate density as a function of redshift. Solid lines indicate posterior medians, and shaded regions outline the 90 percent credible interval. The dashed black lines outline the 90 percent credible interval of the total merger rate inferred by \citet{ligo:gwtc5pop} using their \texttt{Default BBH} model with a \texttt{Madau Dickinson} merger rate evolution. We decompose the total merger rate inferred with our fiducial model (grey) into a contribution from the AGN channel (orangered) and alternative channels (green). For comparison, we plot the rate evolution of the hierarchically merging subpopulation inferred by \citet{Farah:2026} (blue), which we are able to match with an AGN-hosted fraction of $\fagn = 0.07$, considering unobscured AGN brighter than $\lbol{44.5}$.}
    \label{fig:rates}
\end{figure}

In Fig. \ref{fig:rates}, we plot the posterior redshift evolution of the GW merger rate, calculated using the procedure in Appendix C3 of \citet{Essick:2025sel}. Our inferred total merger rate (grey) is consistent with the LVK result (black), but more tightly constrained because our fiducial model marginalises over only one hyperparameter, $\fagn$, whereas the LVK analysis adopts a more flexible population model with 24 free hyperparameters governing the redshift, mass, and spin distributions \citep{ligo:gwtc5pop}. We decompose our total merger rate into two contributions, one from the AGN channel (orangered) and one from the alternative-origin population (green).

If it contributes to the data, the AGN channel becomes more important at higher redshifts, consistent with Fig. \ref{fig:p_AGN}. This is reminiscent of the steeply evolving subpopulation of hierarchically merging BBHs found by \citet{Farah:2026}. They argue that this subpopulation can be explained mostly with a cluster origin, however, in Fig. \ref{fig:rates}, we show that the AGN-channel merger rate evolution is consistent with this subpopulation as well up to $z \approx 1$ for $\fagn = 0.07$. Since we only assume a time-dilation correction in the AGN-channel merger rate evolution, this steep increase of the rate is a natural consequence of the intrinsic spatial distribution of AGN. 
This suggests that the AGN channel could partially contribute to the rate of hierarchical mergers, possibly even dominating below $z=1$. Contributions from higher-luminosity AGN, which evolve steeper in redshift and are located at higher redshift (Fig. \ref{fig:all_zpops}), could enhance the merger rate above $z=1$ to match the steeper redshift evolution of the subpopulation there as well.

Assuming the uncertainty on $\fagn$ scales as the square root of the number of GW detections, we estimate that a precision of $\sigma_{\fagn} = 0.07$ can be reached with $\sim 2 \cdot 10^3$ GWs at the detector sensitivity of GWTC-5.0, at which point $\fagn = 0$ can be distinguished from $\fagn = 0.07$ at $1\sigma_{\fagn}$ precision. This is an upper limit as the completeness and redshift precision of AGN catalogues improves with future spectroscopic surveys, such as that of Euclid \citep{Euclid:overview, Euclid:nisp}, on top of sensitivity improvements of the LVK detectors. The latter will result in GW events that are more informative because of the more precise parameter estimation and the increased detector horizon, allowing the detection of informative high-redshift GW events (see Fig. \ref{fig:sig_v_thresh}). For comparison, our final GW sample contains 256 events with a median luminosity distance of $d_{L} = 2.2_{-1.8}^{+6.2}$ Gpc, while the expected BBH detection rate for LVK O5 is $870_{-480}^{+1100} \, \mathrm{yr^{-1}}$ with a median luminosity distance of $4.607_{-0.082}^{+0.077} \, \mathrm{Gpc}$ \citep{Kiendrebeogo:2023}. We therefore expect to reach the precision required for testing unobscured AGN as the dominant source of the hierarchically merging subpopulation of BBHs below $z=1$ with less than 2.5 years of O5-precision GW observations. This is within the currently scheduled runtime of O5.

\subsection{Relaxing the redshift prior of alternative-origin GWs}\label{subsect:4Dpost}
Our fiducial model constitutes a strong and inflexible prior on the redshift distribution of the alternative-origin GWs. We now relax this prior by jointly sampling $\fagn$ and the three hyperparameters that shape their merger rate evolution ($\gamma$, $z_{\rm peak}$, $\kappa$; Eq. \ref{eq:pnonagn}) using a Markov Chain Monte Carlo routine implemented with the Python package \texttt{emcee} \citep{emcee:2013}. We do this for two sets of hyperpriors: EM-informed Gaussian priors given by the uncertainty in the combined fit of the cosmic SFR of \citet{Strolger:2020}, and the broad uniform priors of LVK \citep{ligo:gwtc5pop}. In both cases, the prior on $\fagn$ is $\mathcal{U}(0,1)$. All priors are listed in Tab. \ref{Tab:hyperpriors}.

\begin{figure}
    \centering
    \includegraphics[width=\linewidth]{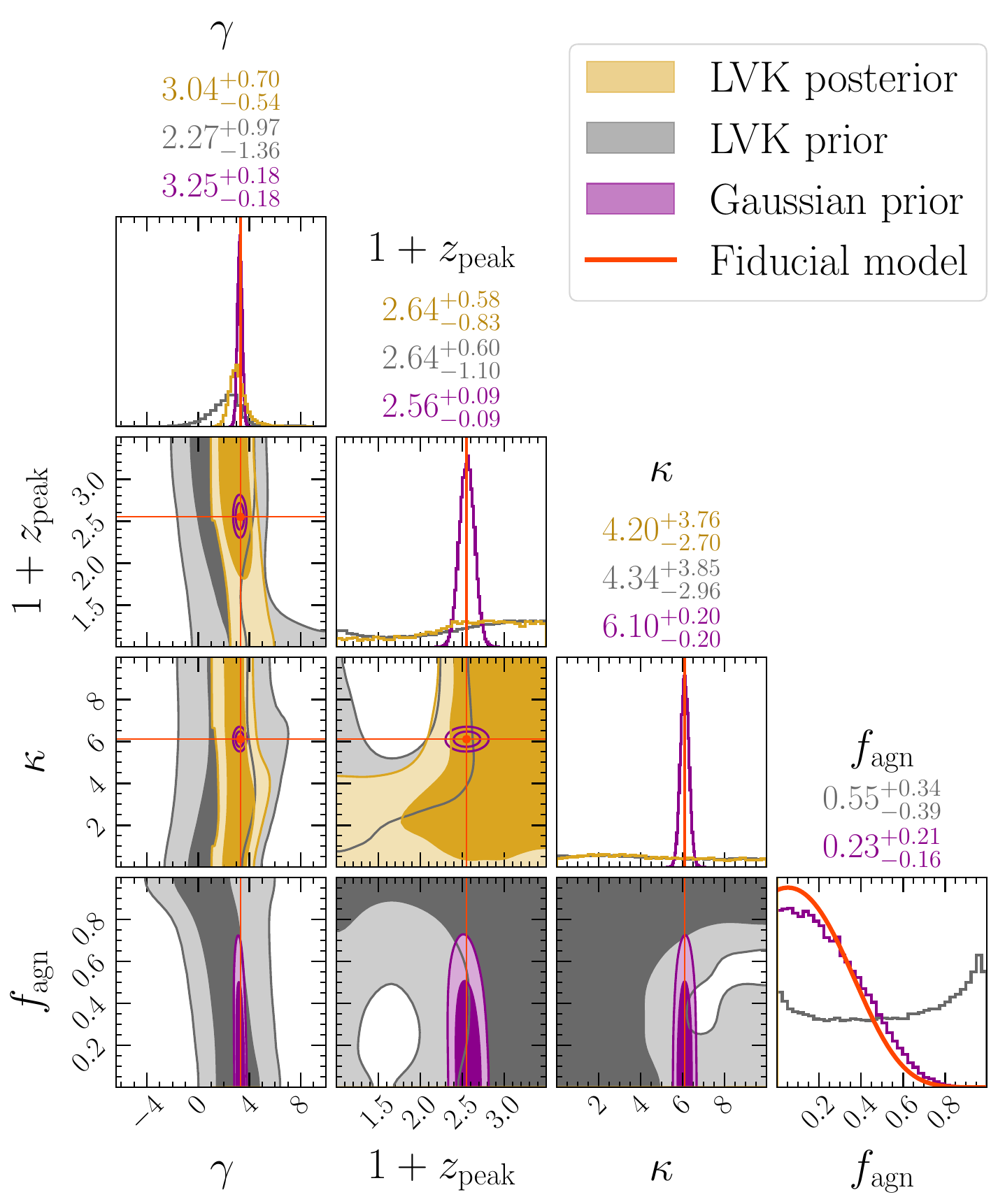}
    \caption{Joint posteriors on the fraction ($\fagn$) of gravitational wave (GW) events coming from unobscured AGN brighter than $\lbol{44.5}$, and the hyperparameters ($\gamma$, $z_{\rm peak}$, $\kappa$) of the merger rate evolution of the alternative-origin GW population. Joint posteriors are shown for three sets of hyperpriors: broad uniform priors used by LVK \citep{ligo:gwtc5pop} (grey), Gaussian priors from a fit to EM data \citep{Strolger:2020} (purple), and our fiducial model where $\fagn$ is the only free parameter (orange). Yellow contours show the hyperposteriors obtained by \citet{ligo:gwtc5pop} under their \texttt{Default BBH} model with a \texttt{Madau-Dickinson} merger rate evolution with redshift. Contours indicate the 68 percent and 95 percent credible levels.
    }
    \label{fig:corner44p5}
\end{figure}

The result is shown in Fig. \ref{fig:corner44p5} for AGN brighter than $\lbol{44.5}$, but there is no qualitative difference between AGN subpopulations. For comparison, we also plot the hyperposteriors on $\gamma$, $z_{\rm peak}$ and $\kappa$ obtained by  \citet{ligo:gwtc5pop} under their \texttt{Default BBH} model with a \texttt{Madau-Dickinson} rate evolution. The 95 percent credible intervals on $\fagn$ for all AGN subpopulations and priors are shown in Tab. \ref{tab:constraints}, along with the corner plots for AGN brighter than $\lbol{45.5}$ (Fig. \ref{fig:corner45p5}) and $\lbol{46.5}$ (\ref{fig:corner46p5}).

The freedom added to the population model by the hyperpriors reduces the strength of our constraints on $\fagn$. 
This is only a slight effect when hyperparameters are sampled within narrow Gaussian priors around our fiducial model, changing our 95 percent CL upper limit on $\fagn$ from $\fagn \leq 0.52$ to $\fagn \leq 0.58$. 
However, the LVK priors result in a near-flat posterior, with a 95 percent credible interval of $\fagn \in [0.04, 0.99]$. Especially the parameter $\kappa$ is unconstrained and degenerate with $\fagn$. The degeneracy is due to this parameter governing the shape of the rate evolution at high redshift, which is the region most associated with an AGN origin (see Fig. \ref{fig:p_AGN}). It is unconstrained since we do not have GW observations at sufficiently high redshifts (see Fig. \ref{fig:data}). In the same way, $\gamma$ is the best-constrained parameter, since it is the power-law slope of the merger rate at low redshift, where most observed GW events lie. It is also the most robust to broad priors, remaining constrained around our Gaussian priors. We find $\gamma = 2.27_{-1.36}^{+0.97}$ using the LVK priors, which is consistent with the LVK posterior of $\gamma = 3.04_{-0.54}^{+0.70}$. However, our posterior on $\gamma$ extends to lower values, because of a degeneracy with $\fagn$. A higher value of $\fagn$ implies a stronger contribution of the steeply evolving AGN-origin GW population, which needs to be compensated by a lower value of $\gamma$ to keep the merger rate evolution of the total GW population consistent with observations. At $\fagn = 0$, we recover the LVK hyperposterior.

Using the LVK priors, we find no evidence for a peak in the merger rate distribution. Overall, we conclude that our results are moderately model dependent. More GW observations at high redshift are required to lift the degeneracies driving this dependence.

\begin{table}\label{Tab:hyperpriors}
\centering
\caption{The priors on the hyperparameters governing the redshift evolution of the alternative-origin BBH merger rate (Eq. \ref{eq:pnonagn}). The fiducial model and Gaussian priors ($\mathcal{N}(\mu,\sigma)$) are based on the fit by \citet{Strolger:2020}. The LVK priors are uniform on the indicated interval, consistent with \citet{ligo:gwtc5pop}.}
\begin{tabular}{llll}
\hline\hline
Hyperparameter     & LVK & Gaussian        & Fiducial \\
\hline
$\fagn$            & $\mathcal{U}(0,1)$         & $\mathcal{U}(0,1)$          & $\mathcal{U}(0,1)$ \\
$\gamma$           & $\mathcal{U}(-10, 10)$     & $\mathcal{N}(3.3, 0.2)$ & $3.3$              \\
$1 + z_{\rm peak}$ & $\mathcal{U}(1.5, 3.5)$  & $\mathcal{N}(2.55, 0.09)$ & $2.55$              \\
$\kappa$           & $\mathcal{U}(0, 10)$     & $\mathcal{N}(6.1, 0.2)$ & $6.1$              \\
\hline
\end{tabular}
\end{table}

\subsection{Assumptions on the AGN-hosted GW population}\label{subsect:AGNchoices}
We have adopted the simplest model for the AGN-hosted GW population, where BBH mergers occur at the same source-frame rate in all AGN. However, this may not be an accurate assumption, as some AGN may be more likely to host a GW event than others, biasing our inference. Indeed, studies of dark-siren cosmology show that incorrect galaxy weighting schemes may lead to statistical biases \citep{Perna:2025, Hanselman:2025}. A usual practice in such cosmological studies is to weight galaxies by a luminosity that traces the physical property governing the BBH merger rate (e.g., \citealt{Fishbach:2019, Gray:2020, gwtc5:2026cosmo}). For example, the B-band luminosity traces the star-formation rate \citep{Singer:2016gtd}, while the K-band luminosity traces the galaxy stellar mass \citep{bell:2003K}.

Such a luminosity-weighting scheme may be of interest for future modelling of the AGN channel as well. Namely, multiple simulations show that intrinsic AGN properties such as the accretion rate, BH mass and disc viscosity can influence the AGN-assisted BBH merger rate \citep{Yang:2019q, Tagawa:2020, Tagawa:2026agn, Vaccaro:2026agn}. Mapping these properties onto a luminosity band is not trivial, since this will be dependent on the specific mechanisms at play within the AGN channel. For example, if in-situ formed BBHs dominate the AGN channel, the merger rate increases with the AGN accretion rate such that luminous AGN would be important GW sources \citep{Yang:2019q}. If instead orbital alignment of pre-existing (binary) BHs is the dominant process by which BBHs are embedded in accretion discs, or if they predominantly merge in migration traps, lower-luminosity AGN are preferred \citep{Yang:2019q, Gilbaum:2024}. In this case, \citet{Ford:2025rev} show that the AGN X-ray luminosity can be a tracer for the growth of nuclear star clusters over time.

Luminosity weighting changes the redshift distribution of AGN-origin GWs, since the redshift distribution of AGN depends on their luminosity. The impact of this dependence on our results is reduced, since we split Quaia into subcatalogues spanning different bolometric luminosity ranges. This limits the dynamic range of AGN luminosities and therefore of the merger rate.

We have assumed that the intrinsic GW parameters of AGN-origin and alternative-origin GWs are both distributed identically to the entire BBH population. If alternative formation channels dominate the BBH population, this may bias $\fagn$ low as our model is tuned to the alternative-origin GW population, and we do not model a preference for AGN-channel signatures such as mass-gap or high-spin BHs. However, the extent to which intrinsic parameters of AGN-channel BBHs differ from those of other formation channels remains unconstrained. Such signatures are expected to arise predominantly in higher-generation mergers, whereas if the AGN channel is dominated by first-generation mergers (e.g., \citealt{McKernan:2024fact, Cook:2024}), its intrinsic parameter distributions may be difficult to distinguish from those of other channels.

Adopting fixed population models such as ours is likely to cause systematic biases, given the large parameter space and uncertainties in modelling the properties of AGN and nuclear star clusters, and how they impact the BBH population. In the future, flexible models that parametrise these effects could be constrained using our statistical framework, providing a direct probe of GW populations, nuclear star clusters, and AGN accretion discs regardless of the value of $\fagn$ \citep{Ford:2025rev}.

\section{Summary and conclusions}\label{sect:conclusions}

Constraining the origin of the GWs observed with the LVK detectors is essential for understanding the formation channels and environments of stellar-mass BBHs. We have adapted the dark-siren framework to distinguish between GWs produced through the AGN channel and those produced through any alternative channel. Informing our likelihood with 256 GWs from BBHs in GWTC-5.0 and the Quaia AGN catalogue, we place constraints on the fraction of the total number of GWs emitted in the Universe that are generated in AGN ($\fagn$). Our method generalises previous spatial-correlation analyses and unifies them with population inference. We now include GW selection effects, redshift-dependent merger rates, full AGN redshift posteriors, a spatially varying EM selection function, and consistent modelling of the AGN distribution. 

From our mock data analysis, we conclude the following:
\begin{itemize}
    \item A perfect AGN catalogue can constrain $\fagn$ as effectively as hundreds of additional O5-sensitivity GW detections without EM information (Fig. \ref{fig:sig_v_thresh}). However, once the error budget on $\fagn$ is dominated by the number of detected GWs, the precision gained with an AGN catalogue is limited. Nevertheless, we argue that an AGN catalogue can help mitigate systematic biases from GW population modelling.    

    \item The constraining power of an AGN catalogue is limited by its completeness and redshift precision. To achieve near-optimal constraints, the catalogue must be complete within GW localisations, and typical AGN redshift uncertainties should be roughly an order of magnitude smaller than those of the GW sample.

    \item Neglecting AGN redshift uncertainties in EM-informed population inference biases $\fagn$ downwards.
\end{itemize}
Our conclusions from analysing real data are:
\begin{itemize}
    \item We obtain 95 percent CL upper limits of $\fagn \leq \{0.52,0.41, 0.61\}$ for unobscured AGN with bolometric luminosities $L_{\rm bol} \geq \{10^{44.5},10^{45.5},10^{46.5}\} \, \mathrm{erg \, s^{-1}}$. The information supplied by Quaia is limited by its redshift precision. Our constraints on $\fagn$ for all AGN populations considered are robust under sampling our alternative-origin GW model in narrow Gaussian priors around our fiducial model. Constraining power is reduced when adopting broad uniform priors instead, indicating our results are moderately model dependent. More GW observations at high redshift are required to lift the degeneracies driving this dependence.

    \item No individual GW event is more likely to have an AGN origin than an alternative origin. The probability of an AGN origin scales with GW redshift. The event GW230704\_212616 has the highest AGN-origin probability, with $p(\mathrm{AGN}) \approx 0.250$ considering AGN with $\lbol{44.5}$. Nevertheless, we find a 98.1 percent probability that at least one GW in the observed dataset originated in an AGN.

    \item The AGN-channel merger rate evolution for AGN with $\lbol{44.5}$ is consistent up to $z \approx 1$ with the recently-inferred \citep{Farah:2026} hierarchically merging subpopulation of BBHs for $\fagn = 0.07$. We expect to reach the level of precision necessary to test the AGN channel as the dominant source of these hierarchical mergers with less than 2.5 years of LVK O5 observations. This is within the currently scheduled runtime of O5.
\end{itemize}

Our statistical framework, combined with future GW and EM data, can be used to directly probe GW populations and the properties of AGN accretion discs and their surrounding nuclear star clusters.

\section*{Acknowledgements}
This work was performed using the compute resources from the Academic Leiden Interdisciplinary Cluster Environment (ALICE) provided by Leiden University.
L.~S.\ is supported by the European Space Agency Research Fellowship Programme. EMR acknowledges support from European Research Council (ERC) grant number: 101002511/project acronym: VEGA P and  the grant VI.C.232.099 of the research programme VICI, which is financed by NWO.
This research has made use of data or software obtained from the Gravitational Wave Open Science Center (gwosc.org), a service of the LIGO Scientific Collaboration, the Virgo Collaboration, and KAGRA. This material is based upon work supported by NSF's LIGO Laboratory which is a major facility fully funded by the National Science Foundation, as well as the Science and Technology Facilities Council (STFC) of the United Kingdom, the Max-Planck-Society (MPS), and the State of Niedersachsen/Germany for support of the construction of Advanced LIGO and construction and operation of the GEO600 detector. Additional support for Advanced LIGO was provided by the Australian Research Council. Virgo is funded, through the European Gravitational Observatory (EGO), by the French Centre National de Recherche Scientifique (CNRS), the Italian Istituto Nazionale di Fisica Nucleare (INFN) and the Dutch Nikhef, with contributions by institutions from Belgium, Germany, Greece, Hungary, Ireland, Japan, Monaco, Poland, Portugal, Spain. KAGRA is supported by Ministry of Education, Culture, Sports, Science and Technology (MEXT), Japan Society for the Promotion of Science (JSPS) in Japan; National Research Foundation (NRF) and Ministry of Science and ICT (MSIT) in Korea; Academia Sinica (AS) and National Science and Technology Council (NSTC) in Taiwan.

\section*{Data Availability}
The code used in this work is publicly available at \url{https://github.com/LucasPouw/darksirenpop}.

\bibliographystyle{mnras}
\bibliography{references} %

@article{Loredo:2004nn,
    author = "Loredo, Thomas J.",
    editor = "Fischer, Rainer and Preuss, Roland and von Toussaint, Udo",
    title = "{Accounting for source uncertainties in analyses of astronomical survey data}",
    eprint = "astro-ph/0409387",
    archivePrefix = "arXiv",
    doi = "10.1063/1.1835214",
    journal = "AIP Conf. Proc.",
    volume = "735",
    number = "1",
    pages = "195--206",
    year = "2004"
}

@ARTICLE{kulkarni:2019,
       author = {{Kulkarni}, Girish and {Worseck}, G{\'a}bor and {Hennawi}, Joseph F.},
        title = "{Evolution of the AGN UV luminosity function from redshift 7.5}",
      journal = {\mnras},
         year = 2019,
        month = sep,
       volume = {488},
       number = {1},
        pages = {1035-1065},
          doi = {10.1093/mnras/stz1493},
archivePrefix = {arXiv},
       eprint = {1807.09774},
 primaryClass = {astro-ph.GA},
       adsurl = {https://ui.adsabs.harvard.edu/abs/2019MNRAS.488.1035K}
}

@ARTICLE{Sidery:2014PP,
       author = {{Sidery}, T. and {Aylott}, B. and {Christensen}, N. and {Farr}, B. and {Farr}, W. and {Feroz}, F. and {Gair}, J. and {Grover}, K. and {Graff}, P. and {Hanna}, C. and {Kalogera}, V. and {Mandel}, I. and {O'Shaughnessy}, R. and {Pitkin}, M. and {Price}, L. and {Raymond}, V. and {R{\"o}ver}, C. and {Singer}, L. and {van der Sluys}, M. and {Smith}, R.~J.~E. and {Vecchio}, A. and {Veitch}, J. and {Vitale}, S.},
        title = "{Reconstructing the sky location of gravitational-wave detected compact binary systems: Methodology for testing and comparison}",
      journal = {\prd},
         year = 2014,
        month = apr,
       volume = {89},
       number = {8},
          eid = {084060},
        pages = {084060},
          doi = {10.1103/PhysRevD.89.084060},
archivePrefix = {arXiv},
       eprint = {1312.6013},
 primaryClass = {astro-ph.IM},
       adsurl = {https://ui.adsabs.harvard.edu/abs/2014PhRvD..89h4060S}
}

@ARTICLE{Bartos:2026spin,
       author = {{Bartos}, I. and {Haiman}, Z.},
        title = "{High-Spin BBH Subpopulation from AGN Accretion}",
      journal = {arXiv e-prints},
         year = 2026,
        month = may,
          eid = {arXiv:2605.09351},
        pages = {arXiv:2605.09351},
          doi = {10.48550/arXiv.2605.09351},
archivePrefix = {arXiv},
       eprint = {2605.09351},
 primaryClass = {astro-ph.HE},
       adsurl = {https://ui.adsabs.harvard.edu/abs/2026arXiv260509351B}
}

@ARTICLE{RomeroShaw:2020PP,
       author = {{Romero-Shaw}, I.~M. and {Talbot}, C. and {Biscoveanu}, S. and {D'Emilio}, V. and {Ashton}, G. and {Berry}, C.~P.~L. and {Coughlin}, S. and {Galaudage}, S. and {Hoy}, C. and {H{\"u}bner}, M. and {Phukon}, K.~S. and {Pitkin}, M. and {Rizzo}, M. and {Sarin}, N. and {Smith}, R. and {Stevenson}, S. and {Vajpeyi}, A. and {Ar{\`e}ne}, M. and {Athar}, K. and {Banagiri}, S. and {Bose}, N. and {Carney}, M. and {Chatziioannou}, K. and {Clark}, J.~A. and {Colleoni}, M. and {Cotesta}, R. and {Edelman}, B. and {Estell{\'e}s}, H. and {Garc{\'\i}a-Quir{\'o}s}, C. and {Ghosh}, Abhirup and {Green}, R. and {Haster}, C.-J. and {Husa}, S. and {Keitel}, D. and {Kim}, A.~X. and {Hernandez-Vivanco}, F. and {Maga{\~n}a Hernandez}, I. and {Karathanasis}, C. and {Lasky}, P.~D. and {De Lillo}, N. and {Lower}, M.~E. and {Macleod}, D. and {Mateu-Lucena}, M. and {Miller}, A. and {Millhouse}, M. and {Morisaki}, S. and {Oh}, S.~H. and {Ossokine}, S. and {Payne}, E. and {Powell}, J. and {Pratten}, G. and {P{\"u}rrer}, M. and {Ramos-Buades}, A. and {Raymond}, V. and {Thrane}, E. and {Veitch}, J. and {Williams}, D. and {Williams}, M.~J. and {Xiao}, L.},
        title = "{Bayesian inference for compact binary coalescences with BILBY: validation and application to the first LIGO-Virgo gravitational-wave transient catalogue}",
      journal = {\mnras},
         year = 2020,
        month = dec,
       volume = {499},
       number = {3},
        pages = {3295-3319},
          doi = {10.1093/mnras/staa2850},
archivePrefix = {arXiv},
       eprint = {2006.00714},
 primaryClass = {astro-ph.IM},
       adsurl = {https://ui.adsabs.harvard.edu/abs/2020MNRAS.499.3295R}
}

@ARTICLE{Runnoe:2012,
       author = {{Runnoe}, Jessie C. and {Brotherton}, Michael S. and {Shang}, Zhaohui},
        title = "{Updating quasar bolometric luminosity corrections}",
      journal = {\mnras},
         year = 2012,
        month = may,
       volume = {422},
       number = {1},
        pages = {478-493},
          doi = {10.1111/j.1365-2966.2012.20620.x},
archivePrefix = {arXiv},
       eprint = {1201.5155},
 primaryClass = {astro-ph.CO},
       adsurl = {https://ui.adsabs.harvard.edu/abs/2012MNRAS.422..478R}
}

@ARTICLE{Tagawa:2026agn,
       author = {{Tagawa}, Hiromichi and {Haiman}, Zolt{\'a}n and {Kocsis}, Bence},
        title = "{Properties of black hole mergers in disks of active galactic nuclei}",
      journal = {arXiv e-prints},
         year = 2026,
        month = apr,
          eid = {arXiv:2604.25994},
        pages = {arXiv:2604.25994},
          doi = {10.48550/arXiv.2604.25994},
archivePrefix = {arXiv},
       eprint = {2604.25994},
 primaryClass = {astro-ph.HE},
       adsurl = {https://ui.adsabs.harvard.edu/abs/2026arXiv260425994T}
}

@ARTICLE{Strolger:2020,
       author = {{Strolger}, Louis-Gregory and {Rodney}, Steven A. and {Pacifici}, Camilla and {Narayan}, Gautham and {Graur}, Or},
        title = "{Delay Time Distributions of Type Ia Supernovae from Galaxy and Cosmic Star Formation Histories}",
      journal = {\apj},
         year = 2020,
        month = feb,
       volume = {890},
       number = {2},
          eid = {140},
        pages = {140},
          doi = {10.3847/1538-4357/ab6a97},
archivePrefix = {arXiv},
       eprint = {2001.05967},
 primaryClass = {astro-ph.GA},
       adsurl = {https://ui.adsabs.harvard.edu/abs/2020ApJ...890..140S}
}

@ARTICLE{Astropy:2022,
       author = {{Astropy Collaboration} and {Price-Whelan}, Adrian M. and {Lim}, Pey Lian and {Earl}, Nicholas and {Starkman}, Nathaniel and {Bradley}, Larry and {Shupe}, David L. and {Patil}, Aarya A. and {Corrales}, Lia and {Brasseur}, C.~E. and {N{\"o}the}, Maximilian and {Donath}, Axel and {Tollerud}, Erik and {Morris}, Brett M. and {Ginsburg}, Adam and {Vaher}, Eero and {Weaver}, Benjamin A. and {Tocknell}, James and {Jamieson}, William and {van Kerkwijk}, Marten H. and {Robitaille}, Thomas P. and {Merry}, Bruce and {Bachetti}, Matteo and {G{\"u}nther}, H. Moritz and {Aldcroft}, Thomas L. and {Alvarado-Montes}, Jaime A. and {Archibald}, Anne M. and {B{\'o}di}, Attila and {Bapat}, Shreyas and {Barentsen}, Geert and {Baz{\'a}n}, Juanjo and {Biswas}, Manish and {Boquien}, M{\'e}d{\'e}ric and {Burke}, D.~J. and {Cara}, Daria and {Cara}, Mihai and {Conroy}, Kyle E. and {Conseil}, Simon and {Craig}, Matthew W. and {Cross}, Robert M. and {Cruz}, Kelle L. and {D'Eugenio}, Francesco and {Dencheva}, Nadia and {Devillepoix}, Hadrien A.~R. and {Dietrich}, J{\"o}rg P. and {Eigenbrot}, Arthur Davis and {Erben}, Thomas and {Ferreira}, Leonardo and {Foreman-Mackey}, Daniel and {Fox}, Ryan and {Freij}, Nabil and {Garg}, Suyog and {Geda}, Robel and {Glattly}, Lauren and {Gondhalekar}, Yash and {Gordon}, Karl D. and {Grant}, David and {Greenfield}, Perry and {Groener}, Austen M. and {Guest}, Steve and {Gurovich}, Sebastian and {Handberg}, Rasmus and {Hart}, Akeem and {Hatfield-Dodds}, Zac and {Homeier}, Derek and {Hosseinzadeh}, Griffin and {Jenness}, Tim and {Jones}, Craig K. and {Joseph}, Prajwel and {Kalmbach}, J. Bryce and {Karamehmetoglu}, Emir and {Ka{\l}uszy{\'n}ski}, Miko{\l}aj and {Kelley}, Michael S.~P. and {Kern}, Nicholas and {Kerzendorf}, Wolfgang E. and {Koch}, Eric W. and {Kulumani}, Shankar and {Lee}, Antony and {Ly}, Chun and {Ma}, Zhiyuan and {MacBride}, Conor and {Maljaars}, Jakob M. and {Muna}, Demitri and {Murphy}, N.~A. and {Norman}, Henrik and {O'Steen}, Richard and {Oman}, Kyle A. and {Pacifici}, Camilla and {Pascual}, Sergio and {Pascual-Granado}, J. and {Patil}, Rohit R. and {Perren}, Gabriel I. and {Pickering}, Timothy E. and {Rastogi}, Tanuj and {Roulston}, Benjamin R. and {Ryan}, Daniel F. and {Rykoff}, Eli S. and {Sabater}, Jose and {Sakurikar}, Parikshit and {Salgado}, Jes{\'u}s and {Sanghi}, Aniket and {Saunders}, Nicholas and {Savchenko}, Volodymyr and {Schwardt}, Ludwig and {Seifert-Eckert}, Michael and {Shih}, Albert Y. and {Jain}, Anany Shrey and {Shukla}, Gyanendra and {Sick}, Jonathan and {Simpson}, Chris and {Singanamalla}, Sudheesh and {Singer}, Leo P. and {Singhal}, Jaladh and {Sinha}, Manodeep and {Sip{\H{o}}cz}, Brigitta M. and {Spitler}, Lee R. and {Stansby}, David and {Streicher}, Ole and {{\v{S}}umak}, Jani and {Swinbank}, John D. and {Taranu}, Dan S. and {Tewary}, Nikita and {Tremblay}, Grant R. and {de Val-Borro}, Miguel and {Van Kooten}, Samuel J. and {Vasovi{\'c}}, Zlatan and {Verma}, Shresth and {de Miranda Cardoso}, Jos{\'e} Vin{\'\i}cius and {Williams}, Peter K.~G. and {Wilson}, Tom J. and {Winkel}, Benjamin and {Wood-Vasey}, W.~M. and {Xue}, Rui and {Yoachim}, Peter and {Zhang}, Chen and {Zonca}, Andrea and {Astropy Project Contributors}},
        title = "{The Astropy Project: Sustaining and Growing a Community-oriented Open-source Project and the Latest Major Release (v5.0) of the Core Package}",
      journal = {\apj},
         year = 2022,
        month = aug,
       volume = {935},
       number = {2},
          eid = {167},
        pages = {167},
          doi = {10.3847/1538-4357/ac7c74},
archivePrefix = {arXiv},
       eprint = {2206.14220},
 primaryClass = {astro-ph.IM},
       adsurl = {https://ui.adsabs.harvard.edu/abs/2022ApJ...935..167A}
}

@ARTICLE{Quaia:2024,
       author = {{Storey-Fisher}, Kate and {Hogg}, David W. and {Rix}, Hans-Walter and {Eilers}, Anna-Christina and {Fabbian}, Giulio and {Blanton}, Michael R. and {Alonso}, David},
        title = "{Quaia, the Gaia-unWISE Quasar Catalog: An All-sky Spectroscopic Quasar Sample}",
      journal = {\apj},
         year = 2024,
        month = mar,
       volume = {964},
       number = {1},
          eid = {69},
        pages = {69},
          doi = {10.3847/1538-4357/ad1328},
archivePrefix = {arXiv},
       eprint = {2306.17749},
 primaryClass = {astro-ph.GA},
       adsurl = {https://ui.adsabs.harvard.edu/abs/2024ApJ...964...69S}
}

@ARTICLE{Thrane:2019,
       author = {{Thrane}, Eric and {Talbot}, Colm},
        title = "{An introduction to Bayesian inference in gravitational-wave astronomy: Parameter estimation, model selection, and hierarchical models}",
      journal = {\pasa},
         year = 2019,
        month = mar,
       volume = {36},
          eid = {e010},
        pages = {e010},
          doi = {10.1017/pasa.2019.2},
archivePrefix = {arXiv},
       eprint = {1809.02293},
 primaryClass = {astro-ph.IM},
       adsurl = {https://ui.adsabs.harvard.edu/abs/2019PASA...36...10T}
}

@ARTICLE{Rowan:2024cap,
       author = {{Rowan}, Connar and {Whitehead}, Henry and {Kocsis}, Bence},
        title = "{Black Hole Merger Rates in AGN: contribution from gas-captured binaries}",
      journal = {arXiv e-prints},
         year = 2024,
        month = dec,
          eid = {arXiv:2412.12086},
        pages = {arXiv:2412.12086},
          doi = {10.48550/arXiv.2412.12086},
archivePrefix = {arXiv},
       eprint = {2412.12086},
 primaryClass = {astro-ph.HE},
       adsurl = {https://ui.adsabs.harvard.edu/abs/2024arXiv241212086R}
}

@ARTICLE{Ostriker:1983,
       author = {{Ostriker}, J.~P.},
        title = "{Viscous drag on an accretion disk due to an embedded stellar system.}",
      journal = {\apj},
         year = 1983,
        month = oct,
       volume = {273},
        pages = {99-104},
          doi = {10.1086/161351},
       adsurl = {https://ui.adsabs.harvard.edu/abs/1983ApJ...273...99O}
}

@ARTICLE{Syer:1991,
       author = {{Syer}, D. and {Clarke}, C.~J. and {Rees}, M.~J.},
        title = "{Star-disc interactions near a massive black hole}",
      journal = {\mnras},
         year = 1991,
        month = jun,
       volume = {250},
        pages = {505-512},
          doi = {10.1093/mnras/250.3.505},
       adsurl = {https://ui.adsabs.harvard.edu/abs/1991MNRAS.250..505S}
}

@ARTICLE{Nasim:2023,
       author = {{Nasim}, Syeda S. and {Fabj}, Gaia and {Caban}, Freddy and {Secunda}, Amy and {Ford}, K.~E. Saavik and {McKernan}, Barry and {Bellovary}, Jillian M. and {Leigh}, Nathan W.~C. and {Lyra}, Wladimir},
        title = "{Aligning Retrograde Nuclear Cluster Orbits with an Active Galactic Nucleus Accretion Disc}",
      journal = {\mnras},
         year = 2023,
        month = jul,
       volume = {522},
       number = {4},
        pages = {5393-5401},
          doi = {10.1093/mnras/stad1295},
archivePrefix = {arXiv},
       eprint = {2207.09540},
 primaryClass = {astro-ph.GA},
       adsurl = {https://ui.adsabs.harvard.edu/abs/2023MNRAS.522.5393N}
}

@ARTICLE{LIGO:2025gwm,
       author = {{The LIGO Scientific Collaboration} and {the Virgo Collaboration} and {the KAGRA Collaboration}},
        title = "{GW231123: a Binary Black Hole Merger with Total Mass 190-265 $M_{\odot}$}",
      journal = {arXiv e-prints},
         year = 2025,
        month = jul,
          eid = {arXiv:2507.08219},
        pages = {arXiv:2507.08219},
          doi = {10.48550/arXiv.2507.08219},
archivePrefix = {arXiv},
       eprint = {2507.08219},
 primaryClass = {astro-ph.HE},
       adsurl = {https://ui.adsabs.harvard.edu/abs/2025arXiv250708219T}
}

@ARTICLE{emcee:2013,
       author = {{Foreman-Mackey}, Daniel and {Hogg}, David W. and {Lang}, Dustin and {Goodman}, Jonathan},
        title = "{emcee: The MCMC Hammer}",
      journal = {\pasp},
         year = 2013,
        month = mar,
       volume = {125},
       number = {925},
        pages = {306},
          doi = {10.1086/670067},
archivePrefix = {arXiv},
       eprint = {1202.3665},
 primaryClass = {astro-ph.IM},
       adsurl = {https://ui.adsabs.harvard.edu/abs/2013PASP..125..306F}
}

@ARTICLE{Euclid:overview,
       author = {{Euclid Collaboration} and {Mellier}, Y. and {Abdurro'uf} and {Acevedo Barroso}, J.~A. and {Ach{\'u}carro}, A. and {Adamek}, J. and {Adam}, R. and {Addison}, G.~E. and {Aghanim}, N. and {Aguena}, M. and {Ajani}, V. and {Akrami}, Y. and {Al-Bahlawan}, A. and {Alavi}, A. and {Albuquerque}, I.~S. and {Alestas}, G. and {Alguero}, G. and {Allaoui}, A. and {Allen}, S.~W. and {Allevato}, V. and {Alonso-Tetilla}, A.~V. and {Altieri}, B. and {Alvarez-Candal}, A. and {Alvi}, S. and {Amara}, A. and {Amendola}, L. and {Amiaux}, J. and {Andika}, I.~T. and {Andreon}, S. and {Andrews}, A. and {Angora}, G. and {Angulo}, R.~E. and {Annibali}, F. and {Anselmi}, A. and {Anselmi}, S. and {Arcari}, S. and {Archidiacono}, M. and {Aric{\`o}}, G. and {Arnaud}, M. and {Arnouts}, S. and {Asgari}, M. and {Asorey}, J. and {Atayde}, L. and {Atek}, H. and {Atrio-Barandela}, F. and {Aubert}, M. and {Aubourg}, E. and {Auphan}, T. and {Auricchio}, N. and {Aussel}, B. and {Aussel}, H. and {Avelino}, P.~P. and {Avgoustidis}, A. and {Avila}, S. and {Awan}, S. and {Azzollini}, R. and {Baccigalupi}, C. and {Bachelet}, E. and {Bacon}, D. and {Baes}, M. and {Bagley}, M.~B. and {Bahr-Kalus}, B. and {Balaguera-Antolinez}, A. and {Balbinot}, E. and {Balcells}, M. and {Baldi}, M. and {Baldry}, I. and {Balestra}, A. and {Ballardini}, M. and {Ballester}, O. and {Balogh}, M. and {Ba{\~n}ados}, E. and {Barbier}, R. and {Bardelli}, S. and {Baron}, M. and {Barreiro}, T. and {Barrena}, R. and {Barriere}, J.-C. and {Barros}, B.~J. and {Barthelemy}, A. and {Bartolo}, N. and {Basset}, A. and {Battaglia}, P. and {Battisti}, A.~J. and {Baugh}, C.~M. and {Baumont}, L. and {Bazzanini}, L. and {Beaulieu}, J.-P. and {Beckmann}, V. and {Belikov}, A.~N. and {Bel}, J. and {Bellagamba}, F. and {Bella}, M. and {Bellini}, E. and {Benabed}, K. and {Bender}, R. and {Benevento}, G. and {Bennett}, C.~L. and {Benson}, K. and {Bergamini}, P. and {Bermejo-Climent}, J.~R. and {Bernardeau}, F. and {Bertacca}, D. and {Berthe}, M. and {Berthier}, J. and {Bethermin}, M. and {Beutler}, F. and {Bevillon}, C. and {Bhargava}, S. and {Bhatawdekar}, R. and {Bianchi}, D. and {Bisigello}, L. and {Biviano}, A. and {Blake}, R.~P. and {Blanchard}, A. and {Blazek}, J. and {Blot}, L. and {Bosco}, A. and {Bodendorf}, C. and {Boenke}, T. and {B{\"o}hringer}, H. and {Boldrini}, P. and {Bolzonella}, M. and {Bonchi}, A. and {Bonici}, M. and {Bonino}, D. and {Bonino}, L. and {Bonvin}, C. and {Bon}, W. and {Booth}, J.~T. and {Borgani}, S. and {Borlaff}, A.~S. and {Borsato}, E. and {Bose}, B. and {Botticella}, M.~T. and {Boucaud}, A. and {Bouche}, F. and {Boucher}, J.~S. and {Boutigny}, D. and {Bouvard}, T. and {Bouwens}, R. and {Bouy}, H. and {Bowler}, R.~A.~A. and {Bozza}, V. and {Bozzo}, E. and {Branchini}, E. and {Brando}, G. and {Brau-Nogue}, S. and {Brekke}, P. and {Bremer}, M.~N. and {Brescia}, M. and {Breton}, M.-A. and {Brinchmann}, J. and {Brinckmann}, T. and {Brockley-Blatt}, C. and {Brodwin}, M. and {Brouard}, L. and {Brown}, M.~L. and {Bruton}, S. and {Bucko}, J. and {Buddelmeijer}, H. and {Buenadicha}, G. and {Buitrago}, F. and {Burger}, P. and {Burigana}, C. and {Busillo}, V. and {Busonero}, D. and {Cabanac}, R. and {Cabayol-Garcia}, L. and {Cagliari}, M.~S. and {Caillat}, A. and {Caillat}, L. and {Calabrese}, M. and {Calabro}, A. and {Calderone}, G. and {Calura}, F. and {Camacho Quevedo}, B. and {Camera}, S. and {Campos}, L. and {Ca{\~n}as-Herrera}, G. and {Candini}, G.~P. and {Cantiello}, M. and {Capobianco}, V. and {Cappellaro}, E. and {Cappelluti}, N. and {Cappi}, A. and {Caputi}, K.~I. and {Cara}, C. and {Carbone}, C. and {Cardone}, V.~F. and {Carella}, E. and {Carlberg}, R.~G. and {Carle}, M. and {Carminati}, L. and {Caro}, F. and {Carrasco}, J.~M. and {Carretero}, J. and {Carrilho}, P. and {Carron Duque}, J. and {Carry}, B.},
        title = "{Euclid: I. Overview of the Euclid mission}",
      journal = {\aap},
         year = 2025,
        month = may,
       volume = {697},
          eid = {A1},
        pages = {A1},
          doi = {10.1051/0004-6361/202450810},
archivePrefix = {arXiv},
       eprint = {2405.13491},
 primaryClass = {astro-ph.CO},
       adsurl = {https://ui.adsabs.harvard.edu/abs/2025A&A...697A...1E}
}

@ARTICLE{Euclid:nisp,
       author = {{Euclid Collaboration} and {Jahnke}, K. and {Gillard}, W. and {Schirmer}, M. and {Ealet}, A. and {Maciaszek}, T. and {Prieto}, E. and {Barbier}, R. and {Bonoli}, C. and {Corcione}, L. and {Dusini}, S. and {Grupp}, F. and {Hormuth}, F. and {Ligori}, S. and {Martin}, L. and {Morgante}, G. and {Padilla}, C. and {Toledo-Moreo}, R. and {Trifoglio}, M. and {Valenziano}, L. and {Bender}, R. and {Castander}, F.~J. and {Garilli}, B. and {Lilje}, P.~B. and {Rix}, H.-W. and {Andersen}, M.~I. and {Auricchio}, N. and {Balestra}, A. and {Barriere}, J.-C. and {Battaglia}, P. and {Berthe}, M. and {Bodendorf}, C. and {Boenke}, T. and {Bon}, W. and {Bonnefoi}, A. and {Caillat}, A. and {Capobianco}, V. and {Carle}, M. and {Casas}, R. and {Cho}, H. and {Costille}, A. and {Ducret}, F. and {Ferriol}, S. and {Franceschi}, E. and {Gimenez}, J.-L. and {Holmes}, W. and {Hornstrup}, A. and {Jhabvala}, M. and {Kohley}, R. and {Kubik}, B. and {Laureijs}, R. and {Le Mignant}, D. and {Lloro}, I. and {Medinaceli}, E. and {Mellier}, Y. and {Polenta}, G. and {Racca}, G.~D. and {Renzi}, A. and {Salvignol}, J.-C. and {Secroun}, A. and {Seidel}, G. and {Seiffert}, M. and {Sirignano}, C. and {Sirri}, G. and {Strada}, P. and {Smadja}, G. and {Stanco}, L. and {Wachter}, S. and {Anselmi}, S. and {Borsato}, E. and {Caillat}, L. and {Cogato}, F. and {Colodro-Conde}, C. and {Crouzet}, P.-E. and {Conforti}, V. and {D'Alessandro}, M. and {Copin}, Y. and {Cuillandre}, J.-C. and {Davies}, J.~E. and {Davini}, S. and {Derosa}, A. and {Diaz}, J.~J. and {Di Domizio}, S. and {Di Ferdinando}, D. and {Farinelli}, R. and {Ferrari}, A.~G. and {Fornari}, F. and {Gabarra}, L. and {Garcia}, R. and {Gutierrez}, C.~M. and {Giacomini}, F. and {Lagier}, P. and {Gianotti}, F. and {Krause}, O. and {Madrid}, F. and {Laudisio}, F. and {Macias-Perez}, J. and {Naletto}, G. and {Niclas}, M. and {Marpaud}, J. and {Mauri}, N. and {da Silva}, R. and {Passalacqua}, F. and {Paterson}, K. and {Patrizii}, L. and {Risso}, I. and {Solheim}, B.~G.~B. and {Scodeggio}, M. and {Stassi}, P. and {Steinwagner}, J. and {Tenti}, M. and {Testera}, G. and {Travaglini}, R. and {Tosi}, S. and {Troja}, A. and {Tubio}, O. and {Valieri}, C. and {Vescovi}, C. and {Ventura}, S. and {Aghanim}, N. and {Altieri}, B. and {Amara}, A. and {Amiaux}, J. and {Andreon}, S. and {Appleton}, P.~N. and {Aussel}, H. and {Baccigalupi}, C. and {Baldi}, M. and {Bardelli}, S. and {Basset}, A. and {Bonchi}, A. and {Bonino}, D. and {Branchini}, E. and {Brescia}, M. and {Brinchmann}, J. and {Camera}, S. and {Carbone}, C. and {Cardone}, V.~F. and {Carretero}, J. and {Casas}, S. and {Castellano}, M. and {Castignani}, G. and {Cavuoti}, S. and {Chabaud}, P.-Y. and {Cimatti}, A. and {Congedo}, G. and {Conselice}, C.~J. and {Conversi}, L. and {Courbin}, F. and {Courtois}, H.~M. and {Crocce}, M. and {Cropper}, M. and {Cuby}, J.-G. and {Da Silva}, A. and {Degaudenzi}, H. and {De Lucia}, G. and {Di Giorgio}, A.~M. and {Dinis}, J. and {Douspis}, M. and {Dubath}, F. and {Duncan}, C.~A.~J. and {Dupac}, X. and {Fabricius}, M. and {Farina}, M. and {Farrens}, S. and {Faustini}, F. and {Fosalba}, P. and {Fotopoulou}, S. and {Fourmanoit}, N. and {Frailis}, M. and {Franzetti}, P. and {Galeotta}, S. and {George}, K. and {Gillis}, B. and {Giocoli}, C. and {G{\'o}mez-Alvarez}, P. and {Granett}, B.~R. and {Grazian}, A. and {Guzzo}, L. and {Hailey}, M. and {Haugan}, S.~V.~H. and {Hoar}, J. and {Hoekstra}, H. and {Hook}, I. and {Hudelot}, P. and {Ili{\'c}}, S. and {Joachimi}, B. and {Keih{\"a}nen}, E. and {Kermiche}, S. and {Kiessling}, A. and {Kilbinger}, M. and {Kitching}, T. and {K{\"u}mmel}, M. and {Kunz}, M. and {Kurki-Suonio}, H. and {Lahav}, O. and {Liebing}, P. and {Lindholm}, V. and {Lorenzo Alvarez}, J. and {Mainetti}, G.},
        title = "{Euclid: III. The NISP Instrument}",
      journal = {\aap},
         year = 2025,
        month = may,
       volume = {697},
          eid = {A3},
        pages = {A3},
          doi = {10.1051/0004-6361/202450786},
archivePrefix = {arXiv},
       eprint = {2405.13493},
 primaryClass = {astro-ph.IM},
       adsurl = {https://ui.adsabs.harvard.edu/abs/2025A&A...697A...3E}
}

@ARTICLE{bell:2003K,
       author = {{Bell}, Eric F. and {McIntosh}, Daniel H. and {Katz}, Neal and {Weinberg}, Martin D.},
        title = "{The Optical and Near-Infrared Properties of Galaxies. I. Luminosity and Stellar Mass Functions}",
      journal = {\apjs},
         year = 2003,
        month = dec,
       volume = {149},
       number = {2},
        pages = {289-312},
          doi = {10.1086/378847},
archivePrefix = {arXiv},
       eprint = {astro-ph/0302543},
 primaryClass = {astro-ph},
       adsurl = {https://ui.adsabs.harvard.edu/abs/2003ApJS..149..289B}
}

@ARTICLE{gwtc5:2026cosmo,
       author = {{The LIGO Scientific Collaboration} and {the Virgo Collaboration} and {the KAGRA Collaboration} and {Abac}, A.~G. and {Abe}, A. and {Abouelfettouh}, I. and {Acernese}, F. and {Ackley}, K. and {Adam}, A. and {Adhicary}, S. and {Adhikari}, D. and {Adhikari}, R.~X. and {Adkins}, V.~K. and {Afroz}, S. and {Agapito}, A. and {Agarwal}, D. and {Agathos}, M. and {Aggarwal}, N. and {Aggarwal}, S. and {Aguiar}, O.~D. and {Ahrend}, I.-L. and {Aiello}, L. and {Ain}, A. and {Ajith}, P. and {Akutsu}, T. and {Albers}, L. and {Ali}, W. and {Al-Kershi}, S. and {Allene}, C. and {Allocca}, A. and {Al-Shammari}, S. and {Alvarez}, J.~A. and {Alvarez-Lopez}, S. and {Amar}, W. and {Amarasinghe}, O. and {Amato}, A. and {Amicucci}, F. and {Amra}, C. and {Anand}, A.~B. and {Anand}, C. and {Ananyeva}, A. and {Anderson}, S.~B. and {Anderson}, W.~G. and {Andia}, M. and {Ando}, M. and {Andrade-Oliveira}, F. and {Andr{\'e}s-Carcasona}, M. and {Andrey}, J.~L. and {Andri{\'c}}, T. and {Anglin}, J. and {Anna}, J. and {Antelis}, J.~M. and {Antier}, S. and {Aoki}, T. and {Aoumi}, M. and {Appavuravther}, E.~Z. and {Appelt}, E.~A. and {Appert}, S. and {Apple}, S.~K. and {Arai}, K. and {Araya}, A. and {Araya}, M.~C. and {Arca Sedda}, M. and {Arciprete}, F. and {Areeda}, J.~S. and {Aritomi}, N. and {Armato}, F. and {Armstrong}, S. and {Arnaud}, N. and {Arogeti}, M. and {Aronson}, S.~M. and {Ashton}, G. and {Aso}, Y. and {Asprea}, L. and {Assiduo}, M. and {Assis de Souza Melo}, S. and {Aston}, S.~M. and {Astone}, P. and {Aswathi}, P.~S. and {Attadio}, F. and {Aubin}, F. and {AultONeal}, K. and {Avallone}, G. and {Avdeev}, N. and {Avila}, E.~A. and {Babak}, S. and {Badger}, C. and {Bae}, S. and {Bagnasco}, S. and {Baimukhametova}, S. and {Baiotti}, L. and {Baka}, T. and {Baker}, K.~A. and {Baker}, T. and {Balbi}, G. and {Baldi}, G. and {Baldicchi}, N. and {Ball}, M. and {Ballardin}, G. and {Ballelli}, M. and {Ballmer}, S.~W. and {Banagiri}, S. and {Banerjee}, B. and {Bankar}, D. and {Baptiste}, T.~M. and {Baral}, P. and {Baratti}, M. and {Barayoga}, J.~C. and {Baric}, K. and {Barish}, B.~C. and {Barker}, D. and {Barman}, N. and {Barone}, F. and {Barr}, B. and {Barrios}, M. and {Barsotti}, L. and {Barsuglia}, M. and {Barta}, D. and {Barton}, M.~A. and {Bartos}, I. and {Basalaev}, A. and {Bassiri}, R. and {Basti}, A. and {Bawaj}, M. and {Bayley}, J.~C. and {Baylor}, A.~C. and {Baynard}, II, P.~A. and {Bazzan}, M. and {Bedakihale}, V.~M. and {Beirnaert}, F. and {Bejger}, M. and {Bell}, A.~S. and {Bellani}, C. and {Bellie}, D.~S. and {Beltran-Martinez}, D. and {Benedetti}, E. and {Benoit}, W. and {Bentara}, I. and {Ben Yaala}, M. and {Bera}, S. and {Bergamin}, F. and {Berger}, B.~K. and {Beroiz}, M. and {Berry}, C.~P.~L. and {Berry}, I. and {Bersanetti}, D. and {Bertheas}, T. and {Bertolini}, A. and {Betzwieser}, J. and {Beveridge}, D. and {Bevins}, N. and {Bezerra-Sobrinho}, J. and {Bhandare}, R. and {Bhatt}, R. and {Bhattacharjee}, A. and {Bhattacharjee}, D. and {Bhattacharyya}, S. and {Bhaumik}, S. and {Biancalana}, V. and {Bianchi}, F. and {Bilenko}, I.~A. and {Bilicki}, M. and {Billingsley}, G. and {Binetti}, A. and {Bini}, S. and {Biot}, S. and {Birnholtz}, O. and {Biscoveanu}, S. and {Bisht}, A. and {Bitossi}, M. and {Bizouard}, M.-A. and {Blaber}, S. and {Blackburn}, J.~K. and {Blagg}, L.~A. and {Blair}, C.~D. and {Blair}, D.~G. and {Bloch}, M. and {Bode}, N. and {Boettner}, N. and {Bogdan}, P. and {Boileau}, G. and {Boldrini}, M. and {Bolingbroke}, G.~N. and {Bonavena}, L.~D. and {Bonhomme}, V.~A. and {Bonilla}, E. and {Bonilla}, M.~S. and {Bonino}, A. and {Bonnand}, R. and {Borchers}, A. and {Borghi}, N. and {Boschi}, V. and {Bose}, S. and {Bossilkov}, V. and {Bothra}, Y. and {Boudon}, A. and {Boybeyi}, T.~D. and {Boyle}, M. and {Bozzi}, A. and {Bradaschia}, C.},
        title = "{GWTC-5.0: Constraints on the Cosmic Expansion Rate and Modified Gravitational-wave Propagation}",
      journal = {arXiv e-prints},
         year = 2026,
        month = may,
          eid = {arXiv:2605.27227},
        pages = {arXiv:2605.27227},
          doi = {10.48550/arXiv.2605.27227},
archivePrefix = {arXiv},
       eprint = {2605.27227},
 primaryClass = {astro-ph.CO},
       adsurl = {https://ui.adsabs.harvard.edu/abs/2026arXiv260527227T}
}

@ARTICLE{Perna:2025,
       author = {{Perna}, G. and {Mastrogiovanni}, S. and {Ricciardone}, A.},
        title = "{Investigating the impact of galaxies' compact binary hosting probability for gravitational wave cosmology}",
      journal = {\aap},
         year = 2025,
        month = jun,
       volume = {698},
          eid = {A128},
        pages = {A128},
          doi = {10.1051/0004-6361/202450840},
archivePrefix = {arXiv},
       eprint = {2405.07904},
 primaryClass = {astro-ph.CO},
       adsurl = {https://ui.adsabs.harvard.edu/abs/2025A&A...698A.128P}
}

@ARTICLE{Hanselman:2025,
       author = {{Hanselman}, Alexandra G. and {Vijaykumar}, Aditya and {Fishbach}, Maya and {Holz}, Daniel E.},
        title = "{Gravitational-wave Dark Siren Cosmology Systematics from Galaxy Weighting}",
      journal = {\apj},
         year = 2025,
        month = jan,
       volume = {979},
       number = {1},
          eid = {9},
        pages = {9},
          doi = {10.3847/1538-4357/ad9393},
archivePrefix = {arXiv},
       eprint = {2405.14818},
 primaryClass = {astro-ph.CO},
       adsurl = {https://ui.adsabs.harvard.edu/abs/2025ApJ...979....9H}
}

@ARTICLE{Yang:2019q,
       author = {{Yang}, Y. and {Bartos}, I. and {Haiman}, Z. and {Kocsis}, B. and {M{\'a}rka}, Z. and {Stone}, N.~C. and {M{\'a}rka}, S.},
        title = "{AGN Disks Harden the Mass Distribution of Stellar-mass Binary Black Hole Mergers}",
      journal = {\apj},
         year = 2019,
        month = may,
       volume = {876},
       number = {2},
          eid = {122},
        pages = {122},
          doi = {10.3847/1538-4357/ab16e3},
archivePrefix = {arXiv},
       eprint = {1903.01405},
 primaryClass = {astro-ph.HE},
       adsurl = {https://ui.adsabs.harvard.edu/abs/2019ApJ...876..122Y}
}

@ARTICLE{Fabj:2020,
       author = {{Fabj}, Gaia and {Nasim}, Syeda S. and {Caban}, Freddy and {Ford}, K.~E. Saavik and {McKernan}, Barry and {Bellovary}, Jillian M.},
        title = "{Aligning Nuclear Cluster Orbits with an Active Galactic Nucleus Accretion Disc}",
      journal = {\mnras},
         year = 2020,
        month = dec,
       volume = {499},
       number = {2},
        pages = {2608-2616},
          doi = {10.1093/mnras/staa3004},
archivePrefix = {arXiv},
       eprint = {2006.11229},
 primaryClass = {astro-ph.GA},
       adsurl = {https://ui.adsabs.harvard.edu/abs/2020MNRAS.499.2608F}
}

@ARTICLE{boekholt:2023,
       author = {{Boekholt}, Tjarda C.~N. and {Rowan}, Connar and {Kocsis}, Bence},
        title = "{On the Jacobi capture origin of binaries with applications to the Earth-Moon system and black holes in galactic nuclei}",
      journal = {\mnras},
         year = 2023,
        month = feb,
       volume = {518},
       number = {4},
        pages = {5653-5669},
          doi = {10.1093/mnras/stac3495},
archivePrefix = {arXiv},
       eprint = {2203.09646},
 primaryClass = {astro-ph.EP},
       adsurl = {https://ui.adsabs.harvard.edu/abs/2023MNRAS.518.5653B}
}

@ARTICLE{Rowan:2023bbh,
       author = {{Rowan}, Connar and {Boekholt}, Tjarda and {Kocsis}, Bence and {Haiman}, Zolt{\'a}n},
        title = "{Black hole binary formation in AGN discs: from isolation to merger}",
      journal = {\mnras},
         year = 2023,
        month = sep,
       volume = {524},
       number = {2},
        pages = {2770-2796},
          doi = {10.1093/mnras/stad1926},
archivePrefix = {arXiv},
       eprint = {2212.06133},
 primaryClass = {astro-ph.GA},
       adsurl = {https://ui.adsabs.harvard.edu/abs/2023MNRAS.524.2770R}
}

@ARTICLE{McKernan:2019ctp,
       author = {{McKernan}, B. and {Ford}, K.~E.~S. and {Bartos}, I. and {Graham}, M.~J. and {Lyra}, W. and {Marka}, S. and {Marka}, Z. and {Ross}, N.~P. and {Stern}, D. and {Yang}, Y.},
        title = "{Ram-pressure Stripping of a Kicked Hill Sphere: Prompt Electromagnetic Emission from the Merger of Stellar Mass Black Holes in an AGN Accretion Disk}",
      journal = {\apjl},
         year = 2019,
        month = oct,
       volume = {884},
       number = {2},
          eid = {L50},
        pages = {L50},
          doi = {10.3847/2041-8213/ab4886},
archivePrefix = {arXiv},
       eprint = {1907.03746},
 primaryClass = {astro-ph.HE},
       adsurl = {https://ui.adsabs.harvard.edu/abs/2019ApJ...884L..50M}
}

@ARTICLE{Bellovary:2016,
       author = {{Bellovary}, Jillian M. and {Mac Low}, Mordecai-Mark and {McKernan}, Barry and {Ford}, K.~E. Saavik},
        title = "{Migration Traps in Disks around Supermassive Black Holes}",
      journal = {\apjl},
         year = 2016,
        month = mar,
       volume = {819},
       number = {2},
          eid = {L17},
        pages = {L17},
          doi = {10.3847/2041-8205/819/2/L17},
archivePrefix = {arXiv},
       eprint = {1511.00005},
 primaryClass = {astro-ph.GA},
       adsurl = {https://ui.adsabs.harvard.edu/abs/2016ApJ...819L..17B}
}

@ARTICLE{Ford:2025rev,
       author = {{Ford}, K.~E. Saavik and {McKernan}, Barry},
        title = "{Using gravitational waves and multi-messenger Astronomy to reverse-engineer the properties of galactic nuclei}",
      journal = {arXiv e-prints},
         year = 2025,
        month = jun,
          eid = {arXiv:2506.08801},
        pages = {arXiv:2506.08801},
          doi = {10.48550/arXiv.2506.08801},
archivePrefix = {arXiv},
       eprint = {2506.08801},
 primaryClass = {astro-ph.HE},
       adsurl = {https://ui.adsabs.harvard.edu/abs/2025arXiv250608801F}
}

@ARTICLE{Duffell:2020,
       author = {{Duffell}, Paul C. and {D'Orazio}, Daniel and {Derdzinski}, Andrea and {Haiman}, Zoltan and {MacFadyen}, Andrew and {Rosen}, Anna L. and {Zrake}, Jonathan},
        title = "{Circumbinary Disks: Accretion and Torque as a Function of Mass Ratio and Disk Viscosity}",
      journal = {\apj},
         year = 2020,
        month = sep,
       volume = {901},
       number = {1},
          eid = {25},
        pages = {25},
          doi = {10.3847/1538-4357/abab95},
archivePrefix = {arXiv},
       eprint = {1911.05506},
 primaryClass = {astro-ph.SR},
       adsurl = {https://ui.adsabs.harvard.edu/abs/2020ApJ...901...25D}
}

@ARTICLE{Munoz:2019,
       author = {{Mu{\~ n}oz}, Diego J. and {Miranda}, Ryan and {Lai}, Dong},
        title = "{Hydrodynamics of Circumbinary Accretion: Angular Momentum Transfer and Binary Orbital Evolution}",
      journal = {\apj},
         year = 2019,
        month = jan,
       volume = {871},
       number = {1},
          eid = {84},
        pages = {84},
          doi = {10.3847/1538-4357/aaf867},
archivePrefix = {arXiv},
       eprint = {1810.04676},
 primaryClass = {astro-ph.HE},
       adsurl = {https://ui.adsabs.harvard.edu/abs/2019ApJ...871...84M}
}

@ARTICLE{Ishibashi:2024,
       author = {{Ishibashi}, W. and {Gr{\"o}bner}, M.},
        title = "{Gravitational wave mergers of accreting binary black holes in AGN discs}",
      journal = {\mnras},
         year = 2024,
        month = apr,
       volume = {529},
       number = {2},
        pages = {883-892},
          doi = {10.1093/mnras/stae569},
archivePrefix = {arXiv},
       eprint = {2412.01925},
 primaryClass = {astro-ph.HE},
       adsurl = {https://ui.adsabs.harvard.edu/abs/2024MNRAS.529..883I}
}

@ARTICLE{Gilbaum:2022,
       author = {{Gilbaum}, Shmuel and {Stone}, Nicholas C.},
        title = "{Feedback-dominated Accretion Flows}",
      journal = {\apj},
         year = 2022,
        month = apr,
       volume = {928},
       number = {2},
          eid = {191},
        pages = {191},
          doi = {10.3847/1538-4357/ac4ded},
archivePrefix = {arXiv},
       eprint = {2107.07519},
 primaryClass = {astro-ph.HE},
       adsurl = {https://ui.adsabs.harvard.edu/abs/2022ApJ...928..191G}
}

@ARTICLE{Grishin:2024,
       author = {{Grishin}, Evgeni and {Gilbaum}, Shmuel and {Stone}, Nicholas C.},
        title = "{The effect of thermal torques on AGN disc migration traps and gravitational wave populations}",
      journal = {\mnras},
         year = 2024,
        month = may,
       volume = {530},
       number = {2},
        pages = {2114-2132},
          doi = {10.1093/mnras/stae828},
archivePrefix = {arXiv},
       eprint = {2307.07546},
 primaryClass = {astro-ph.HE},
       adsurl = {https://ui.adsabs.harvard.edu/abs/2024MNRAS.530.2114G}
}

@ARTICLE{Paardekooper:2006,
       author = {{Paardekooper}, S. -J. and {Mellema}, G.},
        title = "{Halting type I planet migration in non-isothermal disks}",
      journal = {\aap},
         year = 2006,
        month = nov,
       volume = {459},
       number = {1},
        pages = {L17-L20},
          doi = {10.1051/0004-6361:20066304},
archivePrefix = {arXiv},
       eprint = {astro-ph/0608658},
 primaryClass = {astro-ph},
       adsurl = {https://ui.adsabs.harvard.edu/abs/2006A&A...459L..17P}
}

@ARTICLE{Paardekooper:2010,
       author = {{Paardekooper}, S. -J. and {Baruteau}, C. and {Crida}, A. and {Kley}, W.},
        title = "{A torque formula for non-isothermal type I planetary migration - I. Unsaturated horseshoe drag}",
      journal = {\mnras},
         year = 2010,
        month = jan,
       volume = {401},
       number = {3},
        pages = {1950-1964},
          doi = {10.1111/j.1365-2966.2009.15782.x},
archivePrefix = {arXiv},
       eprint = {0909.4552},
 primaryClass = {astro-ph.EP},
       adsurl = {https://ui.adsabs.harvard.edu/abs/2010MNRAS.401.1950P}
}

@ARTICLE{Secunda:2019tra,
       author = {{Secunda}, Amy and {Bellovary}, Jillian and {Mac Low}, Mordecai-Mark and {Ford}, K.~E. Saavik and {McKernan}, Barry and {Leigh}, Nathan W.~C. and {Lyra}, Wladimir and {S{\'a}ndor}, Zsolt},
        title = "{Orbital Migration of Interacting Stellar Mass Black Holes in Disks around Supermassive Black Holes}",
      journal = {\apj},
         year = 2019,
        month = jun,
       volume = {878},
       number = {2},
          eid = {85},
        pages = {85},
          doi = {10.3847/1538-4357/ab20ca},
archivePrefix = {arXiv},
       eprint = {1807.02859},
 primaryClass = {astro-ph.HE},
       adsurl = {https://ui.adsabs.harvard.edu/abs/2019ApJ...878...85S}
}

@ARTICLE{McKernan:2012tra,
       author = {{McKernan}, B. and {Ford}, K.~E.~S. and {Lyra}, W. and {Perets}, H.~B.},
        title = "{Intermediate mass black holes in AGN discs - I. Production and growth}",
      journal = {\mnras},
         year = 2012,
        month = sep,
       volume = {425},
       number = {1},
        pages = {460-469},
          doi = {10.1111/j.1365-2966.2012.21486.x},
archivePrefix = {arXiv},
       eprint = {1206.2309},
 primaryClass = {astro-ph.GA},
       adsurl = {https://ui.adsabs.harvard.edu/abs/2012MNRAS.425..460M}
}

@ARTICLE{Tagawa:2023ctp,
       author = {{Tagawa}, Hiromichi and {Kimura}, Shigeo S. and {Haiman}, Zolt{\'a}n and {Perna}, Rosalba and {Bartos}, Imre},
        title = "{Observable Signature of Merging Stellar-mass Black Holes in Active Galactic Nuclei}",
      journal = {\apj},
         year = 2023,
        month = jun,
       volume = {950},
       number = {1},
          eid = {13},
        pages = {13},
          doi = {10.3847/1538-4357/acc4bb},
archivePrefix = {arXiv},
       eprint = {2301.07111},
 primaryClass = {astro-ph.HE},
       adsurl = {https://ui.adsabs.harvard.edu/abs/2023ApJ...950...13T}
}

@ARTICLE{Kim:2008,
       author = {{Kim}, Hyosun and {Kim}, Woong-Tae and {S{\'a}nchez-Salcedo}, F.~J.},
        title = "{Dynamical Friction of Double Perturbers in a Gaseous Medium}",
      journal = {\apjl},
         year = 2008,
        month = may,
       volume = {679},
       number = {1},
        pages = {L33},
          doi = {10.1086/589149},
archivePrefix = {arXiv},
       eprint = {0804.2010},
 primaryClass = {astro-ph},
       adsurl = {https://ui.adsabs.harvard.edu/abs/2008ApJ...679L..33K}
}

@ARTICLE{Smith:2026,
       author = {{Smith}, William J. and {Ruiz-Rocha}, Krystal and {Holley-Bockelmann}, Kelly and {Mapelli}, Michela and {Jani}, Karan},
        title = "{Probing Binary Black Hole Formation Channels through Cosmic Large-scale Structure}",
      journal = {\apjl},
         year = 2026,
        month = mar,
       volume = {999},
       number = {1},
          eid = {L10},
        pages = {L10},
          doi = {10.3847/2041-8213/ae41ae},
archivePrefix = {arXiv},
       eprint = {2510.19249},
 primaryClass = {gr-qc},
       adsurl = {https://ui.adsabs.harvard.edu/abs/2026ApJ...999L..10S}
}

@ARTICLE{Vaccaro:2026agn,
       author = {{Vaccaro}, Maria Paola and {Mapelli}, Michela and {Trani}, Alessandro Alberto and {Liu}, Boyuan},
        title = "{AGN-driven BBH mergers: Black hole populations and hierarchical growth across the AGN parameter space}",
      journal = {arXiv e-prints},
         year = 2026,
        month = jun,
          eid = {arXiv:2606.10823},
        pages = {arXiv:2606.10823},
          doi = {10.48550/arXiv.2606.10823},
archivePrefix = {arXiv},
       eprint = {2606.10823},
 primaryClass = {astro-ph.GA},
       adsurl = {https://ui.adsabs.harvard.edu/abs/2026arXiv260610823V}
}

@ARTICLE{Vaccaro:2026trap,
       author = {{Vaccaro}, Maria Paola and {Seif}, Yannick and {Mapelli}, Michela},
        title = "{The role of migration traps in the formation of binary black holes in AGN disks}",
      journal = {\aap},
         year = 2026,
        month = apr,
       volume = {708},
          eid = {A171},
        pages = {A171},
          doi = {10.1051/0004-6361/202556761},
archivePrefix = {arXiv},
       eprint = {2508.03637},
 primaryClass = {astro-ph.HE},
       adsurl = {https://ui.adsabs.harvard.edu/abs/2026A&A...708A.171V}
}

@ARTICLE{Vaccaro:2024,
       author = {{Vaccaro}, Maria Paola and {Mapelli}, Michela and {P{\'e}rigois}, Carole and {Barone}, Dario and {Artale}, Maria Celeste and {Dall'Amico}, Marco and {Iorio}, Giuliano and {Torniamenti}, Stefano},
        title = "{Impact of gas hardening on the population properties of hierarchical black hole mergers in active galactic nucleus disks}",
      journal = {\aap},
         year = 2024,
        month = may,
       volume = {685},
          eid = {A51},
        pages = {A51},
          doi = {10.1051/0004-6361/202348509},
archivePrefix = {arXiv},
       eprint = {2311.18548},
 primaryClass = {astro-ph.HE},
       adsurl = {https://ui.adsabs.harvard.edu/abs/2024A&A...685A..51V}
}

@ARTICLE{McKernan:2024fact,
       author = {{McKernan}, Barry and {Ford}, K.~E. Saavik and {Cook}, Harry E. and {Delfavero}, Vera and {Nathaniel}, Kaila and {Postiglione}, Jake and {Ray}, Shawn and {O'Shaughnessy}, Richard},
        title = "{McFACTS I: Testing the LVK AGN channel with Monte Carlo For AGN Channel Testing \& Simulation (McFACTS)}",
      journal = {arXiv e-prints},
         year = 2024,
        month = oct,
          eid = {arXiv:2410.16515},
        pages = {arXiv:2410.16515},
          doi = {10.48550/arXiv.2410.16515},
archivePrefix = {arXiv},
       eprint = {2410.16515},
 primaryClass = {astro-ph.HE},
       adsurl = {https://ui.adsabs.harvard.edu/abs/2024arXiv241016515M}
}

@ARTICLE{Wang:2025bsi,
       author = {{Wang}, Mengye and {Ma}, Yiqiu and {Li}, Hui and {Wu}, Qingwen and {Li}, Ya-Ping and {Lei}, Xiangli and {Wu}, Jiancheng},
        title = "{Simulation of Binary-single Interactions in AGN Disk. I. Gas-enhanced Binary Orbital Hardening}",
      journal = {\apj},
         year = 2025,
        month = apr,
       volume = {983},
       number = {2},
          eid = {114},
        pages = {114},
          doi = {10.3847/1538-4357/adbf8e},
archivePrefix = {arXiv},
       eprint = {2501.10703},
 primaryClass = {astro-ph.HE},
       adsurl = {https://ui.adsabs.harvard.edu/abs/2025ApJ...983..114W}
}

@ARTICLE{Whitehead:2024,
       author = {{Whitehead}, Henry and {Rowan}, Connar and {Boekholt}, Tjarda and {Kocsis}, Bence},
        title = "{Gas assisted binary black hole formation in AGN discs}",
      journal = {\mnras},
         year = 2024,
        month = jul,
       volume = {531},
       number = {4},
        pages = {4656-4680},
          doi = {10.1093/mnras/stae1430},
archivePrefix = {arXiv},
       eprint = {2309.11561},
 primaryClass = {astro-ph.GA},
       adsurl = {https://ui.adsabs.harvard.edu/abs/2024MNRAS.531.4656W}
}

@ARTICLE{Rowan:2024gas,
       author = {{Rowan}, Connar and {Whitehead}, Henry and {Boekholt}, Tjarda and {Kocsis}, Bence and {Haiman}, Zolt{\'a}n},
        title = "{Black hole binaries in AGN accretion discs - II. Gas effects on black hole satellite scatterings}",
      journal = {\mnras},
         year = 2024,
        month = feb,
       volume = {527},
       number = {4},
        pages = {10448-10468},
          doi = {10.1093/mnras/stad3641},
archivePrefix = {arXiv},
       eprint = {2309.14433},
 primaryClass = {astro-ph.HE},
       adsurl = {https://ui.adsabs.harvard.edu/abs/2024MNRAS.52710448R}
}

@ARTICLE{Lang:2014wise,
       author = {{Lang}, Dustin},
        title = "{unWISE: Unblurred Coadds of the WISE Imaging}",
      journal = {\aj},
         year = 2014,
        month = may,
       volume = {147},
       number = {5},
          eid = {108},
        pages = {108},
          doi = {10.1088/0004-6256/147/5/108},
archivePrefix = {arXiv},
       eprint = {1405.0308},
 primaryClass = {astro-ph.IM},
       adsurl = {https://ui.adsabs.harvard.edu/abs/2014AJ....147..108L}
}

@ARTICLE{Meisner:2019wise,
       author = {{Meisner}, A.~M. and {Lang}, D. and {Schlafly}, E.~F. and {Schlegel}, D.~J.},
        title = "{unWISE Coadds: The Five-year Data Set}",
      journal = {\pasp},
         year = 2019,
        month = dec,
       volume = {131},
       number = {1006},
        pages = {124504},
          doi = {10.1088/1538-3873/ab3df4},
archivePrefix = {arXiv},
       eprint = {1909.05444},
 primaryClass = {astro-ph.IM},
       adsurl = {https://ui.adsabs.harvard.edu/abs/2019PASP..131l4504M}
}

@ARTICLE{GaiaDR3,
       author = {{Gaia Collaboration} and {Vallenari}, A. and {Brown}, A.~G.~A. and {Prusti}, T. and {de Bruijne}, J.~H.~J. and {Arenou}, F. and {Babusiaux}, C. and {Biermann}, M. and {Creevey}, O.~L. and {Ducourant}, C. and {Evans}, D.~W. and {Eyer}, L. and {Guerra}, R. and {Hutton}, A. and {Jordi}, C. and {Klioner}, S.~A. and {Lammers}, U.~L. and {Lindegren}, L. and {Luri}, X. and {Mignard}, F. and {Panem}, C. and {Pourbaix}, D. and {Randich}, S. and {Sartoretti}, P. and {Soubiran}, C. and {Tanga}, P. and {Walton}, N.~A. and {Bailer-Jones}, C.~A.~L. and {Bastian}, U. and {Drimmel}, R. and {Jansen}, F. and {Katz}, D. and {Lattanzi}, M.~G. and {van Leeuwen}, F. and {Bakker}, J. and {Cacciari}, C. and {Casta{\~n}eda}, J. and {De Angeli}, F. and {Fabricius}, C. and {Fouesneau}, M. and {Fr{\'e}mat}, Y. and {Galluccio}, L. and {Guerrier}, A. and {Heiter}, U. and {Masana}, E. and {Messineo}, R. and {Mowlavi}, N. and {Nicolas}, C. and {Nienartowicz}, K. and {Pailler}, F. and {Panuzzo}, P. and {Riclet}, F. and {Roux}, W. and {Seabroke}, G.~M. and {Sordo}, R. and {Th{\'e}venin}, F. and {Gracia-Abril}, G. and {Portell}, J. and {Teyssier}, D. and {Altmann}, M. and {Andrae}, R. and {Audard}, M. and {Bellas-Velidis}, I. and {Benson}, K. and {Berthier}, J. and {Blomme}, R. and {Burgess}, P.~W. and {Busonero}, D. and {Busso}, G. and {C{\'a}novas}, H. and {Carry}, B. and {Cellino}, A. and {Cheek}, N. and {Clementini}, G. and {Damerdji}, Y. and {Davidson}, M. and {de Teodoro}, P. and {Nu{\~n}ez Campos}, M. and {Delchambre}, L. and {Dell'Oro}, A. and {Esquej}, P. and {Fern{\'a}ndez-Hern{\'a}ndez}, J. and {Fraile}, E. and {Garabato}, D. and {Garc{\'\i}a-Lario}, P. and {Gosset}, E. and {Haigron}, R. and {Halbwachs}, J.-L. and {Hambly}, N.~C. and {Harrison}, D.~L. and {Hern{\'a}ndez}, J. and {Hestroffer}, D. and {Hodgkin}, S.~T. and {Holl}, B. and {Jan{\ss}en}, K. and {Jevardat de Fombelle}, G. and {Jordan}, S. and {Krone-Martins}, A. and {Lanzafame}, A.~C. and {L{\"o}ffler}, W. and {Marchal}, O. and {Marrese}, P.~M. and {Moitinho}, A. and {Muinonen}, K. and {Osborne}, P. and {Pancino}, E. and {Pauwels}, T. and {Recio-Blanco}, A. and {Reyl{\'e}}, C. and {Riello}, M. and {Rimoldini}, L. and {Roegiers}, T. and {Rybizki}, J. and {Sarro}, L.~M. and {Siopis}, C. and {Smith}, M. and {Sozzetti}, A. and {Utrilla}, E. and {van Leeuwen}, M. and {Abbas}, U. and {{\'A}brah{\'a}m}, P. and {Abreu Aramburu}, A. and {Aerts}, C. and {Aguado}, J.~J. and {Ajaj}, M. and {Aldea-Montero}, F. and {Altavilla}, G. and {{\'A}lvarez}, M.~A. and {Alves}, J. and {Anders}, F. and {Anderson}, R.~I. and {Anglada Varela}, E. and {Antoja}, T. and {Baines}, D. and {Baker}, S.~G. and {Balaguer-N{\'u}{\~n}ez}, L. and {Balbinot}, E. and {Balog}, Z. and {Barache}, C. and {Barbato}, D. and {Barros}, M. and {Barstow}, M.~A. and {Bartolom{\'e}}, S. and {Bassilana}, J.-L. and {Bauchet}, N. and {Becciani}, U. and {Bellazzini}, M. and {Berihuete}, A. and {Bernet}, M. and {Bertone}, S. and {Bianchi}, L. and {Binnenfeld}, A. and {Blanco-Cuaresma}, S. and {Blazere}, A. and {Boch}, T. and {Bombrun}, A. and {Bossini}, D. and {Bouquillon}, S. and {Bragaglia}, A. and {Bramante}, L. and {Breedt}, E. and {Bressan}, A. and {Brouillet}, N. and {Brugaletta}, E. and {Bucciarelli}, B. and {Burlacu}, A. and {Butkevich}, A.~G. and {Buzzi}, R. and {Caffau}, E. and {Cancelliere}, R. and {Cantat-Gaudin}, T. and {Carballo}, R. and {Carlucci}, T. and {Carnerero}, M.~I. and {Carrasco}, J.~M. and {Casamiquela}, L. and {Castellani}, M. and {Castro-Ginard}, A. and {Chaoul}, L. and {Charlot}, P. and {Chemin}, L. and {Chiaramida}, V. and {Chiavassa}, A. and {Chornay}, N. and {Comoretto}, G. and {Contursi}, G. and {Cooper}, W.~J. and {Cornez}, T. and {Cowell}, S. and {Crifo}, F. and {Cropper}, M. and {Crosta}, M. and {Crowley}, C. and {Dafonte}, C. and {Dapergolas}, A. and {David}, M. and {David}, P. and {de Laverny}, P. and {De Luise}, F. and {De March}, R.},
        title = "{Gaia Data Release 3. Summary of the content and survey properties}",
      journal = {\aap},
         year = 2023,
        month = jun,
       volume = {674},
          eid = {A1},
        pages = {A1},
          doi = {10.1051/0004-6361/202243940},
archivePrefix = {arXiv},
       eprint = {2208.00211},
 primaryClass = {astro-ph.GA},
       adsurl = {https://ui.adsabs.harvard.edu/abs/2023A&A...674A...1G}
}

@ARTICLE{Dodici:2024,
       author = {{Dodici}, Mark and {Tremaine}, Scott},
        title = "{Studying Binary Formation under Dynamical Friction Using Hill's Problem}",
      journal = {\apj},
         year = 2024,
        month = sep,
       volume = {972},
       number = {2},
          eid = {193},
        pages = {193},
          doi = {10.3847/1538-4357/ad5cf2},
archivePrefix = {arXiv},
       eprint = {2404.08138},
 primaryClass = {astro-ph.GA},
       adsurl = {https://ui.adsabs.harvard.edu/abs/2024ApJ...972..193D}
}

@ARTICLE{Rowan:2025gcb,
       author = {{Rowan}, Connar and {Whitehead}, Henry and {Fabj}, Gaia and {Saini}, Pankaj and {Kocsis}, Bence and {Pessah}, Martin and {Samsing}, Johan},
        title = "{Prompt gravitational-wave mergers aided by gas in active galactic nuclei: the hydrodynamics of binary-single black hole scatterings}",
      journal = {\mnras},
         year = 2025,
        month = may,
       volume = {539},
       number = {2},
        pages = {1501-1515},
          doi = {10.1093/mnras/staf547},
archivePrefix = {arXiv},
       eprint = {2501.09017},
 primaryClass = {astro-ph.GA},
       adsurl = {https://ui.adsabs.harvard.edu/abs/2025MNRAS.539.1501R}
}

@ARTICLE{delaurentiis:2023,
       author = {{DeLaurentiis}, Stanislav and {Epstein-Martin}, Marguerite and {Haiman}, Zolt{\'a}n},
        title = "{Gas dynamical friction as a binary formation mechanism in AGN discs}",
      journal = {\mnras},
         year = 2023,
        month = jul,
       volume = {523},
       number = {1},
        pages = {1126-1139},
          doi = {10.1093/mnras/stad1412},
archivePrefix = {arXiv},
       eprint = {2212.02650},
 primaryClass = {astro-ph.HE},
       adsurl = {https://ui.adsabs.harvard.edu/abs/2023MNRAS.523.1126D}
}

@ARTICLE{Whitehead:2025,
       author = {{Whitehead}, Henry and {Rowan}, Connar and {Kocsis}, Bence},
        title = "{Hydrodynamic simulations of black hole evolution in AGN discs II: inclination damping for partially embedded satellites}",
      journal = {arXiv e-prints},
         year = 2025,
        month = may,
          eid = {arXiv:2505.23899},
        pages = {arXiv:2505.23899},
          doi = {10.48550/arXiv.2505.23899},
archivePrefix = {arXiv},
       eprint = {2505.23899},
 primaryClass = {astro-ph.HE},
       adsurl = {https://ui.adsabs.harvard.edu/abs/2025arXiv250523899W}
}

@ARTICLE{Tagawa:2020,
       author = {{Tagawa}, Hiromichi and {Haiman}, Zolt{\'a}n and {Kocsis}, Bence},
        title = "{Formation and Evolution of Compact-object Binaries in AGN Disks}",
      journal = {\apj},
         year = 2020,
        month = jul,
       volume = {898},
       number = {1},
          eid = {25},
        pages = {25},
          doi = {10.3847/1538-4357/ab9b8c},
archivePrefix = {arXiv},
       eprint = {1912.08218},
 primaryClass = {astro-ph.GA},
       adsurl = {https://ui.adsabs.harvard.edu/abs/2020ApJ...898...25T}
}

@ARTICLE{Generozov:2023,
       author = {{Generozov}, A. and {Perets}, H.~B.},
        title = "{Capture of stars into gaseous discs around massive black holes: alignment, circularization, and growth}",
      journal = {\mnras},
         year = 2023,
        month = jun,
       volume = {522},
       number = {2},
        pages = {1763-1778},
          doi = {10.1093/mnras/stad1016},
archivePrefix = {arXiv},
       eprint = {2212.11301},
 primaryClass = {astro-ph.GA},
       adsurl = {https://ui.adsabs.harvard.edu/abs/2023MNRAS.522.1763G}
}

@ARTICLE{Li:2025a,
       author = {{Li}, Guo-Peng and {Fan}, Xi-Long},
        title = "{Multimessenger hierarchical triple merger gravitational-wave event pair GW190514-GW190521 inside AGN J124942.3 + 344929}",
      journal = {arXiv e-prints},
         year = 2025,
        month = mar,
          eid = {arXiv:2503.16864},
        pages = {arXiv:2503.16864},
          doi = {10.48550/arXiv.2503.16864},
archivePrefix = {arXiv},
       eprint = {2503.16864},
 primaryClass = {astro-ph.HE},
       adsurl = {https://ui.adsabs.harvard.edu/abs/2025arXiv250316864L}
}

@ARTICLE{Barbieri:2026,
       author = {{Barbieri}, Riccardo and {Kalomenopoulos}, Marios and {Tagliazucchi}, Matteo and {Gair}, Jonathan and {Khochfar}, Sadegh},
        title = "{Not all those missing are lost: leveraging galaxy clustering in incomplete catalogs to unleash dark sirens cosmology}",
      journal = {arXiv e-prints},
         year = 2026,
        month = aug,
          eid = {arXiv:2608.30599},
        pages = {arXiv:2608.30599},
          doi = {10.48550/arXiv.2608.30599},
archivePrefix = {arXiv},
       eprint = {2608.30599},
 primaryClass = {astro-ph.CO},
       adsurl = {https://ui.adsabs.harvard.edu/abs/2026arXiv260830599B}
}

@ARTICLE{Ghosh:2026,
       author = {{Ghosh}, Tathagata and {More}, Surhud},
        title = "{Recovering Large-Scale Clustering of Missing Galaxies for Galaxy--Gravitational-Wave Cross-correlations}",
      journal = {arXiv e-prints},
         year = 2026,
        month = sep,
          eid = {arXiv:2609.07348},
        pages = {arXiv:2609.07348},
archivePrefix = {arXiv},
       eprint = {2609.07348},
 primaryClass = {astro-ph.CO},
       adsurl = {https://ui.adsabs.harvard.edu/abs/2026arXiv260907348G}
}

@ARTICLE{Bellomo:2026,
       author = {{Bellomo}, Nicola and {Bosi}, Michele and {Libanore}, Sarah and {Liguori}, Michele and {Mapelli}, Michela and {Semenzato}, Federico and {Torniamenti}, Stefano},
        title = "{Breaking binary formation mechanism degeneracies with gravitational wave clustering}",
      journal = {arXiv e-prints},
         year = 2026,
        month = sep,
          eid = {arXiv:2609.09083},
        pages = {arXiv:2609.09083},
archivePrefix = {arXiv},
       eprint = {2609.09083},
 primaryClass = {gr-qc},
       adsurl = {https://ui.adsabs.harvard.edu/abs/2026arXiv260909083B}
}

@ARTICLE{Mandel:2026pe,
       author = {{Mandel}, Ilya},
        title = "{What is the Most Massive Gravitational-wave Source?}",
      journal = {\apjl},
         year = 2026,
        month = jan,
       volume = {996},
       number = {1},
          eid = {L4},
        pages = {L4},
          doi = {10.3847/2041-8213/ae278d},
archivePrefix = {arXiv},
       eprint = {2509.05885},
 primaryClass = {astro-ph.HE},
       adsurl = {https://ui.adsabs.harvard.edu/abs/2026ApJ...996L...4M}
}

@ARTICLE{Mould:2026pe,
       author = {{Mould}, Matthew and {Tenorio}, Rodrigo and {Gerosa}, Davide},
        title = "{Gravitational-wave astronomy requires population-informed parameter estimation}",
      journal = {\prd},
         year = 2026,
        month = jul,
       volume = {114},
       number = {2},
          eid = {L021302},
        pages = {L021302},
          doi = {10.1103/blhd-ktbf},
archivePrefix = {arXiv},
       eprint = {2604.15885},
 primaryClass = {gr-qc},
       adsurl = {https://ui.adsabs.harvard.edu/abs/2026PhRvD.114b1302M}
}

@ARTICLE{WuShen:2022sdss,
       author = {{Wu}, Qiaoya and {Shen}, Yue},
        title = "{A Catalog of Quasar Properties from Sloan Digital Sky Survey Data Release 16}",
      journal = {\apjs},
         year = 2022,
        month = dec,
       volume = {263},
       number = {2},
          eid = {42},
        pages = {42},
          doi = {10.3847/1538-4365/ac9ead},
archivePrefix = {arXiv},
       eprint = {2209.03987},
 primaryClass = {astro-ph.GA},
       adsurl = {https://ui.adsabs.harvard.edu/abs/2022ApJS..263...42W}
}

@ARTICLE{GWOSC:2026,
       author = {{The LIGO Scientific Collaboration} and {the Virgo Collaboration} and {the KAGRA Collaboration} and {Abac}, A.~G. and {Abe}, A. and {Abouelfettouh}, I. and {Acernese}, F. and {Ackley}, K. and {Adam}, A. and {Adhicary}, S. and {Adhikari}, D. and {Adhikari}, R.~X. and {Adkins}, V.~K. and {Afroz}, S. and {Agapito}, A. and {Agarwal}, D. and {Agathos}, M. and {Aggarwal}, N. and {Aggarwal}, S. and {Aguiar}, O.~D. and {Ahrend}, I.-L. and {Aiello}, L. and {Ain}, A. and {Ajith}, P. and {Akutsu}, T. and {Albers}, L. and {Ali}, W. and {Al-Kershi}, S. and {Allene}, C. and {Allocca}, A. and {Al-Shammari}, S. and {Alvarez}, J.~A. and {Alvarez-Lopez}, S. and {Amar}, W. and {Amarasinghe}, O. and {Amato}, A. and {Amicucci}, F. and {Amra}, C. and {Anand}, A.~B. and {Anand}, C. and {Ananyeva}, A. and {Anderson}, S.~B. and {Anderson}, W.~G. and {Andia}, M. and {Ando}, M. and {Andrade-Oliveira}, F. and {Andr{\'e}s-Carcasona}, M. and {Andrey}, J.~L. and {Andri{\'c}}, T. and {Anglin}, J. and {Anna}, J. and {Antelis}, J.~M. and {Antier}, S. and {Aoki}, T. and {Aoumi}, M. and {Appavuravther}, E.~Z. and {Appelt}, E.~A. and {Appert}, S. and {Apple}, S.~K. and {Arai}, K. and {Araya}, A. and {Araya}, M.~C. and {Arca Sedda}, M. and {Arciprete}, F. and {Areeda}, J.~S. and {Aritomi}, N. and {Armato}, F. and {Armstrong}, S. and {Arnaud}, N. and {Arogeti}, M. and {Aronson}, S.~M. and {Ashton}, G. and {Aso}, Y. and {Asprea}, L. and {Assiduo}, M. and {Assis de Souza Melo}, S. and {Aston}, S.~M. and {Astone}, P. and {Aswathi}, P.~S. and {Attadio}, F. and {Aubin}, F. and {AultONeal}, K. and {Avallone}, G. and {Avdeev}, N. and {Avila}, E.~A. and {Babak}, S. and {Badger}, C. and {Bae}, S. and {Bagnasco}, S. and {Baimukhametova}, S. and {Baiotti}, L. and {Baka}, T. and {Baker}, K.~A. and {Baker}, T. and {Balbi}, G. and {Baldi}, G. and {Baldicchi}, N. and {Ball}, M. and {Ballardin}, G. and {Ballelli}, M. and {Ballmer}, S.~W. and {Banagiri}, S. and {Banerjee}, B. and {Bankar}, D. and {Baptiste}, T.~M. and {Baral}, P. and {Baratti}, M. and {Barayoga}, J.~C. and {Baric}, K. and {Barish}, B.~C. and {Barker}, D. and {Barman}, N. and {Barone}, F. and {Barr}, B. and {Barrios}, M. and {Barsotti}, L. and {Barsuglia}, M. and {Barta}, D. and {Barton}, M.~A. and {Bartos}, I. and {Basalaev}, A. and {Bassiri}, R. and {Basti}, A. and {Bawaj}, M. and {Bayley}, J.~C. and {Baylor}, A.~C. and {Baynard}, II, P.~A. and {Bazzan}, M. and {Bedakihale}, V.~M. and {Beirnaert}, F. and {Bejger}, M. and {Bell}, A.~S. and {Bellani}, C. and {Bellie}, D.~S. and {Beltran-Martinez}, D. and {Benedetti}, E. and {Benoit}, W. and {Bentara}, I. and {Ben Yaala}, M. and {Bera}, S. and {Bergamin}, F. and {Berger}, B.~K. and {Beroiz}, M. and {Berry}, C.~P.~L. and {Berry}, I. and {Bersanetti}, D. and {Bertheas}, T. and {Bertolini}, A. and {Betzwieser}, J. and {Beveridge}, D. and {Bevins}, N. and {Bezerra-Sobrinho}, J. and {Bhandare}, R. and {Bhatt}, R. and {Bhattacharjee}, A. and {Bhattacharjee}, D. and {Bhattacharyya}, S. and {Bhaumik}, S. and {Biancalana}, V. and {Bianchi}, F. and {Bilenko}, I.~A. and {Bilicki}, M. and {Billingsley}, G. and {Binetti}, A. and {Bini}, S. and {Biot}, S. and {Birnholtz}, O. and {Biscoveanu}, S. and {Bisht}, A. and {Bitossi}, M. and {Bizouard}, M.-A. and {Blaber}, S. and {Blackburn}, J.~K. and {Blagg}, L.~A. and {Blair}, C.~D. and {Blair}, D.~G. and {Bloch}, M. and {Bode}, N. and {Boettner}, N. and {Bogdan}, P. and {Boileau}, G. and {Boldrini}, M. and {Bolingbroke}, G.~N. and {Bonavena}, L.~D. and {Bonhomme}, V.~A. and {Bonilla}, E. and {Bonilla}, M.~S. and {Bonino}, A. and {Bonnand}, R. and {Borchers}, A. and {Borghi}, N. and {Boschi}, V. and {Bose}, S. and {Bossilkov}, V. and {Bothra}, Y. and {Boudon}, A. and {Boybeyi}, T.~D. and {Boyle}, M. and {Bozzi}, A. and {Bradaschia}, C.},
        title = "{Open Data from LIGO, Virgo, and KAGRA through the Second Part of the Fourth Observing Run}",
      journal = {arXiv e-prints},
         year = 2026,
        month = may,
          eid = {arXiv:2605.27090},
        pages = {arXiv:2605.27090},
          doi = {10.48550/arXiv.2605.27090},
archivePrefix = {arXiv},
       eprint = {2605.27090},
 primaryClass = {gr-qc},
       adsurl = {https://ui.adsabs.harvard.edu/abs/2026arXiv260527090T}
}

@ARTICLE{lvk:2026gwtc5,
       author = {{The LIGO Scientific Collaboration} and {the Virgo Collaboration} and {the KAGRA Collaboration} and {Abac}, A.~G. and {Abe}, A. and {Abouelfettouh}, I. and {Acernese}, F. and {Ackley}, K. and {Adam}, A. and {Adhicary}, S. and {Adhikari}, D. and {Adhikari}, R.~X. and {Adkins}, V.~K. and {Afroz}, S. and {Agapito}, A. and {Agarwal}, D. and {Agathos}, M. and {Aggarwal}, N. and {Aggarwal}, S. and {Aguiar}, O.~D. and {Ahrend}, I.-L. and {Aiello}, L. and {Ain}, A. and {Ajith}, P. and {Akutsu}, T. and {Albers}, L. and {Ali}, W. and {Al-Kershi}, S. and {Allene}, C. and {Allocca}, A. and {Al-Shammari}, S. and {Alvarez}, J.~A. and {Alvarez-Lopez}, S. and {Amar}, W. and {Amarasinghe}, O. and {Amato}, A. and {Amicucci}, F. and {Amra}, C. and {Anand}, A.~B. and {Anand}, C. and {Ananyeva}, A. and {Anderson}, S.~B. and {Anderson}, W.~G. and {Andia}, M. and {Ando}, M. and {Andrade-Oliveira}, F. and {Andr{\'e}s-Carcasona}, M. and {Andrey}, J.~L. and {Andri{\'c}}, T. and {Anglin}, J. and {Anna}, J. and {Antelis}, J.~M. and {Antier}, S. and {Aoki}, T. and {Aoumi}, M. and {Appavuravther}, E.~Z. and {Appelt}, E.~A. and {Appert}, S. and {Apple}, S.~K. and {Arai}, K. and {Araya}, A. and {Araya}, M.~C. and {Arca Sedda}, M. and {Arciprete}, F. and {Areeda}, J.~S. and {Aritomi}, N. and {Armato}, F. and {Armstrong}, S. and {Arnaud}, N. and {Arogeti}, M. and {Aronson}, S.~M. and {Ashton}, G. and {Aso}, Y. and {Asprea}, L. and {Assiduo}, M. and {Assis de Souza Melo}, S. and {Aston}, S.~M. and {Astone}, P. and {Aswathi}, P.~S. and {Attadio}, F. and {Aubin}, F. and {AultONeal}, K. and {Avallone}, G. and {Avdeev}, N. and {Avila}, E.~A. and {Babak}, S. and {Badger}, C. and {Bae}, S. and {Bagnasco}, S. and {Baimukhametova}, S. and {Baiotti}, L. and {Baka}, T. and {Baker}, K.~A. and {Baker}, T. and {Balbi}, G. and {Baldi}, G. and {Baldicchi}, N. and {Ball}, M. and {Ballardin}, G. and {Ballelli}, M. and {Ballmer}, S.~W. and {Banagiri}, S. and {Banerjee}, B. and {Bankar}, D. and {Baptiste}, T.~M. and {Baral}, P. and {Baratti}, M. and {Barayoga}, J.~C. and {Baric}, K. and {Barish}, B.~C. and {Barker}, D. and {Barman}, N. and {Barone}, F. and {Barr}, B. and {Barrios}, M. and {Barsotti}, L. and {Barsuglia}, M. and {Barta}, D. and {Barton}, M.~A. and {Bartos}, I. and {Basalaev}, A. and {Bassiri}, R. and {Basti}, A. and {Bawaj}, M. and {Bayley}, J.~C. and {Baylor}, A.~C. and {Baynard}, II, P.~A. and {Bazzan}, M. and {Bedakihale}, V.~M. and {Beirnaert}, F. and {Bejger}, M. and {Bell}, A.~S. and {Bellani}, C. and {Bellie}, D.~S. and {Beltran-Martinez}, D. and {Benedetti}, E. and {Benoit}, W. and {Bentara}, I. and {Ben Yaala}, M. and {Bera}, S. and {Bergamin}, F. and {Berger}, B.~K. and {Beroiz}, M. and {Berry}, C.~P.~L. and {Berry}, I. and {Bersanetti}, D. and {Bertheas}, T. and {Bertolini}, A. and {Betzwieser}, J. and {Beveridge}, D. and {Bevins}, N. and {Bezerra-Sobrinho}, J. and {Bhandare}, R. and {Bhatt}, R. and {Bhattacharjee}, A. and {Bhattacharjee}, D. and {Bhattacharyya}, S. and {Bhaumik}, S. and {Biancalana}, V. and {Bianchi}, F. and {Bilenko}, I.~A. and {Bilicki}, M. and {Billingsley}, G. and {Binetti}, A. and {Bini}, S. and {Biot}, S. and {Birnholtz}, O. and {Biscoveanu}, S. and {Bisht}, A. and {Bitossi}, M. and {Bizouard}, M.-A. and {Blaber}, S. and {Blackburn}, J.~K. and {Blagg}, L.~A. and {Blair}, C.~D. and {Blair}, D.~G. and {Bloch}, M. and {Bode}, N. and {Boettner}, N. and {Bogdan}, P. and {Boileau}, G. and {Boldrini}, M. and {Bolingbroke}, G.~N. and {Bonavena}, L.~D. and {Bonhomme}, V.~A. and {Bonilla}, E. and {Bonilla}, M.~S. and {Bonino}, A. and {Bonnand}, R. and {Borchers}, A. and {Borghi}, N. and {Boschi}, V. and {Bose}, S. and {Bossilkov}, V. and {Bothra}, Y. and {Boudon}, A. and {Boybeyi}, T.~D. and {Boyle}, M. and {Bozzi}, A. and {Bradaschia}, C.},
        title = "{GWTC-5.0: Observations from the Second Part of the Fourth LIGO-Virgo-KAGRA Observing Run and Updates to the Gravitational-Wave Transient Catalog}",
      journal = {arXiv e-prints},
         year = 2026,
        month = may,
          eid = {arXiv:2605.27225},
        pages = {arXiv:2605.27225},
          doi = {10.48550/arXiv.2605.27225},
archivePrefix = {arXiv},
       eprint = {2605.27225},
 primaryClass = {gr-qc},
       adsurl = {https://ui.adsabs.harvard.edu/abs/2026arXiv260527225T}
}

@ARTICLE{Farah:2026,
       author = {{Farah}, Amanda M. and {Vijaykumar}, Aditya and {Fishbach}, Maya},
        title = "{The Steep Redshift Evolution of the Hierarchical Binary Black Hole Merger Rate May Cause the z-{\ensuremath{\chi}}$_{eff}$ Correlation}",
      journal = {\apjl},
         year = 2026,
        month = apr,
       volume = {1001},
       number = {2},
          eid = {L40},
        pages = {L40},
          doi = {10.3847/2041-8213/ae4e19},
archivePrefix = {arXiv},
       eprint = {2601.03456},
 primaryClass = {astro-ph.HE},
       adsurl = {https://ui.adsabs.harvard.edu/abs/2026ApJ..1001L..40F}
}

@ARTICLE{Colleoni:2025xphm,
       author = {{Colleoni}, Marta and {Ramis Vidal}, Felip A. and {Garc{\'\i}a-Quir{\'o}s}, Cecilio and {Ak{\c{c}}ay}, Sarp and {Bera}, Sayantani},
        title = "{Fast frequency-domain gravitational waveforms for precessing binaries with a new twist}",
      journal = {\prd},
         year = 2025,
        month = may,
       volume = {111},
       number = {10},
          eid = {104019},
        pages = {104019},
          doi = {10.1103/PhysRevD.111.104019},
archivePrefix = {arXiv},
       eprint = {2412.16721},
 primaryClass = {gr-qc},
       adsurl = {https://ui.adsabs.harvard.edu/abs/2025PhRvD.111j4019C}
}

@ARTICLE{VanSon:2025Z,
       author = {{van Son}, L.~A.~C. and {Roy}, S.~K. and {Mandel}, I. and {Farr}, W.~M. and {Lam}, A. and {Merritt}, J. and {Broekgaarden}, F.~S. and {Sander}, A.~A.~C. and {Andrews}, J.~J.},
        title = "{Not Just Winds: Why Models Find That Binary Black Hole Formation Is Metallicity-dependent, while Binary Neutron Star Formation Is Not}",
      journal = {\apj},
         year = 2025,
        month = feb,
       volume = {979},
       number = {2},
          eid = {209},
        pages = {209},
          doi = {10.3847/1538-4357/ada14a},
archivePrefix = {arXiv},
       eprint = {2411.02484},
 primaryClass = {astro-ph.HE},
       adsurl = {https://ui.adsabs.harvard.edu/abs/2025ApJ...979..209V}
}

@ARTICLE{Padhyegurjar:2026dtd,
       author = {{Padhyegurjar}, Shaunak and {Mukherjee}, Suvodip},
        title = "{The First Detection of Sub-Populations in the Delay-Time Distribution of Binary Black Holes in GWTC-4 of LIGO-Virgo-KAGRA}",
      journal = {arXiv e-prints},
         year = 2026,
        month = jun,
          eid = {arXiv:2606.02318},
        pages = {arXiv:2606.02318},
          doi = {10.48550/arXiv.2606.02318},
archivePrefix = {arXiv},
       eprint = {2606.02318},
 primaryClass = {astro-ph.HE},
       adsurl = {https://ui.adsabs.harvard.edu/abs/2026arXiv260602318P}
}

@ARTICLE{Fishbach:2021tdel,
       author = {{Fishbach}, Maya and {Kalogera}, Vicky},
        title = "{The Time Delay Distribution and Formation Metallicity of LIGO-Virgos Binary Black Holes}",
      journal = {\apjl},
         year = 2021,
        month = jun,
       volume = {914},
       number = {2},
          eid = {L30},
        pages = {L30},
          doi = {10.3847/2041-8213/ac05c4},
archivePrefix = {arXiv},
       eprint = {2105.06491},
 primaryClass = {astro-ph.HE},
       adsurl = {https://ui.adsabs.harvard.edu/abs/2021ApJ...914L..30F}
}

@ARTICLE{Yang:2020zevo,
       author = {{Yang}, Y. and {Bartos}, I. and {Haiman}, Z. and {Kocsis}, B. and {M{\'a}rka}, S. and {Tagawa}, H.},
        title = "{Cosmic Evolution of Stellar-mass Black Hole Merger Rate in Active Galactic Nuclei}",
      journal = {\apj},
         year = 2020,
        month = jun,
       volume = {896},
       number = {2},
          eid = {138},
        pages = {138},
          doi = {10.3847/1538-4357/ab91b4},
archivePrefix = {arXiv},
       eprint = {2003.08564},
 primaryClass = {astro-ph.HE},
       adsurl = {https://ui.adsabs.harvard.edu/abs/2020ApJ...896..138Y}
}

@ARTICLE{Fishbach:2018lowZ,
       author = {{Fishbach}, Maya and {Holz}, Daniel E. and {Farr}, Will M.},
        title = "{Does the Black Hole Merger Rate Evolve with Redshift?}",
      journal = {\apjl},
         year = 2018,
        month = aug,
       volume = {863},
       number = {2},
          eid = {L41},
        pages = {L41},
          doi = {10.3847/2041-8213/aad800},
archivePrefix = {arXiv},
       eprint = {1805.10270},
 primaryClass = {astro-ph.HE},
       adsurl = {https://ui.adsabs.harvard.edu/abs/2018ApJ...863L..41F}
}

@ARTICLE{Mandel:2022rev,
       author = {{Mandel}, Ilya and {Broekgaarden}, Floor S.},
        title = "{Rates of compact object coalescences}",
      journal = {Living Reviews in Relativity},
         year = 2022,
        month = dec,
       volume = {25},
       number = {1},
          eid = {1},
        pages = {1},
          doi = {10.1007/s41114-021-00034-3},
archivePrefix = {arXiv},
       eprint = {2107.14239},
 primaryClass = {astro-ph.HE},
       adsurl = {https://ui.adsabs.harvard.edu/abs/2022LRR....25....1M}
}

@INCOLLECTION{Mapelli:2021,
       author = {{Mapelli}, Michela},
        title = "{Formation Channels of Single and Binary Stellar-Mass Black Holes}",
    booktitle = {Handbook of Gravitational Wave Astronomy},
         year = 2021,
       editor = {{Bambi}, Cosimo and {Katsanevas}, Stavros and {Kokkotas}, Konstantinos D.},
          eid = {16},
        pages = {16},
          doi = {10.1007/978-981-15-4702-7_16-1},
       adsurl = {https://ui.adsabs.harvard.edu/abs/2021hgwa.bookE..16M}
}

@ARTICLE{Bartos:2017orb,
       author = {{Bartos}, Imre and {Kocsis}, Bence and {Haiman}, Zolt{\'a}n and {M{\'a}rka}, Szabolcs},
        title = "{Rapid and Bright Stellar-mass Binary Black Hole Mergers in Active Galactic Nuclei}",
      journal = {\apj},
         year = 2017,
        month = feb,
       volume = {835},
       number = {2},
          eid = {165},
        pages = {165},
          doi = {10.3847/1538-4357/835/2/165},
archivePrefix = {arXiv},
       eprint = {1602.03831},
 primaryClass = {astro-ph.HE},
       adsurl = {https://ui.adsabs.harvard.edu/abs/2017ApJ...835..165B}
}

@ARTICLE{Stone:2017insitu,
       author = {{Stone}, Nicholas C. and {Metzger}, Brian D. and {Haiman}, Zolt{\'a}n},
        title = "{Assisted inspirals of stellar mass black holes embedded in AGN discs: solving the `final au problem'}",
      journal = {\mnras},
         year = 2017,
        month = jan,
       volume = {464},
       number = {1},
        pages = {946-954},
          doi = {10.1093/mnras/stw2260},
archivePrefix = {arXiv},
       eprint = {1602.04226},
 primaryClass = {astro-ph.GA},
       adsurl = {https://ui.adsabs.harvard.edu/abs/2017MNRAS.464..946S}
}

@ARTICLE{EpsMart:2025,
       author = {{Epstein-Martin}, Marguerite and {Tagawa}, Hiromichi and {Haiman}, Zolt{\'a}n and {Perna}, Rosalba},
        title = "{Time-dependent models of AGN discs with radiation from embedded stellar-mass black holes}",
      journal = {\mnras},
         year = 2025,
        month = mar,
       volume = {537},
       number = {4},
        pages = {3396-3420},
          doi = {10.1093/mnras/staf237},
archivePrefix = {arXiv},
       eprint = {2405.09380},
 primaryClass = {astro-ph.HE},
       adsurl = {https://ui.adsabs.harvard.edu/abs/2025MNRAS.537.3396E}
}

@ARTICLE{Pratten:2021,
       author = {{Pratten}, Geraint and {Garc{\'\i}a-Quir{\'o}s}, Cecilio and {Colleoni}, Marta and {Ramos-Buades}, Antoni and {Estell{\'e}s}, H{\'e}ctor and {Mateu-Lucena}, Maite and {Jaume}, Rafel and {Haney}, Maria and {Keitel}, David and {Thompson}, Jonathan E. and {Husa}, Sascha},
        title = "{Computationally efficient models for the dominant and subdominant harmonic modes of precessing binary black holes}",
      journal = {\prd},
         year = 2021,
        month = may,
       volume = {103},
       number = {10},
          eid = {104056},
        pages = {104056},
          doi = {10.1103/PhysRevD.103.104056},
archivePrefix = {arXiv},
       eprint = {2004.06503},
 primaryClass = {gr-qc},
       adsurl = {https://ui.adsabs.harvard.edu/abs/2021PhRvD.103j4056P}
}

@ARTICLE{ligo:gwtc5pop,
       author = {{The LIGO Scientific Collaboration} and {the Virgo Collaboration} and {the KAGRA Collaboration}},
        title = "{GWTC-5.0: Population Properties of Merging Compact Binaries}",
      journal = {arXiv e-prints},
         year = 2026,
        month = may,
          eid = {arXiv:2605.27226},
        pages = {arXiv:2605.27226},
          doi = {10.48550/arXiv.2605.27226},
archivePrefix = {arXiv},
       eprint = {2605.27226},
 primaryClass = {astro-ph.HE},
       adsurl = {https://ui.adsabs.harvard.edu/abs/2026arXiv260527226T}
}

@ARTICLE{Planck15,
       author = {{Planck Collaboration} and {Ade}, P.~A.~R. and {Aghanim}, N. and {Arnaud}, M. and {Ashdown}, M. and {Aumont}, J. and {Baccigalupi}, C. and {Banday}, A.~J. and {Barreiro}, R.~B. and {Bartlett}, J.~G. and {Bartolo}, N. and {Battaner}, E. and {Battye}, R. and {Benabed}, K. and {Beno{\^\i}t}, A. and {Benoit-L{\'e}vy}, A. and {Bernard}, J. -P. and {Bersanelli}, M. and {Bielewicz}, P. and {Bock}, J.~J. and {Bonaldi}, A. and {Bonavera}, L. and {Bond}, J.~R. and {Borrill}, J. and {Bouchet}, F.~R. and {Boulanger}, F. and {Bucher}, M. and {Burigana}, C. and {Butler}, R.~C. and {Calabrese}, E. and {Cardoso}, J. -F. and {Catalano}, A. and {Challinor}, A. and {Chamballu}, A. and {Chary}, R. -R. and {Chiang}, H.~C. and {Chluba}, J. and {Christensen}, P.~R. and {Church}, S. and {Clements}, D.~L. and {Colombi}, S. and {Colombo}, L.~P.~L. and {Combet}, C. and {Coulais}, A. and {Crill}, B.~P. and {Curto}, A. and {Cuttaia}, F. and {Danese}, L. and {Davies}, R.~D. and {Davis}, R.~J. and {de Bernardis}, P. and {de Rosa}, A. and {de Zotti}, G. and {Delabrouille}, J. and {D{\'e}sert}, F. -X. and {Di Valentino}, E. and {Dickinson}, C. and {Diego}, J.~M. and {Dolag}, K. and {Dole}, H. and {Donzelli}, S. and {Dor{\'e}}, O. and {Douspis}, M. and {Ducout}, A. and {Dunkley}, J. and {Dupac}, X. and {Efstathiou}, G. and {Elsner}, F. and {En{\ss}lin}, T.~A. and {Eriksen}, H.~K. and {Farhang}, M. and {Fergusson}, J. and {Finelli}, F. and {Forni}, O. and {Frailis}, M. and {Fraisse}, A.~A. and {Franceschi}, E. and {Frejsel}, A. and {Galeotta}, S. and {Galli}, S. and {Ganga}, K. and {Gauthier}, C. and {Gerbino}, M. and {Ghosh}, T. and {Giard}, M. and {Giraud-H{\'e}raud}, Y. and {Giusarma}, E. and {Gjerl{\o}w}, E. and {Gonz{\'a}lez-Nuevo}, J. and {G{\'o}rski}, K.~M. and {Gratton}, S. and {Gregorio}, A. and {Gruppuso}, A. and {Gudmundsson}, J.~E. and {Hamann}, J. and {Hansen}, F.~K. and {Hanson}, D. and {Harrison}, D.~L. and {Helou}, G. and {Henrot-Versill{\'e}}, S. and {Hern{\'a}ndez-Monteagudo}, C. and {Herranz}, D. and {Hildebrandt}, S.~R. and {Hivon}, E. and {Hobson}, M. and {Holmes}, W.~A. and {Hornstrup}, A. and {Hovest}, W. and {Huang}, Z. and {Huffenberger}, K.~M. and {Hurier}, G. and {Jaffe}, A.~H. and {Jaffe}, T.~R. and {Jones}, W.~C. and {Juvela}, M. and {Keih{\"a}nen}, E. and {Keskitalo}, R. and {Kisner}, T.~S. and {Kneissl}, R. and {Knoche}, J. and {Knox}, L. and {Kunz}, M. and {Kurki-Suonio}, H. and {Lagache}, G. and {L{\"a}hteenm{\"a}ki}, A. and {Lamarre}, J. -M. and {Lasenby}, A. and {Lattanzi}, M. and {Lawrence}, C.~R. and {Leahy}, J.~P. and {Leonardi}, R. and {Lesgourgues}, J. and {Levrier}, F. and {Lewis}, A. and {Liguori}, M. and {Lilje}, P.~B. and {Linden-V{\o}rnle}, M. and {L{\'o}pez-Caniego}, M. and {Lubin}, P.~M. and {Mac{\'\i}as-P{\'e}rez}, J.~F. and {Maggio}, G. and {Maino}, D. and {Mandolesi}, N. and {Mangilli}, A. and {Marchini}, A. and {Maris}, M. and {Martin}, P.~G. and {Martinelli}, M. and {Mart{\'\i}nez-Gonz{\'a}lez}, E. and {Masi}, S. and {Matarrese}, S. and {McGehee}, P. and {Meinhold}, P.~R. and {Melchiorri}, A. and {Melin}, J. -B. and {Mendes}, L. and {Mennella}, A. and {Migliaccio}, M. and {Millea}, M. and {Mitra}, S. and {Miville-Desch{\^e}nes}, M. -A. and {Moneti}, A. and {Montier}, L. and {Morgante}, G. and {Mortlock}, D. and {Moss}, A. and {Munshi}, D. and {Murphy}, J.~A. and {Naselsky}, P. and {Nati}, F. and {Natoli}, P. and {Netterfield}, C.~B. and {N{\o}rgaard-Nielsen}, H.~U. and {Noviello}, F. and {Novikov}, D. and {Novikov}, I. and {Oxborrow}, C.~A. and {Paci}, F. and {Pagano}, L. and {Pajot}, F. and {Paladini}, R. and {Paoletti}, D. and {Partridge}, B. and {Pasian}, F. and {Patanchon}, G. and {Pearson}, T.~J. and {Perdereau}, O. and {Perotto}, L. and {Perrotta}, F. and {Pettorino}, V. and {Piacentini}, F. and {Piat}, M. and {Pierpaoli}, E. and {Pietrobon}, D. and {Plaszczynski}, S. and {Pointecouteau}, E. and {Polenta}, G. and {Popa}, L. and {Pratt}, G.~W. and {Pr{\'e}zeau}, G.},
        title = "{Planck 2015 results. XIII. Cosmological parameters}",
      journal = {\aap},
         year = 2016,
        month = sep,
       volume = {594},
          eid = {A13},
        pages = {A13},
          doi = {10.1051/0004-6361/201525830},
archivePrefix = {arXiv},
       eprint = {1502.01589},
 primaryClass = {astro-ph.CO},
       adsurl = {https://ui.adsabs.harvard.edu/abs/2016A&A...594A..13P}
}

@ARTICLE{Lyke:2020,
       author = {{Lyke}, Brad W. and {Higley}, Alexandra N. and {McLane}, J.~N. and {Schurhammer}, Danielle P. and {Myers}, Adam D. and {Ross}, Ashley J. and {Dawson}, Kyle and {Chabanier}, Sol{\`e}ne and {Martini}, Paul and {Busca}, Nicol{\'a}s G. and {Mas des Bourboux}, H{\'e}lion du and {Salvato}, Mara and {Streblyanska}, Alina and {Zarrouk}, Pauline and {Burtin}, Etienne and {Anderson}, Scott F. and {Bautista}, Julian and {Bizyaev}, Dmitry and {Brandt}, W.~N. and {Brinkmann}, Jonathan and {Brownstein}, Joel R. and {Comparat}, Johan and {Green}, Paul and {de la Macorra}, Axel and {Mu{\~n}oz Guti{\'e}rrez}, Andrea and {Hou}, Jiamin and {Newman}, Jeffrey A. and {Palanque-Delabrouille}, Nathalie and {P{\^a}ris}, Isabelle and {Percival}, Will J. and {Petitjean}, Patrick and {Rich}, James and {Rossi}, Graziano and {Schneider}, Donald P. and {Smith}, Alexander and {Vivek}, M. and {Weaver}, Benjamin Alan},
        title = "{The Sloan Digital Sky Survey Quasar Catalog: Sixteenth Data Release}",
      journal = {\apjs},
         year = 2020,
        month = sep,
       volume = {250},
       number = {1},
          eid = {8},
        pages = {8},
          doi = {10.3847/1538-4365/aba623},
archivePrefix = {arXiv},
       eprint = {2007.09001},
 primaryClass = {astro-ph.GA},
       adsurl = {https://ui.adsabs.harvard.edu/abs/2020ApJS..250....8L}
}

@ARTICLE{Madau:2014,
       author = {{Madau}, Piero and {Dickinson}, Mark},
        title = "{Cosmic Star-Formation History}",
      journal = {\araa},
         year = 2014,
        month = aug,
       volume = {52},
        pages = {415-486},
          doi = {10.1146/annurev-astro-081811-125615},
archivePrefix = {arXiv},
       eprint = {1403.0007},
 primaryClass = {astro-ph.CO},
       adsurl = {https://ui.adsabs.harvard.edu/abs/2014ARA&A..52..415M}
}

@ARTICLE{Graham:2020em,
       author = {{Graham}, M.~J. and {Ford}, K.~E.~S. and {McKernan}, B. and {Ross}, N.~P. and {Stern}, D. and {Burdge}, K. and {Coughlin}, M. and {Djorgovski}, S.~G. and {Drake}, A.~J. and {Duev}, D. and {Kasliwal}, M. and {Mahabal}, A.~A. and {van Velzen}, S. and {Belecki}, J. and {Bellm}, E.~C. and {Burruss}, R. and {Cenko}, S.~B. and {Cunningham}, V. and {Helou}, G. and {Kulkarni}, S.~R. and {Masci}, F.~J. and {Prince}, T. and {Reiley}, D. and {Rodriguez}, H. and {Rusholme}, B. and {Smith}, R.~M. and {Soumagnac}, M.~T.},
        title = "{Candidate Electromagnetic Counterpart to the Binary Black Hole Merger Gravitational-Wave Event S190521g$^{*}$}",
      journal = {\prl},
         year = 2020,
        month = jun,
       volume = {124},
       number = {25},
          eid = {251102},
        pages = {251102},
          doi = {10.1103/PhysRevLett.124.251102},
archivePrefix = {arXiv},
       eprint = {2006.14122},
 primaryClass = {astro-ph.HE},
       adsurl = {https://ui.adsabs.harvard.edu/abs/2020PhRvL.124y1102G}
}

@ARTICLE{Moncrieff:2025,
       author = {{Moncrieff}, Jordan W.~N. and {Panther}, Fiona H.},
        title = "{Connecting the hierarchically merging binary black hole population to their host galaxies}",
      journal = {\mnras},
         year = 2025,
        month = oct,
       volume = {543},
       number = {2},
        pages = {1833-1841},
          doi = {10.1093/mnras/staf1582},
archivePrefix = {arXiv},
       eprint = {2508.18704},
 primaryClass = {astro-ph.CO},
       adsurl = {https://ui.adsabs.harvard.edu/abs/2025MNRAS.543.1833M}
}

@ARTICLE{Kiendrebeogo:2023,
       author = {{Kiendrebeogo}, R. Weizmann and {Farah}, Amanda M. and {Foley}, Emily M. and {Gray}, Abigail and {Kunert}, Nina and {Puecher}, Anna and {Toivonen}, Andrew and {VandenBerg}, R. Oliver and {Anand}, Shreya and {Ahumada}, Tom{\'a}s and {Karambelkar}, Viraj and {Coughlin}, Michael W. and {Dietrich}, Tim and {Kam}, S. Zacharie and {Pang}, Peter T.~H. and {Singer}, Leo P. and {Sravan}, Niharika},
        title = "{Updated Observing Scenarios and Multimessenger Implications for the International Gravitational-wave Networks O4 and O5}",
      journal = {\apj},
         year = 2023,
        month = dec,
       volume = {958},
       number = {2},
          eid = {158},
        pages = {158},
          doi = {10.3847/1538-4357/acfcb1},
archivePrefix = {arXiv},
       eprint = {2306.09234},
 primaryClass = {astro-ph.HE},
       adsurl = {https://ui.adsabs.harvard.edu/abs/2023ApJ...958..158K}
}

@ARTICLE{Palmese:2021flare,
       author = {{Palmese}, A. and {Fishbach}, M. and {Burke}, C.~J. and {Annis}, J. and {Liu}, X.},
        title = "{Do LIGO/Virgo Black Hole Mergers Produce AGN Flares? The Case of GW190521 and Prospects for Reaching a Confident Association}",
      journal = {\apjl},
         year = 2021,
        month = jun,
       volume = {914},
       number = {2},
          eid = {L34},
        pages = {L34},
          doi = {10.3847/2041-8213/ac0883},
archivePrefix = {arXiv},
       eprint = {2103.16069},
 primaryClass = {astro-ph.HE},
       adsurl = {https://ui.adsabs.harvard.edu/abs/2021ApJ...914L..34P}
}

@ARTICLE{Cabrera:2024em,
       author = {{Cabrera}, Tom{\'a}s and {Palmese}, Antonella and {Hu}, Lei and {O'Connor}, Brendan and {Ford}, K.~E. Saavik and {McKernan}, Barry and {Andreoni}, Igor and {Ahumada}, Tom{\'a}s and {Amsellem}, Ariel and {Busmann}, Malte and {Clark}, Peter and {Coughlin}, Michael W. and {Dadiani}, Ekaterine and {Diaz}, Veronica and {Graham}, Matthew J. and {Gruen}, Daniel and {Kunnumkai}, Keerthi and {Postiglione}, Jake and {Riffeser}, Arno and {Sommer}, Julian S. and {Valdes}, Francisco},
        title = "{Searching for electromagnetic emission in an AGN from the gravitational wave binary black hole merger candidate S230922g}",
      journal = {\prd},
         year = 2024,
        month = dec,
       volume = {110},
       number = {12},
          eid = {123029},
        pages = {123029},
          doi = {10.1103/PhysRevD.110.123029},
archivePrefix = {arXiv},
       eprint = {2407.10698},
 primaryClass = {astro-ph.HE},
       adsurl = {https://ui.adsabs.harvard.edu/abs/2024PhRvD.110l3029C}
}

@ARTICLE{Cabrera:2026em,
       author = {{Cabrera}, T. and {Palmese}, A. and {Fishbach}, M.},
        title = "{Multimessenger Constraints on LIGO/Virgo/KAGRA Gravitational-wave Binary Black Holes Merging in Active Galactic Nucleus Disks}",
      journal = {\apj},
         year = 2026,
        month = apr,
       volume = {1000},
       number = {2},
          eid = {234},
        pages = {234},
          doi = {10.3847/1538-4357/ae4d37},
archivePrefix = {arXiv},
       eprint = {2510.20767},
 primaryClass = {astro-ph.HE},
       adsurl = {https://ui.adsabs.harvard.edu/abs/2026ApJ..1000..234C}
}

@ARTICLE{Zhang:2025em,
       author = {{Zhang}, Shu-Rui and {Wang}, Yu and {Yuan}, Ye-Fei and {Tagawa}, Hiromichi and {Wei}, Yun-Feng and {Li}, Liang and {Cai}, Rong-Gen},
        title = "{S241125n: Binary Black Hole Merger Produces Short GRB in AGN Disk}",
      journal = {arXiv e-prints},
         year = 2025,
        month = may,
          eid = {arXiv:2505.10395},
        pages = {arXiv:2505.10395},
          doi = {10.48550/arXiv.2505.10395},
archivePrefix = {arXiv},
       eprint = {2505.10395},
 primaryClass = {astro-ph.HE},
       adsurl = {https://ui.adsabs.harvard.edu/abs/2025arXiv250510395Z}
}

@INCOLLECTION{Vitale:2020aaz,
       author = {{Vitale}, Salvatore and {Gerosa}, Davide and {Farr}, Will M. and {Taylor}, Stephen R.},
        title = "{Inferring the Properties of a Population of Compact Binaries in Presence of Selection Effects}",
    booktitle = {Handbook of Gravitational Wave Astronomy},
         year = 2022,
       editor = {{Bambi}, Cosimo and {Katsanevas}, Stavros and {Kokkotas}, Konstantinos D.},
          eid = {45},
        pages = {45},
          doi = {10.1007/978-981-15-4702-7_45-1},
       adsurl = {https://ui.adsabs.harvard.edu/abs/2022hgwa.bookE..45V}
}

@ARTICLE{Gayathri:2023bha,
       author = {{Gayathri}, V. and {Wysocki}, Daniel and {Yang}, Y. and {Delfavero}, Vera and {O'Shaughnessy}, R. and {Haiman}, Z. and {Tagawa}, H. and {Bartos}, I.},
        title = "{Gravitational Wave Source Populations: Disentangling an AGN Component}",
      journal = {\apjl},
         year = 2023,
        month = mar,
       volume = {945},
       number = {2},
          eid = {L29},
        pages = {L29},
          doi = {10.3847/2041-8213/acbfb8},
archivePrefix = {arXiv},
       eprint = {2301.04187},
 primaryClass = {gr-qc},
       adsurl = {https://ui.adsabs.harvard.edu/abs/2023ApJ...945L..29G}
}

@ARTICLE{Bartos:2017abc,
       author = {{Bartos}, I. and {Haiman}, Z. and {Marka}, Z. and {Metzger}, B.~D. and {Stone}, N.~C. and {Marka}, S.},
        title = "{Gravitational-wave localization alone can probe origin of stellar-mass black hole mergers}",
      journal = {Nature Communications},
         year = 2017,
        month = oct,
       volume = {8},
          eid = {831},
        pages = {831},
          doi = {10.1038/s41467-017-00851-7},
archivePrefix = {arXiv},
       eprint = {1701.02328},
 primaryClass = {astro-ph.HE},
       adsurl = {https://ui.adsabs.harvard.edu/abs/2017NatCo...8..831B}
}

@ARTICLE{Mandel:2018mve,
       author = {{Mandel}, Ilya and {Farr}, Will M. and {Gair}, Jonathan R.},
        title = "{Extracting distribution parameters from multiple uncertain observations with selection biases}",
      journal = {\mnras},
         year = 2019,
        month = jun,
       volume = {486},
       number = {1},
        pages = {1086-1093},
          doi = {10.1093/mnras/stz896},
archivePrefix = {arXiv},
       eprint = {1809.02063},
 primaryClass = {physics.data-an},
       adsurl = {https://ui.adsabs.harvard.edu/abs/2019MNRAS.486.1086M}
}

@ARTICLE{Veronesi:2022,
       author = {{Veronesi}, Niccol{\`o} and {Rossi}, Elena Maria and {van Velzen}, Sjoert and {Buscicchio}, Riccardo},
        title = "{Detectability of a spatial correlation between stellar mass black hole mergers and active galactic nuclei in the local Universe}",
      journal = {\mnras},
         year = 2022,
        month = aug,
       volume = {514},
       number = {2},
        pages = {2092-2097},
          doi = {10.1093/mnras/stac1346},
archivePrefix = {arXiv},
       eprint = {2203.05907},
 primaryClass = {astro-ph.HE},
       adsurl = {https://ui.adsabs.harvard.edu/abs/2022MNRAS.514.2092V}
}

@ARTICLE{Veronesi:2023,
       author = {{Veronesi}, Niccol{\`o} and {Rossi}, Elena Maria and {van Velzen}, Sjoert},
        title = "{The most luminous AGN do not produce the majority of the detected stellar-mass black hole binary mergers in the local Universe}",
      journal = {\mnras},
         year = 2023,
        month = dec,
       volume = {526},
       number = {4},
        pages = {6031-6040},
          doi = {10.1093/mnras/stad3157},
archivePrefix = {arXiv},
       eprint = {2306.09415},
 primaryClass = {astro-ph.HE},
       adsurl = {https://ui.adsabs.harvard.edu/abs/2023MNRAS.526.6031V}
}

@ARTICLE{Veronesi:2024,
       author = {{Veronesi}, Niccol{\`o} and {van Velzen}, Sjoert and {Rossi}, Elena Maria},
        title = "{AGN flares as counterparts to the mergers detected by LIGO and Virgo: a novel spatial correlation analysis}",
      journal = {arXiv e-prints},
         year = 2024,
        month = may,
          eid = {arXiv:2405.05318},
        pages = {arXiv:2405.05318},
          doi = {10.48550/arXiv.2405.05318},
archivePrefix = {arXiv},
       eprint = {2405.05318},
 primaryClass = {astro-ph.HE},
       adsurl = {https://ui.adsabs.harvard.edu/abs/2024arXiv240505318V}
}

@ARTICLE{Veronesi:2025,
       author = {{Veronesi}, Niccol{\`o} and {van Velzen}, Sjoert and {Rossi}, Elena Maria and {Storey-Fisher}, Kate},
        title = "{Constraining the AGN formation channel for detected black hole binary mergers up to z = 1.5 with the Quaia catalogue}",
      journal = {\mnras},
         year = 2025,
        month = jan,
       volume = {536},
       number = {1},
        pages = {375-386},
          doi = {10.1093/mnras/stae2575},
archivePrefix = {arXiv},
       eprint = {2407.21568},
 primaryClass = {astro-ph.HE},
       adsurl = {https://ui.adsabs.harvard.edu/abs/2025MNRAS.536..375V}
}

@ARTICLE{Zhu:2025,
       author = {{Zhu}, Liang-Gui and {Chen}, Xian},
        title = "{Evidence of a fraction of LIGO/Virgo/KAGRA events coming from active galactic nuclei}",
      journal = {arXiv e-prints},
         year = 2025,
        month = may,
          eid = {arXiv:2505.02924},
        pages = {arXiv:2505.02924},
          doi = {10.48550/arXiv.2505.02924},
archivePrefix = {arXiv},
       eprint = {2505.02924},
 primaryClass = {astro-ph.HE},
       adsurl = {https://ui.adsabs.harvard.edu/abs/2025arXiv250502924Z}
}

@ARTICLE{Gupte:2025pecc,
       author = {{Gupte}, Nihar and {Ramos-Buades}, Antoni and {Buonanno}, Alessandra and {Gair}, Jonathan and {Coleman Miller}, M. and {Dax}, Maximilian and {Green}, Stephen R. and {P{\"u}rrer}, Michael and {Wildberger}, Jonas and {Macke}, Jakob and {Romero-Shaw}, Isobel M. and {Sch{\"o}lkopf}, Bernhard},
        title = "{Evidence for eccentricity in the population of binary black holes observed by LIGO-Virgo-KAGRA}",
      journal = {\prd},
         year = 2025,
        month = nov,
       volume = {112},
       number = {10},
          eid = {104045},
        pages = {104045},
          doi = {10.1103/vpyp-nvfp},
archivePrefix = {arXiv},
       eprint = {2404.14286},
 primaryClass = {gr-qc},
       adsurl = {https://ui.adsabs.harvard.edu/abs/2025PhRvD.112j4045G}
}

@ARTICLE{McKernan:2022spin,
       author = {{McKernan}, B. and {Ford}, K.~E.~S. and {Callister}, T. and {Farr}, W.~M. and {O'Shaughnessy}, R. and {Smith}, R. and {Thrane}, E. and {Vajpeyi}, A.},
        title = "{LIGO-Virgo correlations between mass ratio and effective inspiral spin: testing the active galactic nuclei channel}",
      journal = {\mnras},
         year = 2022,
        month = aug,
       volume = {514},
       number = {3},
        pages = {3886-3893},
          doi = {10.1093/mnras/stac1570},
archivePrefix = {arXiv},
       eprint = {2107.07551},
 primaryClass = {astro-ph.HE},
       adsurl = {https://ui.adsabs.harvard.edu/abs/2022MNRAS.514.3886M}
}

@ARTICLE{Cook:2024,
       author = {{Cook}, Harrison E. and {McKernan}, Barry and {Ford}, K.~E. Saavik and {Delfavero}, Vera and {Nathaniel}, Kaila and {Postiglione}, Jake and {Ray}, Shawn and {O'Shaughnessy}, Richard},
        title = "{McFACTS II: Mass Ratio--Effective Spin Relationship of Black Hole Mergers in the AGN Channel}",
      journal = {arXiv e-prints},
         year = 2024,
        month = nov,
          eid = {arXiv:2411.10590},
        pages = {arXiv:2411.10590},
          doi = {10.48550/arXiv.2411.10590},
archivePrefix = {arXiv},
       eprint = {2411.10590},
 primaryClass = {astro-ph.HE},
       adsurl = {https://ui.adsabs.harvard.edu/abs/2024arXiv241110590C}
}

@ARTICLE{Gilbaum:2024,
       author = {{Gilbaum}, Shmuel and {Grishin}, Evgeni and {Stone}, Nicholas C. and {Mandel}, Ilya},
        title = "{How to Escape from a Trap: Outcomes of Repeated Black Hole Mergers in AGN}",
      journal = {arXiv e-prints},
         year = 2024,
        month = oct,
          eid = {arXiv:2410.19904},
        pages = {arXiv:2410.19904},
          doi = {10.48550/arXiv.2410.19904},
archivePrefix = {arXiv},
       eprint = {2410.19904},
 primaryClass = {astro-ph.HE},
       adsurl = {https://ui.adsabs.harvard.edu/abs/2024arXiv241019904G}
}

@ARTICLE{Fishbach:2019,
       author = {{Fishbach}, M. and {Gray}, R. and {Maga{\~n}a Hernandez}, I. and {Qi}, H. and {Sur}, A. and {Acernese}, F. and {Aiello}, L. and {Allocca}, A. and {Aloy}, M.~A. and {Amato}, A. and {Antier}, S. and {Ar{\`e}ne}, M. and {Arnaud}, N. and {Ascenzi}, S. and {Astone}, P. and {Aubin}, F. and {Babak}, S. and {Bacon}, P. and {Badaracco}, F. and {Bader}, M.~K.~M. and {Baldaccini}, F. and {Ballardin}, G. and {Barone}, F. and {Barsuglia}, M. and {Barta}, D. and {Basti}, A. and {Bawaj}, M. and {Bazzan}, M. and {Bejger}, M. and {Belahcene}, I. and {Bernuzzi}, S. and {Bersanetti}, D. and {Bertolini}, A. and {Bitossi}, M. and {Bizouard}, M.~A. and {Blair}, C.~D. and {Bloemen}, S. and {Boer}, M. and {Bogaert}, G. and {Bondu}, F. and {Bonnand}, R. and {Boom}, B.~A. and {Boschi}, V. and {Bouffanais}, Y. and {Bozzi}, A. and {Bradaschia}, C. and {Brady}, P.~R. and {Branchesi}, M. and {Briant}, T. and {Brighenti}, F. and {Brillet}, A. and {Brisson}, V. and {Bulik}, T. and {Bulten}, H.~J. and {Buskulic}, D. and {Buy}, C. and {Cagnoli}, G. and {Calloni}, E. and {Canepa}, M. and {Capocasa}, E. and {Carbognani}, F. and {Carullo}, G. and {Casanueva Diaz}, J. and {Casentini}, C. and {Caudill}, S. and {Cavalier}, F. and {Cavalieri}, R. and {Cella}, G. and {Cerd{\'a}-Dur{\'a}n}, P. and {Cerretani}, G. and {Cesarini}, E. and {Chaibi}, O. and {Chassande-Mottin}, E. and {Chatziioannou}, K. and {Chen}, H.~Y. and {Chincarini}, A. and {Chiummo}, A. and {Christensen}, N. and {Chua}, S. and {Ciani}, G. and {Ciolfi}, R. and {Cipriano}, F. and {Cirone}, A. and {Cleva}, F. and {Coccia}, E. and {Cohadon}, P. -F. and {Cohen}, D. and {Conti}, L. and {Cordero-Carri{\'o}n}, I. and {Cortese}, S. and {Coughlin}, M.~W. and {Coulon}, J. -P. and {Croquette}, M. and {Cuoco}, E. and {D{\'a}lya}, G. and {D'Antonio}, S. and {Datrier}, L.~E.~H. and {Dattilo}, V. and {Davier}, M. and {Degallaix}, J. and {De Laurentis}, M. and {Del{\'e}glise}, S. and {Del Pozzo}, W. and {Denys}, M. and {De Pietri}, R. and {De Rosa}, R. and {De Rossi}, C. and {DeSalvo}, R. and {Dietrich}, T. and {Di Fiore}, L. and {Di Giovanni}, M. and {Di Girolamo}, T. and {Di Lieto}, A. and {Di Pace}, S. and {Di Palma}, I. and {Di Renzo}, F. and {Doctor}, Z. and {Drago}, M. and {Ducoin}, J. -G. and {Eisenmann}, M. and {Essick}, R.~C. and {Estevez}, D. and {Fafone}, V. and {Farinon}, S. and {Farr}, W.~M. and {Feng}, F. and {Ferrante}, I. and {Ferrini}, F. and {Fidecaro}, F. and {Fiori}, I. and {Fiorucci}, D. and {Flaminio}, R. and {Font}, J.~A. and {Fournier}, J. -D. and {Frasca}, S. and {Frasconi}, F. and {Frey}, V. and {Gair}, J.~R. and {Gammaitoni}, L. and {Garufi}, F. and {Gemme}, G. and {Genin}, E. and {Gennai}, A. and {George}, D. and {Germain}, V. and {Ghosh}, A. and {Giacomazzo}, B. and {Giazotto}, A. and {Giordano}, G. and {Gonzalez Castro}, J.~M. and {Gosselin}, M. and {Gouaty}, R. and {Grado}, A. and {Granata}, M. and {Greco}, G. and {Groot}, P. and {Gruning}, P. and {Guidi}, G.~M. and {Guo}, Y. and {Halim}, O. and {Harms}, J. and {Haster}, C. -J. and {Heidmann}, A. and {Heitmann}, H. and {Hello}, P. and {Hemming}, G. and {Hendry}, M. and {Hinderer}, T. and {Hoak}, D. and {Hofman}, D. and {Holz}, D.~E. and {Hreibi}, A. and {Huet}, D. and {Idzkowski}, B. and {Iess}, A. and {Intini}, G. and {Isac}, J. -M. and {Jacqmin}, T. and {Jaranowski}, P. and {Jonker}, R.~J.~G. and {Katsanevas}, S. and {Katsavounidis}, E. and {K{\'e}f{\'e}lian}, F. and {Khan}, I. and {Koekoek}, G. and {Koley}, S. and {Kowalska}, I. and {Kr{\'o}lak}, A. and {Kutynia}, A. and {Lange}, J. and {Lartaux-Vollard}, A. and {Lazzaro}, C. and {Leaci}, P. and {Letendre}, N. and {Li}, T.~G.~F. and {Linde}, F. and {Longo}, A. and {Lorenzini}, M. and {Loriette}, V. and {Losurdo}, G. and {Lumaca}, D. and {Macas}, R. and {Macquet}, A. and {Majorana}, E. and {Maksimovic}, I. and {Man}, N. and {Mantovani}, M. and {Marchesoni}, F. and {Markakis}, C. and {Marquina}, A. and {Martelli}, F. and {Massera}, E. and {Masserot}, A. and {Mastrogiovanni}, S. and {Meidam}, J. and {Mereni}, L. and {Merzougui}, M. and {Messenger}, C. and {Metzdorff}, R. and {Michel}, C. and {Milano}, L. and {Miller}, A. and {Minazzoli}, O. and {Minenkov}, Y. and {Montani}, M. and {Morisaki}, S. and {Mours}, B. and {Nagar}, A. and {Nardecchia}, I. and {Naticchioni}, L. and {Nelemans}, G. and {Nichols}, D. and {Nocera}, F. and {Obergaulinger}, M. and {Pagano}, G. and {Palomba}, C. and {Pannarale}, F. and {Paoletti}, F. and {Paoli}, A. and {Pasqualetti}, A. and {Passaquieti}, R. and {Passuello}, D. and {Patil}, M. and {Patricelli}, B. and {Pedurand}, R. and {Perreca}, A. and {Piccinni}, O.~J. and {Pichot}, M. and {Piergiovanni}, F. and {Pillant}, G. and {Pinard}, L. and {Poggiani}, R. and {Popolizio}, P. and {Prodi}, G.~A. and {Punturo}, M. and {Puppo}, P. and {Radulescu}, N. and {Raffai}, P. and {Rapagnani}, P. and {Raymond}, V. and {Razzano}, M. and {Regimbau}, T. and {Rei}, L. and {Ricci}, F. and {Rocchi}, A. and {Rolland}, L. and {Romanelli}, M. and {Romano}, R. and {Rosi{\'n}ska}, D. and {Ruggi}, P. and {Salconi}, L. and {Samajdar}, A. and {Sanchis-Gual}, N. and {Sassolas}, B. and {Schutz}, B.~F. and {Sentenac}, D. and {Sequino}, V. and {Sieniawska}, M. and {Singh}, N. and {Singhal}, A. and {Sorrentino}, F. and {Stachie}, C. and {Steer}, D.~A. and {Stratta}, G. and {Swinkels}, B.~L. and {Tacca}, M. and {Tamanini}, N. and {Tiwari}, S. and {Tonelli}, M. and {Torres-Forn{\'e}}, A. and {Travasso}, F. and {Tringali}, M.~C. and {Trovato}, A. and {Trozzo}, L. and {Tsang}, K.~W. and {van Bakel}, N. and {van Beuzekom}, M. and {van den Brand}, J.~F.~J. and {Van Den Broeck}, C. and {van der Schaaf}, L. and {van Heijningen}, J.~V. and {Vardaro}, M. and {Vas{\'u}th}, M. and {Vedovato}, G. and {Veitch}, J. and {Verkindt}, D. and {Vetrano}, F. and {Vicer{\'e}}, A. and {Vinet}, J. -Y. and {Vocca}, H. and {Walet}, R. and {Wang}, G. and {Wang}, Y.~F. and {Was}, M. and {Williamson}, A.~R. and {Yvert}, M. and {Zadro{\.z}ny}, A. and {Zelenova}, T. and {Zendri}, J. -P. and {Zimmerman}, A.~B.},
        title = "{A Standard Siren Measurement of the Hubble Constant from GW170817 without the Electromagnetic Counterpart}",
      journal = {\apjl},
         year = 2019,
        month = jan,
       volume = {871},
       number = {1},
          eid = {L13},
        pages = {L13},
          doi = {10.3847/2041-8213/aaf96e},
archivePrefix = {arXiv},
       eprint = {1807.05667},
 primaryClass = {astro-ph.CO},
       adsurl = {https://ui.adsabs.harvard.edu/abs/2019ApJ...871L..13F}
}

@ARTICLE{Gray:2020,
       author = {{Gray}, Rachel and {Hernandez}, Ignacio Maga{\~n}a and {Qi}, Hong and {Sur}, Ankan and {Brady}, Patrick R. and {Chen}, Hsin-Yu and {Farr}, Will M. and {Fishbach}, Maya and {Gair}, Jonathan R. and {Ghosh}, Archisman and {Holz}, Daniel E. and {Mastrogiovanni}, Simone and {Messenger}, Christopher and {Steer}, Dani{\`e}le A. and {Veitch}, John},
        title = "{Cosmological inference using gravitational wave standard sirens: A mock data analysis}",
      journal = {\prd},
         year = 2020,
        month = jun,
       volume = {101},
       number = {12},
          eid = {122001},
        pages = {122001},
          doi = {10.1103/PhysRevD.101.122001},
archivePrefix = {arXiv},
       eprint = {1908.06050},
 primaryClass = {gr-qc},
       adsurl = {https://ui.adsabs.harvard.edu/abs/2020PhRvD.101l2001G}
}

@ARTICLE{Gray:2022,
       author = {{Gray}, R. and {Messenger}, C. and {Veitch}, J.},
        title = "{A pixelated approach to galaxy catalogue incompleteness: improving the dark siren measurement of the Hubble constant}",
      journal = {\mnras},
         year = 2022,
        month = may,
       volume = {512},
       number = {1},
        pages = {1127-1140},
          doi = {10.1093/mnras/stac366},
archivePrefix = {arXiv},
       eprint = {2111.04629},
 primaryClass = {astro-ph.CO},
       adsurl = {https://ui.adsabs.harvard.edu/abs/2022MNRAS.512.1127G}
}

@ARTICLE{Gray:2023,
       author = {{Gray}, Rachel and {Beirnaert}, Freija and {Karathanasis}, Christos and {Revenu}, Beno{\^\i}t and {Turski}, Cezary and {Chen}, Anson and {Baker}, Tessa and {Vallejo}, Sergio and {Romano}, Antonio Enea and {Ghosh}, Tathagata and {Ghosh}, Archisman and {Leyde}, Konstantin and {Mastrogiovanni}, Simone and {More}, Surhud},
        title = "{Joint cosmological and gravitational-wave population inference using dark sirens and galaxy catalogues}",
      journal = {\jcap},
         year = 2023,
        month = dec,
       volume = {2023},
       number = {12},
          eid = {023},
        pages = {023},
          doi = {10.1088/1475-7516/2023/12/023},
archivePrefix = {arXiv},
       eprint = {2308.02281},
 primaryClass = {astro-ph.CO},
       adsurl = {https://ui.adsabs.harvard.edu/abs/2023JCAP...12..023G}
}

@ARTICLE{Li:2024,
       author = {{Li}, Guo-Peng and {Fan}, Xi-Long},
        title = "{The origin channels of hierarchical binary black hole mergers in the LIGO-Virgo-KAGRA O1, O2, and O3 runs}",
      journal = {arXiv e-prints},
         year = 2024,
        month = nov,
          eid = {arXiv:2411.09195},
        pages = {arXiv:2411.09195},
          doi = {10.48550/arXiv.2411.09195},
archivePrefix = {arXiv},
       eprint = {2411.09195},
 primaryClass = {astro-ph.HE},
       adsurl = {https://ui.adsabs.harvard.edu/abs/2024arXiv241109195L}
}

@ARTICLE{Abbott:2021gwtc2,
       author = {{Abbott}, R. and {Abbott}, T.~D. and {Abraham}, S. and {Acernese}, F. and {Ackley}, K. and {Adams}, A. and {Adams}, C. and {Adhikari}, R.~X. and {Adya}, V.~B. and {Affeldt}, C. and {Agathos}, M. and {Agatsuma}, K. and {Aggarwal}, N. and {Aguiar}, O.~D. and {Aiello}, L. and {Ain}, A. and {Ajith}, P. and {Akcay}, S. and {Allen}, G. and {Allocca}, A. and {Altin}, P.~A. and {Amato}, A. and {Anand}, S. and {Ananyeva}, A. and {Anderson}, S.~B. and {Anderson}, W.~G. and {Angelova}, S.~V. and {Ansoldi}, S. and {Antelis}, J.~M. and {Antier}, S. and {Appert}, S. and {Arai}, K. and {Araya}, M.~C. and {Areeda}, J.~S. and {Ar{\`e}ne}, M. and {Arnaud}, N. and {Aronson}, S.~M. and {Arun}, K.~G. and {Asali}, Y. and {Ascenzi}, S. and {Ashton}, G. and {Aston}, S.~M. and {Astone}, P. and {Aubin}, F. and {Aufmuth}, P. and {AultONeal}, K. and {Austin}, C. and {Avendano}, V. and {Babak}, S. and {Badaracco}, F. and {Bader}, M.~K.~M. and {Bae}, S. and {Baer}, A.~M. and {Bagnasco}, S. and {Baird}, J. and {Ball}, M. and {Ballardin}, G. and {Ballmer}, S.~W. and {Bals}, A. and {Balsamo}, A. and {Baltus}, G. and {Banagiri}, S. and {Bankar}, D. and {Bankar}, R.~S. and {Barayoga}, J.~C. and {Barbieri}, C. and {Barish}, B.~C. and {Barker}, D. and {Barneo}, P. and {Barnum}, S. and {Barone}, F. and {Barr}, B. and {Barsotti}, L. and {Barsuglia}, M. and {Barta}, D. and {Bartlett}, J. and {Bartos}, I. and {Bassiri}, R. and {Basti}, A. and {Bawaj}, M. and {Bayley}, J.~C. and {Bazzan}, M. and {Becher}, B.~R. and {B{\'e}csy}, B. and {Bedakihale}, V.~M. and {Bejger}, M. and {Belahcene}, I. and {Beniwal}, D. and {Benjamin}, M.~G. and {Bennett}, T.~F. and {Bentley}, J.~D. and {Bergamin}, F. and {Berger}, B.~K. and {Bergmann}, G. and {Bernuzzi}, S. and {Berry}, C.~P.~L. and {Bersanetti}, D. and {Bertolini}, A. and {Betzwieser}, J. and {Bhandare}, R. and {Bhandari}, A.~V. and {Bhattacharjee}, D. and {Bidler}, J. and {Bilenko}, I.~A. and {Billingsley}, G. and {Birney}, R. and {Birnholtz}, O. and {Biscans}, S. and {Bischi}, M. and {Biscoveanu}, S. and {Bisht}, A. and {Bitossi}, M. and {Bizouard}, M. -A. and {Blackburn}, J.~K. and {Blackman}, J. and {Blair}, C.~D. and {Blair}, D.~G. and {Blair}, R.~M. and {Blanch}, O. and {Bobba}, F. and {Bode}, N. and {Boer}, M. and {Boetzel}, Y. and {Bogaert}, G. and {Boldrini}, M. and {Bondu}, F. and {Bonilla}, E. and {Bonnand}, R. and {Booker}, P. and {Boom}, B.~A. and {Bork}, R. and {Boschi}, V. and {Bose}, S. and {Bossilkov}, V. and {Boudart}, V. and {Bouffanais}, Y. and {Bozzi}, A. and {Bradaschia}, C. and {Brady}, P.~R. and {Bramley}, A. and {Branchesi}, M. and {Brau}, J.~E. and {Breschi}, M. and {Briant}, T. and {Briggs}, J.~H. and {Brighenti}, F. and {Brillet}, A. and {Brinkmann}, M. and {Brockill}, P. and {Brooks}, A.~F. and {Brooks}, J. and {Brown}, D.~D. and {Brunett}, S. and {Bruno}, G. and {Bruntz}, R. and {Buikema}, A. and {Bulik}, T. and {Bulten}, H.~J. and {Buonanno}, A. and {Buscicchio}, R. and {Buskulic}, D. and {Byer}, R.~L. and {Cabero}, M. and {Cadonati}, L. and {Caesar}, M. and {Cagnoli}, G. and {Cahillane}, C. and {Calder{\'o}n Bustillo}, J. and {Callaghan}, J.~D. and {Callister}, T.~A. and {Calloni}, E. and {Camp}, J.~B. and {Canepa}, M. and {Cannon}, K.~C. and {Cao}, H. and {Cao}, J. and {Carapella}, G. and {Carbognani}, F. and {Carney}, M.~F. and {Carpinelli}, M. and {Carullo}, G. and {Carver}, T.~L. and {Casanueva Diaz}, J. and {Casentini}, C. and {Caudill}, S. and {Cavagli{\`a}}, M. and {Cavalier}, F. and {Cavalieri}, R. and {Cella}, G. and {Cerd{\'a}-Dur{\'a}n}, P. and {Cesarini}, E. and {Chaibi}, W. and {Chakravarti}, K. and {Chan}, C. -L. and {Chan}, C. and {Chandra}, K. and {Chanial}, P. and {Chao}, S. and {Charlton}, P. and {Chase}, E.~A.},
        title = "{GWTC-2: Compact Binary Coalescences Observed by LIGO and Virgo during the First Half of the Third Observing Run}",
      journal = {Physical Review X},
         year = 2021,
        month = apr,
       volume = {11},
       number = {2},
          eid = {021053},
        pages = {021053},
          doi = {10.1103/PhysRevX.11.021053},
archivePrefix = {arXiv},
       eprint = {2010.14527},
 primaryClass = {gr-qc},
       adsurl = {https://ui.adsabs.harvard.edu/abs/2021PhRvX..11b1053A}
}

@ARTICLE{Abbott:2024gwtc21,
       author = {{Abbott}, R. and {Abbott}, T.~D. and {Acernese}, F. and {Ackley}, K. and {Adams}, C. and {Adhikari}, N. and {Adhikari}, R.~X. and {Adya}, V.~B. and {Affeldt}, C. and {Agarwal}, D. and {Agathos}, M. and {Agatsuma}, K. and {Aggarwal}, N. and {Aguiar}, O.~D. and {Aiello}, L. and {Ain}, A. and {Ajith}, P. and {Albanesi}, S. and {Allocca}, A. and {Altin}, P.~A. and {Amato}, A. and {Anand}, C. and {Anand}, S. and {Ananyeva}, A. and {Anderson}, S.~B. and {Anderson}, W.~G. and {Andrade}, T. and {Andres}, N. and {Andri{\'c}}, T. and {Angelova}, S.~V. and {Ansoldi}, S. and {Antelis}, J.~M. and {Antier}, S. and {Appert}, S. and {Arai}, K. and {Araya}, M.~C. and {Areeda}, J.~S. and {Ar{\`e}ne}, M. and {Arnaud}, N. and {Aronson}, S.~M. and {Arun}, K.~G. and {Asali}, Y. and {Ashton}, G. and {Assiduo}, M. and {Aston}, S.~M. and {Astone}, P. and {Aubin}, F. and {Austin}, C. and {Babak}, S. and {Badaracco}, F. and {Bader}, M.~K.~M. and {Badger}, C. and {Bae}, S. and {Baer}, A.~M. and {Bagnasco}, S. and {Bai}, Y. and {Baird}, J. and {Ball}, M. and {Ballardin}, G. and {Ballmer}, S.~W. and {Balsamo}, A. and {Baltus}, G. and {Banagiri}, S. and {Bankar}, D. and {Barayoga}, J.~C. and {Barbieri}, C. and {Barish}, B.~C. and {Barker}, D. and {Barneo}, P. and {Barone}, F. and {Barr}, B. and {Barsotti}, L. and {Barsuglia}, M. and {Barta}, D. and {Bartlett}, J. and {Barton}, M.~A. and {Bartos}, I. and {Bassiri}, R. and {Basti}, A. and {Bawaj}, M. and {Bayley}, J.~C. and {Baylor}, A.~C. and {Bazzan}, M. and {B{\'e}csy}, B. and {Bedakihale}, V.~M. and {Bejger}, M. and {Belahcene}, I. and {Benedetto}, V. and {Beniwal}, D. and {Bennett}, T.~F. and {Bentley}, J.~D. and {BenYaala}, M. and {Bergamin}, F. and {Berger}, B.~K. and {Bernuzzi}, S. and {Berry}, C.~P.~L. and {Bersanetti}, D. and {Bertolini}, A. and {Betzwieser}, J. and {Beveridge}, D. and {Bhandare}, R. and {Bhardwaj}, U. and {Bhattacharjee}, D. and {Bhaumik}, S. and {Bilenko}, I.~A. and {Billingsley}, G. and {Bini}, S. and {Birney}, R. and {Birnholtz}, O. and {Biscans}, S. and {Bischi}, M. and {Biscoveanu}, S. and {Bisht}, A. and {Biswas}, B. and {Bitossi}, M. and {Bizouard}, M. -A. and {Blackburn}, J.~K. and {Blair}, C.~D. and {Blair}, D.~G. and {Blair}, R.~M. and {Bobba}, F. and {Bode}, N. and {Boer}, M. and {Bogaert}, G. and {Boldrini}, M. and {Bonavena}, L.~D. and {Bondu}, F. and {Bonilla}, E. and {Bonnand}, R. and {Booker}, P. and {Boom}, B.~A. and {Bork}, R. and {Boschi}, V. and {Bose}, N. and {Bose}, S. and {Bossilkov}, V. and {Boudart}, V. and {Bouffanais}, Y. and {Bozzi}, A. and {Bradaschia}, C. and {Brady}, P.~R. and {Bramley}, A. and {Branch}, A. and {Branchesi}, M. and {Brau}, J.~E. and {Breschi}, M. and {Briant}, T. and {Briggs}, J.~H. and {Brillet}, A. and {Brinkmann}, M. and {Brockill}, P. and {Brooks}, A.~F. and {Brooks}, J. and {Brown}, D.~D. and {Brunett}, S. and {Bruno}, G. and {Bruntz}, R. and {Bryant}, J. and {Bulik}, T. and {Bulten}, H.~J. and {Buonanno}, A. and {Buscicchio}, R. and {Buskulic}, D. and {Buy}, C. and {Byer}, R.~L. and {Cadonati}, L. and {Cagnoli}, G. and {Cahillane}, C. and {Bustillo}, J. Calder{\'o}n and {Callaghan}, J.~D. and {Callister}, T.~A. and {Calloni}, E. and {Cameron}, J. and {Camp}, J.~B. and {Canepa}, M. and {Canevarolo}, S. and {Cannavacciuolo}, M. and {Cannon}, K.~C. and {Cao}, H. and {Capote}, E. and {Carapella}, G. and {Carbognani}, F. and {Carlin}, J.~B. and {Carney}, M.~F. and {Carpinelli}, M. and {Carrillo}, G. and {Carullo}, G. and {Carver}, T.~L. and {Diaz}, J. Casanueva and {Casentini}, C. and {Castaldi}, G. and {Caudill}, S. and {Cavagli{\`a}}, M. and {Cavalier}, F. and {Cavalieri}, R. and {Ceasar}, M. and {Cella}, G. and {Cerd{\'a}-Dur{\'a}n}, P. and {Cesarini}, E. and {Chaibi}, W.},
        title = "{GWTC-2.1: Deep extended catalog of compact binary coalescences observed by LIGO and Virgo during the first half of the third observing run}",
      journal = {\prd},
         year = 2024,
        month = jan,
       volume = {109},
       number = {2},
          eid = {022001},
        pages = {022001},
          doi = {10.1103/PhysRevD.109.022001},
archivePrefix = {arXiv},
       eprint = {2108.01045},
 primaryClass = {gr-qc},
       adsurl = {https://ui.adsabs.harvard.edu/abs/2024PhRvD.109b2001A}
}

@ARTICLE{Kritos:2024,
       author = {{Kritos}, Konstantinos and {Reali}, Luca and {Gerosa}, Davide and {Berti}, Emanuele},
        title = "{Minimum gas mass accreted by spinning intermediate-mass black holes in stellar clusters}",
      journal = {\prd},
         year = 2024,
        month = dec,
       volume = {110},
       number = {12},
          eid = {123017},
        pages = {123017},
          doi = {10.1103/PhysRevD.110.123017},
archivePrefix = {arXiv},
       eprint = {2409.15439},
 primaryClass = {astro-ph.HE},
       adsurl = {https://ui.adsabs.harvard.edu/abs/2024PhRvD.110l3017K}
}

@ARTICLE{Abbott:2023gwtc3,
       author = {{Abbott}, R. and {Abbott}, T.~D. and {Acernese}, F. and {Ackley}, K. and {Adams}, C. and {Adhikari}, N. and {Adhikari}, R.~X. and {Adya}, V.~B. and {Affeldt}, C. and {Agarwal}, D. and {Agathos}, M. and {Agatsuma}, K. and {Aggarwal}, N. and {Aguiar}, O.~D. and {Aiello}, L. and {Ain}, A. and {Ajith}, P. and {Akcay}, S. and {Akutsu}, T. and {Albanesi}, S. and {Allocca}, A. and {Altin}, P.~A. and {Amato}, A. and {Anand}, C. and {Anand}, S. and {Ananyeva}, A. and {Anderson}, S.~B. and {Anderson}, W.~G. and {Ando}, M. and {Andrade}, T. and {Andres}, N. and {Andri{\'c}}, T. and {Angelova}, S.~V. and {Ansoldi}, S. and {Antelis}, J.~M. and {Antier}, S. and {Appert}, S. and {Arai}, Koji and {Arai}, Koya and {Arai}, Y. and {Araki}, S. and {Araya}, A. and {Araya}, M.~C. and {Areeda}, J.~S. and {Ar{\`e}ne}, M. and {Aritomi}, N. and {Arnaud}, N. and {Arogeti}, M. and {Aronson}, S.~M. and {Arun}, K.~G. and {Asada}, H. and {Asali}, Y. and {Ashton}, G. and {Aso}, Y. and {Assiduo}, M. and {Aston}, S.~M. and {Astone}, P. and {Aubin}, F. and {Austin}, C. and {Babak}, S. and {Badaracco}, F. and {Bader}, M.~K.~M. and {Badger}, C. and {Bae}, S. and {Bae}, Y. and {Baer}, A.~M. and {Bagnasco}, S. and {Bai}, Y. and {Baiotti}, L. and {Baird}, J. and {Bajpai}, R. and {Ball}, M. and {Ballardin}, G. and {Ballmer}, S.~W. and {Balsamo}, A. and {Baltus}, G. and {Banagiri}, S. and {Bankar}, D. and {Barayoga}, J.~C. and {Barbieri}, C. and {Barish}, B.~C. and {Barker}, D. and {Barneo}, P. and {Barone}, F. and {Barr}, B. and {Barsotti}, L. and {Barsuglia}, M. and {Barta}, D. and {Bartlett}, J. and {Barton}, M.~A. and {Bartos}, I. and {Bassiri}, R. and {Basti}, A. and {Bawaj}, M. and {Bayley}, J.~C. and {Baylor}, A.~C. and {Bazzan}, M. and {B{\'e}csy}, B. and {Bedakihale}, V.~M. and {Bejger}, M. and {Belahcene}, I. and {Benedetto}, V. and {Beniwal}, D. and {Bennett}, T.~F. and {Bentley}, J.~D. and {Benyaala}, M. and {Bergamin}, F. and {Berger}, B.~K. and {Bernuzzi}, S. and {Berry}, C.~P.~L. and {Bersanetti}, D. and {Bertolini}, A. and {Betzwieser}, J. and {Beveridge}, D. and {Bhandare}, R. and {Bhardwaj}, U. and {Bhattacharjee}, D. and {Bhaumik}, S. and {Bilenko}, I.~A. and {Billingsley}, G. and {Bini}, S. and {Birney}, R. and {Birnholtz}, O. and {Biscans}, S. and {Bischi}, M. and {Biscoveanu}, S. and {Bisht}, A. and {Biswas}, B. and {Bitossi}, M. and {Bizouard}, M. -A. and {Blackburn}, J.~K. and {Blair}, C.~D. and {Blair}, D.~G. and {Blair}, R.~M. and {Bobba}, F. and {Bode}, N. and {Boer}, M. and {Bogaert}, G. and {Boldrini}, M. and {Bonavena}, L.~D. and {Bondu}, F. and {Bonilla}, E. and {Bonnand}, R. and {Booker}, P. and {Boom}, B.~A. and {Bork}, R. and {Boschi}, V. and {Bose}, N. and {Bose}, S. and {Bossilkov}, V. and {Boudart}, V. and {Bouffanais}, Y. and {Bozzi}, A. and {Bradaschia}, C. and {Brady}, P.~R. and {Bramley}, A. and {Branch}, A. and {Branchesi}, M. and {Brandt}, J. and {Brau}, J.~E. and {Breschi}, M. and {Briant}, T. and {Briggs}, J.~H. and {Brillet}, A. and {Brinkmann}, M. and {Brockill}, P. and {Brooks}, A.~F. and {Brooks}, J. and {Brown}, D.~D. and {Brunett}, S. and {Bruno}, G. and {Bruntz}, R. and {Bryant}, J. and {Bulik}, T. and {Bulten}, H.~J. and {Buonanno}, A. and {Buscicchio}, R. and {Buskulic}, D. and {Buy}, C. and {Byer}, R.~L. and {Davies}, G.~S. Cabourn and {Cadonati}, L. and {Cagnoli}, G. and {Cahillane}, C. and {Bustillo}, J. Calder{\'o}n and {Callaghan}, J.~D. and {Callister}, T.~A. and {Calloni}, E. and {Cameron}, J. and {Camp}, J.~B. and {Canepa}, M. and {Canevarolo}, S. and {Cannavacciuolo}, M. and {Cannon}, K.~C. and {Cao}, H. and {Cao}, Z. and {Capocasa}, E. and {Capote}, E. and {Carapella}, G. and {Carbognani}, F.},
        title = "{GWTC-3: Compact Binary Coalescences Observed by LIGO and Virgo during the Second Part of the Third Observing Run}",
      journal = {Physical Review X},
         year = 2023,
        month = oct,
       volume = {13},
       number = {4},
          eid = {041039},
        pages = {041039},
          doi = {10.1103/PhysRevX.13.041039},
archivePrefix = {arXiv},
       eprint = {2111.03606},
 primaryClass = {gr-qc},
       adsurl = {https://ui.adsabs.harvard.edu/abs/2023PhRvX..13d1039A}
}

@ARTICLE{Schutz:1986,
       author = {{Schutz}, B.~F.},
        title = "{Determining the Hubble constant from gravitational wave observations}",
      journal = {\nat},
         year = 1986,
        month = sep,
       volume = {323},
       number = {6086},
        pages = {310-311},
          doi = {10.1038/323310a0},
       adsurl = {https://ui.adsabs.harvard.edu/abs/1986Natur.323..310S}
}

@ARTICLE{Pozzo:2012,
       author = {{Del Pozzo}, Walter},
        title = "{Inference of cosmological parameters from gravitational waves: Applications to second generation interferometers}",
      journal = {\prd},
         year = 2012,
        month = aug,
       volume = {86},
       number = {4},
          eid = {043011},
        pages = {043011},
          doi = {10.1103/PhysRevD.86.043011},
archivePrefix = {arXiv},
       eprint = {1108.1317},
 primaryClass = {astro-ph.CO},
       adsurl = {https://ui.adsabs.harvard.edu/abs/2012PhRvD..86d3011D}
}

@ARTICLE{Farr:2019,
       author = {{Farr}, Will M. and {Fishbach}, Maya and {Ye}, Jiani and {Holz}, Daniel E.},
        title = "{A Future Percent-level Measurement of the Hubble Expansion at Redshift 0.8 with Advanced LIGO}",
      journal = {\apjl},
         year = 2019,
        month = oct,
       volume = {883},
       number = {2},
          eid = {L42},
        pages = {L42},
          doi = {10.3847/2041-8213/ab4284},
archivePrefix = {arXiv},
       eprint = {1908.09084},
 primaryClass = {astro-ph.CO},
       adsurl = {https://ui.adsabs.harvard.edu/abs/2019ApJ...883L..42F}
}

@ARTICLE{Ezquiaga:2022,
       author = {{Ezquiaga}, Jose Mar{\'\i}a and {Holz}, Daniel E.},
        title = "{Spectral Sirens: Cosmology from the Full Mass Distribution of Compact Binaries}",
      journal = {\prl},
         year = 2022,
        month = aug,
       volume = {129},
       number = {6},
          eid = {061102},
        pages = {061102},
          doi = {10.1103/PhysRevLett.129.061102},
archivePrefix = {arXiv},
       eprint = {2202.08240},
 primaryClass = {astro-ph.CO},
       adsurl = {https://ui.adsabs.harvard.edu/abs/2022PhRvL.129f1102E}
}

@ARTICLE{Palmese:2025,
       author = {{Palmese}, Antonella and {Mastrogiovanni}, Simone},
        title = "{Gravitational Wave Cosmology}",
      journal = {arXiv e-prints},
         year = 2025,
        month = jan,
          eid = {arXiv:2502.00239},
        pages = {arXiv:2502.00239},
          doi = {10.48550/arXiv.2502.00239},
archivePrefix = {arXiv},
       eprint = {2502.00239},
 primaryClass = {astro-ph.CO},
       adsurl = {https://ui.adsabs.harvard.edu/abs/2025arXiv250200239P}
}

@ARTICLE{Chernoff:1993,
       author = {{Chernoff}, David F. and {Finn}, Lee S.},
        title = "{Gravitational Radiation, Inspiraling Binaries, and Cosmology}",
      journal = {\apjl},
         year = 1993,
        month = jul,
       volume = {411},
        pages = {L5},
          doi = {10.1086/186898},
archivePrefix = {arXiv},
       eprint = {gr-qc/9304020},
 primaryClass = {gr-qc},
       adsurl = {https://ui.adsabs.harvard.edu/abs/1993ApJ...411L...5C}
}

@ARTICLE{Taylor:2012,
       author = {{Taylor}, Stephen R. and {Gair}, Jonathan R. and {Mandel}, Ilya},
        title = "{Cosmology using advanced gravitational-wave detectors alone}",
      journal = {\prd},
         year = 2012,
        month = jan,
       volume = {85},
       number = {2},
          eid = {023535},
        pages = {023535},
          doi = {10.1103/PhysRevD.85.023535},
archivePrefix = {arXiv},
       eprint = {1108.5161},
 primaryClass = {gr-qc},
       adsurl = {https://ui.adsabs.harvard.edu/abs/2012PhRvD..85b3535T}
}

@ARTICLE{Mastrogiovanni:2023icaro,
       author = {{Mastrogiovanni}, Simone and {Laghi}, Danny and {Gray}, Rachel and {Santoro}, Giada Caneva and {Ghosh}, Archisman and {Karathanasis}, Christos and {Leyde}, Konstantin and {Steer}, Dani{\`e}le A. and {Perri{\`e}s}, St{\'e}phane and {Pierra}, Gr{\'e}goire},
        title = "{Joint population and cosmological properties inference with gravitational waves standard sirens and galaxy surveys}",
      journal = {\prd},
         year = 2023,
        month = aug,
       volume = {108},
       number = {4},
          eid = {042002},
        pages = {042002},
          doi = {10.1103/PhysRevD.108.042002},
archivePrefix = {arXiv},
       eprint = {2305.10488},
 primaryClass = {astro-ph.CO},
       adsurl = {https://ui.adsabs.harvard.edu/abs/2023PhRvD.108d2002M}
}

@ARTICLE{Tomar:2026adk,
       author = {{Tomar}, Yashvardhan and {Hopkins}, Philip F. and {Kremer}, Kyle},
        title = "{Thick Disks, Thin Hopes: Suppressed Capture and Merger Rates in AGN}",
      journal = {arXiv e-prints},
         year = 2026,
        month = jan,
          eid = {arXiv:2601.02487},
        pages = {arXiv:2601.02487},
          doi = {10.48550/arXiv.2601.02487},
archivePrefix = {arXiv},
       eprint = {2601.02487},
 primaryClass = {astro-ph.HE},
       adsurl = {https://ui.adsabs.harvard.edu/abs/2026arXiv260102487T}
}

\appendix
\section{Derivations}

\subsection{GW redshift populations}\label{subsect:zpops}

The spatial population prior (Eq. \ref{eq:ppop_full}) contains two terms: the AGN-origin GW population, $p_{\rm pop}(z, \Omega| \Lambda, A)$, and the alternative-origin GW population, $p_{\rm pop}(z, \Omega| \Lambda, \conj{A})$. In this section, we will derive expressions for each of these terms. Since we assume a fixed AGN-origin GW population, we drop its dependence on $\Lambda$ from here. We add a conditioning on the label $s$, which indicates that the GW is of astrophysical origin, as opposed to a glitch. This condition is implicit in Sect. \ref{sect:method} for conciseness, but it is necessary here to model the redshift dependence of the merger rate.

GWs produced through the AGN-channel necessarily occur in an AGN. Therefore, observations of AGN positions provide a prior on the GW redshift and sky position. 
However, EM observations are not complete. Following the approach of \citet{Gray:2022, Gray:2023}, this is taken into account by marginalizing the population prior over the cases that the GW host is ($G$) or is not ($\conj{G}$) in the catalogue (however, see \citealt{Barbieri:2026} and \citealt{Ghosh:2026} for an alternative method). Since the EM selection effects depend strongly on AGN luminosity, we marginalise over $L_{\rm bol}$ as well:
\begin{align}
    p_{\rm pop}(z, \Omega | s, A) 
    &= p(G | s, A) \int \dd L_{\rm bol} \, p(z, \Omega, L_{\rm bol}|G, s, A) \notag \\
    &\quad + p(\conj{G} | s, A) \int \dd L_{\rm bol} \, p(z, \Omega, L_{\rm bol}|\conj{G}, s, A) \, .
\end{align}
Separating out the dependence on $s$, this reduces to
\begin{align}\label{eq:agn_origin_ppop}
    p_{\rm pop}(z, \Omega | s, A) 
    &\propto p(s | z, A) \Bigg[ p(G | A) \int \dd L_{\rm bol} \, p(z, \Omega, L_{\rm bol}|G, A) \notag \\
    &\quad + \int \dd L_{\rm bol} \, p(\conj{G}|z, \Omega, L_{\rm bol}, A) p(z, \Omega, L_{\rm bol}| A) \Bigg] \, ,
\end{align}
where we assumed the merger rate to be independent of sky position and AGN luminosity, such that $p(s|z, \Omega, L_{\rm bol}, A) = p(s|z, A)$ (see \citealt{Gray:2023} Sect. 2.1 for a detailed derivation). 
We now proceed with the derivations of the functions appearing in Eq. \ref{eq:agn_origin_ppop}.

The in-catalogue part of the population prior, $p(z, \Omega, L_{\rm bol} | G, A)$, is constructed from the EM catalogue, since it is conditioned on $G$. We assume that the AGN sky position error is negligible compared to the GW localisation areas. Additionally, we neglect the measurement uncertainty on the bolometric luminosity, such that the luminosity-marginalised redshift posterior of AGN $k$ is $\int \dd L_{\rm bol} \, p_{\rm agn}(z, L_{\rm bol} | \hat{z}_{k}, \hat{L}_{\mathrm{bol}, k}) = p_{\rm agn}(z | \hat{z}_{k}, \hat{L}_{\mathrm{bol}, k})$. Combined, we obtain
\begin{align}
    \int \dd L_{\rm bol} \, p(z, \Omega, L_{\rm bol} &| G, A) \notag \\
    &= \frac{1}{N_{\rm agn}} \sum_{k=1}^{N_{\rm agn}} p_{\rm agn}(z | \hat{z}_{k}, \hat{L}_{\mathrm{bol}, k})\delta(\Omega - \hat \Omega_k) \, .
\end{align}

The EM selection effects enter the redshift prior in the out-of-catalogue term, through $p(\conj{G}|z, \Omega, L_{\rm bol}, A)$. Since $G$ is the complement of $\conj{G}$, we can write $p(\conj{G} | z, \Omega, L_{\rm bol}, A) = 1 - p(G | z, \Omega, L_{\rm bol}, A)$.
Integrating over bolometric luminosity then gives
\begin{align}
    \int \dd L_{\rm bol} \, p(\conj{G} | z, \Omega, L_{\rm bol}, A) &p(z, \Omega, L_{\rm bol} | A) \notag \\
    &= \left( 1 - p(G | z, \Omega, A) \right) p(z, \Omega| A) \, ,
\end{align}
where $p(G | z, \Omega, A) \equiv P_{\rm det}^{\rm EM}(z, \Omega)$ is the AGN detection probability, or equivalently, the completeness of the AGN catalogue as a function of redshift and sky position. We show how the completeness is calculated in Sect. \ref{sect:completeness}. Since $p(z, \Omega| A)$ is not conditioned on $\conj{G}$, it represents the overall distribution of all AGN redshifts, for which we adopt the observationally constrained QLF of \citet{kulkarni:2019}. Since this is our prior on the spatial AGN distribution, we obtain $p(z, \Omega| A) = \tfrac{\pi_{\rm agn}(z)}{4\pi}$.

Since $p(G | A)$ is not conditioned on the redshift or sky position, it represents the spatial average of the completeness,
\begin{align}
    p(G | A) 
    &= \iint \dd z \dd \Omega \, P_{\rm det}^{\rm EM}(z, \Omega) \frac{\pi_{\rm agn}(z)}{4\pi} \equiv \left \langle P_{\rm det}^{\rm EM} \right \rangle \, .
\end{align}

Finally, we model the redshift evolution of the merger rate, $p(s|z, A)$. We assume all AGN to have an equal probability of hosting a GW event. Therefore, only the time dilation between the source frame and the detector frame contributes to this term, adding a factor $\frac{1}{1 + z}$. The total AGN-origin GW population prior is given in Eq. \ref{eq:pagn}.

The alternative-origin population prior can be split in a part that models the merger rate evolution and a part that indicates the spatial distribution of GW sources: $p_{\rm pop}(z, \Omega | \Lambda, s, \conj{A}) \propto p(s|z, \Lambda, \conj{A}) p(z, \Omega | \conj{A})$. We assume the GW sources to follow a uniform-in-comoving-volume distribution ($\propto \dd V_{\rm c}/\dd z$). The rate is modelled as $p(s|z, \Lambda, \conj{A}) \propto \frac{\psi(z | \Lambda)}{1 + z}$, with $\psi(z | \Lambda)$ being the Madau-Dickinson cosmic star-formation rate (SFR) parametrisation \citep{Madau:2014}. Combined, the alternative-origin GW population prior is given by
\begin{equation}
    p_{\rm pop}(z, \Omega | \Lambda, \conj{A}, s) \propto \frac{1}{4 \pi} \frac{\psi(z | \Lambda)}{1 + z} \frac{\dd V_{\rm c}}{\dd z} \, .
\end{equation}

\subsection{The spatial-correlation likelihood}\label{apx:V23}
In this section we derive the spatial-correlation likelihood presented in \citet{Veronesi:2023}. To this end, we restate our population prior as a function of comoving volume ($V_{\rm c}$). Uniform-in-comoving-volume distributions are then equal to a constant, denoted $1/V$. The AGN positions are assumed to be Dirac-delta functions in comoving volume at the measured position $\hat{V}_{k}$. The spatial distribution of AGN is assumed to be uniform in comoving volume. Note that this is inconsistent with the completeness estimation, where the AGN distribution is assumed to follow the QLF. Finally, we neglect the redshift-evolution of the GW merger rate. In total, we obtain the population prior
\begin{align}
p_{\rm pop}(V_{\rm c}, \theta'&|\Lambda, \fagn) = p_{\rm pop}^{\rm ref}(\theta' | \Lambda) \Bigg\{ \notag \\
& \fagn \left \langle P_{\rm det}^{\rm EM} \right \rangle \frac{1}{N_{\rm agn}} \sum_{k=1}^{N_{\rm agn}} \delta(V_{\rm c} - \hat{V}_{\mathrm{c}, k}) \notag \\
&+ \big(1 - \fagn P_{\rm det}^{\rm EM}(V_{\rm c})\big) \frac{1}{V} \Bigg\} .
\end{align}

Filling this expression into the hierarchical Bayesian likelihood (Eq. \ref{eq:hierarchical_llh}), and further assuming $p_{\rm pop}^{\rm ref}(\theta' | \Lambda) = \pi_{\rm PE}(\theta')$, $\pi_{\rm PE}(V_{\rm c}) = 1/V$, and a GW detection efficiency that is independent of $\fagn$, we get
\begin{align}
    p(D|\Lambda) 
    &\propto \prod_{i=1}^{N_{\rm GW}} p(d_i) \int \dd \theta \, \frac{p_{\rm GW}(\theta|d_i)}{\pi_{\rm PE}(\theta)} p_{\rm pop}(\theta | \Lambda) \, \\
    &\propto \prod_{i=1}^{N_{\rm GW}} p(d_i) \int \dd V_{\rm c} \, \frac{p_{\rm GW}(V_{\rm c}|d_i)}{\pi_{\rm PE}(V_{\rm c})} p_{\rm pop}(V_{\rm c} | \Lambda) \\
    &\propto \prod_{i=1}^{N_{\rm GW}} p(d_i)V \Bigg[ \fagn \left \langle P_{\rm det}^{\rm EM} \right \rangle \frac{\sum_{k=1}^{N_{\rm agn}} p_{\rm GW} \left (\hat{V}_{\mathrm{c}, k} \right)}{N_{\rm agn}} \notag \\
    &\qquad\qquad\quad + \left(1 - \fagn \left \langle P_{\rm det}^{\rm EM} \right \rangle \right) \frac{1}{V} \Bigg ] \, ,
\end{align}
where in the final equality we have used that the posterior $p_{\rm GW}(V_{\rm c}|d_i)$ is normalised in comoving volume, and we have assumed the completeness does not change significantly within the GW localisation volume over which it is integrated: $P_{\rm det}^{\rm EM}(V_{\rm c}) \approx \left \langle P_{\rm det}^{\rm EM} \right \rangle$.

\section{Biases from the inconsistent handling of EM data}\label{subsect:bias}

Parameter inference is biased when the likelihood is inconsistent with the data generation process. In this section, we quantify the bias introduced in $\fagn$ estimates by two inconsistencies in modelling the AGN catalogue. The first is neglecting AGN redshift errors (Sect. \ref{subsubsect:neglect_err}). The second is estimating the redshift selection function from the data, either by binning the mean redshifts or by considering the full AGN redshift posteriors (Sect. \ref{subsubsect:agn_selfunc}).

\subsection{Neglecting AGN redshift errors}\label{subsubsect:neglect_err}

\begin{figure}
    \centering
    \includegraphics[width=\linewidth]{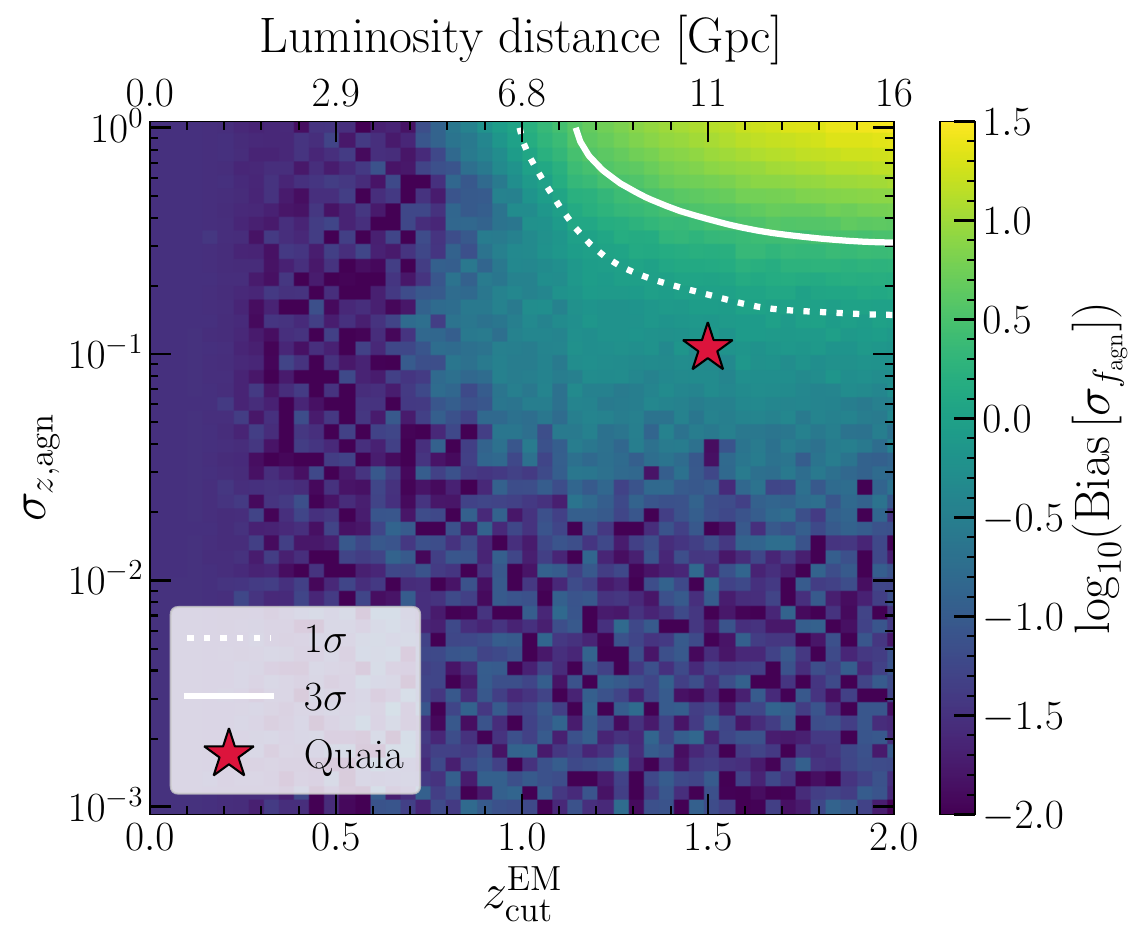}
    \caption{The median bias introduced by neglecting AGN redshift errors as a function of the AGN redshift error ($\sigma_{z,\rm agn}$) and the redshift cut of the AGN catalogue ($z_{\rm cut}^{\rm EM}$). The bias is calculated as the number of standard deviations the biased posterior median is away from the injected value of $\fagn$. It is expressed in terms of the standard deviation of the biased posterior. The white dotted (solid) line demarcates the region where the bias reaches $1\sigma_{\fagn}$($3\sigma_{\fagn}$).
    The star indicates the median redshift error of AGN with bolometric luminosities $\lbol{46.5}$ in Quaia, corresponding to the sample used by \citet{Veronesi:2025}, who used $z_{\rm cut}^{\rm EM}=1.5$. Neglecting these redshift errors results in an underestimation of $\fagn$ by $\approx 0.5\sigma_{\fagn}$.
    }
    \label{fig:bias_zerror}
\end{figure}

Past efforts to constrain $\fagn$ by combining the spatial GW posteriors and an AGN catalogue have always involved neglecting the redshift errors of the AGN \citep{Veronesi:2023, Veronesi:2025, Zhu:2025}. The reasoning has been that GW redshift errors are large in comparison to AGN redshift errors. With Fig. \ref{fig:bias_zerror} we show the bias this may introduce in EM-informed population inference. This plot is made with the same mock data as was used in making Fig. \ref{fig:cat_quality_effect}, only now we wrongly assume in our likelihood that the AGN redshift errors are zero. We do not extrapolate these results to $10^5$ injections.

Neglecting AGN redshift errors always causes $\fagn$ to be underestimated. This can be understood in the limit of very large redshift errors, where the mean observed redshift may lie far from the true redshift. In this case, the mean is not a good probe for the true GW redshift. Effectively, the AGN catalogue will look like a random set of points, uncorrelated to the GW localisations. This causes an underestimation of the evidence for the AGN channel (Eq. \ref{eq:hypothesis_evidence}), driving the posterior on $\fagn$ towards zero. Incorporating the full AGN redshift posteriors restores the overlap between the AGN and GW localisation volumes, and is therefore essential for an unbiased inference.

Interestingly, the largest bias arises from incorrectly modelling AGN that provide the least information on $\fagn$, namely those with $z > z_{\rm thresh}^{\rm GW}$ and large $\sigma_{z,\rm agn}$. Although these sources lie beyond the GW horizon, redshift uncertainties can scatter some of them below $z_{\rm thresh}^{\rm GW}$, contaminating the AGN sample in this regime. 

The bias decreases as the AGN redshift measurements become more precise. This is consistent with the argument that the AGN redshift errors may be neglected once they become small compared to the GW redshift errors. The average bias in \citet{Veronesi:2025} from assuming perfect redshift errors in Quaia is $\approx 0.5\sigma_{\fagn}$. As GW localisations improve in the future, this bias should be re-evaluated. Currently, our GW and AGN samples have comparable median redshift errors of $\sigma_z \approx 0.1$. 

Below $z_{\rm cut}^{\rm EM}\approx0.3$, the bias remains unchanged because AGN are too sparse to significantly affect the $\fagn$ posterior.

\subsection{Measuring the AGN redshift selection function from the data}\label{subsubsect:agn_selfunc}

The fraction of observable events from a population, or detection efficiency, can be calculated from Eq. \ref{eq:detection_efficiency}. This equation requires a specification of the population priors from which events (either GW or EM) are generated, and of the detection probability. The latter depends on the likelihood function for the data realisation of the event, and a detectability threshold set by the detector. In many cases, this cannot be calculated analytically. For GW observations, the numerical calculation of detection efficiencies involve simulating the injection and detection of GW events. This is only possible because the detection process of GWs can be modelled accurately. This is often not the case for EM catalogues. For example, Quaia is the intersection of Gaia and unWISE, to which additional colour selections have been applied. Modelling this selection function would pose a significant challenge, as it would require modelling the detection sensitivities of multiple instruments that operate at different wavelengths, as well as specifying population models of colours of astronomical objects. As a result, EM selection functions are often determined from the data. Even though this is fundamentally an inconsistent approach, there are applications where the error introduced is negligible.

We consider two ways of estimating the redshift selection function of Quaia. One method is binning the mean observed AGN redshifts and comparing to the expected number of AGN in that bin, as done by \citet{Veronesi:2025}. In this work we adopt an alternative solution: we sum the full AGN redshift posteriors and convert this to a number density at all redshifts, and compare it to the expected number density (see Sect. \ref{sect:completeness}). 

We expect our method to do best when considering an AGN population with many members and large redshift errors, as this will result in a smooth redshift distribution. For example, consider a flux-limited catalogue of an AGN population that has just a few AGN located at high redshifts. The absence of AGN at low redshifts will be interpreted as a low completeness in these regions. However, the detection probability should reflect the probability of detecting an AGN if it were to be located at a given redshift. If the catalogue is flux-limited, the completeness should have been high at low redshifts, but the data realisation of rare AGN caused the completeness to approach zero. Likewise, if the redshift errors of AGN are small, the estimated selection function will approach a sum of Dirac delta functions, resulting in a completeness of zero in several redshift intervals. This effect is less severe the more AGN posteriors are summed, leaving a smaller part of redshift-space unoccupied. In this case, a hybrid approach may be beneficial, where the number densities are binned after summing the redshift posteriors, but we do not explore this further here.

We conclude that, using our method, the completeness estimation of the $\lbol{46.5}$ AGN population is most at risk of being biased, especially at low redshifts, where there are few of these AGN. Fortunately, in our framework the completeness ($P_{\rm det}^{\rm EM}(z)$) is always multiplied by the AGN redshift population prior ($\pi_{\rm agn}(z)$). The population prior suppresses the completeness in regions where the expected number of detected AGN is low. These are the redshift intervals where the error in the estimated completeness function would be largest. Therefore, even though the relative error in the quantity $P_{\rm det}^{\rm EM}(z)\pi_{\rm agn}(z)$ could be large, the absolute error is suppressed. This ensures that our method for calculating the detection efficiency leads to negligible biases in our final results.

\begin{figure}
    \centering
    \includegraphics[width=\linewidth]{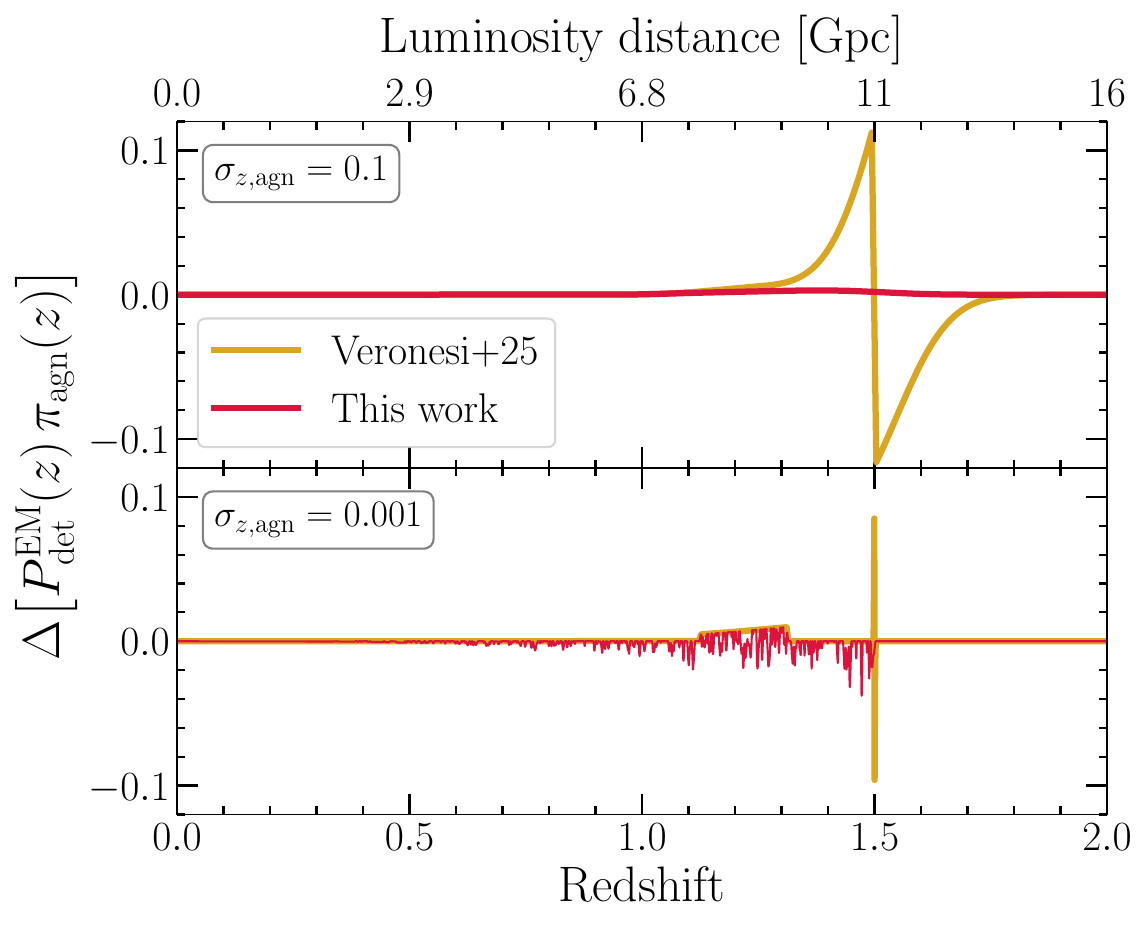}
    \caption{The absolute error of the product of the AGN detection probability ($P_{\rm det}^{\rm EM}$), or completeness, and the AGN redshift population distribution ($\pi_{\rm agn}$) as a function of redshift. We compare the accuracy of this quantity when approximating $P_{\rm det}^{\rm EM}$ in two ways. The gold solid lines show the result of binning the mean observed AGN redshifts and comparing to the expected number of AGN in that bin, cf. \citet{Veronesi:2025}. The red solid lines show the result of summing the full AGN redshift posteriors and comparing to the expected number density of AGN at all redshifts, which is the method we adopt in this work. The upper panel shows the absolute error when all AGN have an absolute redshift error of $\sigma_{z, \rm agn} = 0.1$, whereas the bottom panel shows $\sigma_{z, \rm agn} = 0.001$.}
    \label{fig:pdet_times_ppop}
\end{figure}

We show this principle in Fig. \ref{fig:pdet_times_ppop} for two values of the absolute AGN redshift error, generating the incomplete catalogue as described in Sect. \ref{subsubsect:mockagn}. The top panel shows that the method employed in this work performs better in the presence of redshift uncertainties with a magnitude typical of Quaia ($\sigma_{z, \rm agn} = 0.1$). The bottom panel shows that binning is preferred when AGN redshift measurements approach spectroscopic precision. This is especially true when the number of AGN is small enough to leave visible gaps in the redshift distribution. The error at $z = 1.5$ using the V25 approach is due to the assumption of a sharp transition to zero completeness, whereas the actual completeness is smoothed around this value due to the AGN redshift errors. Accordingly, the error spike is more narrow in the bottom panel because the AGN redshift errors are smaller.

\section{Additional figures and tables}\label{additional_fig_tab}

\begin{figure}
    \centering
    \includegraphics[width=\linewidth]{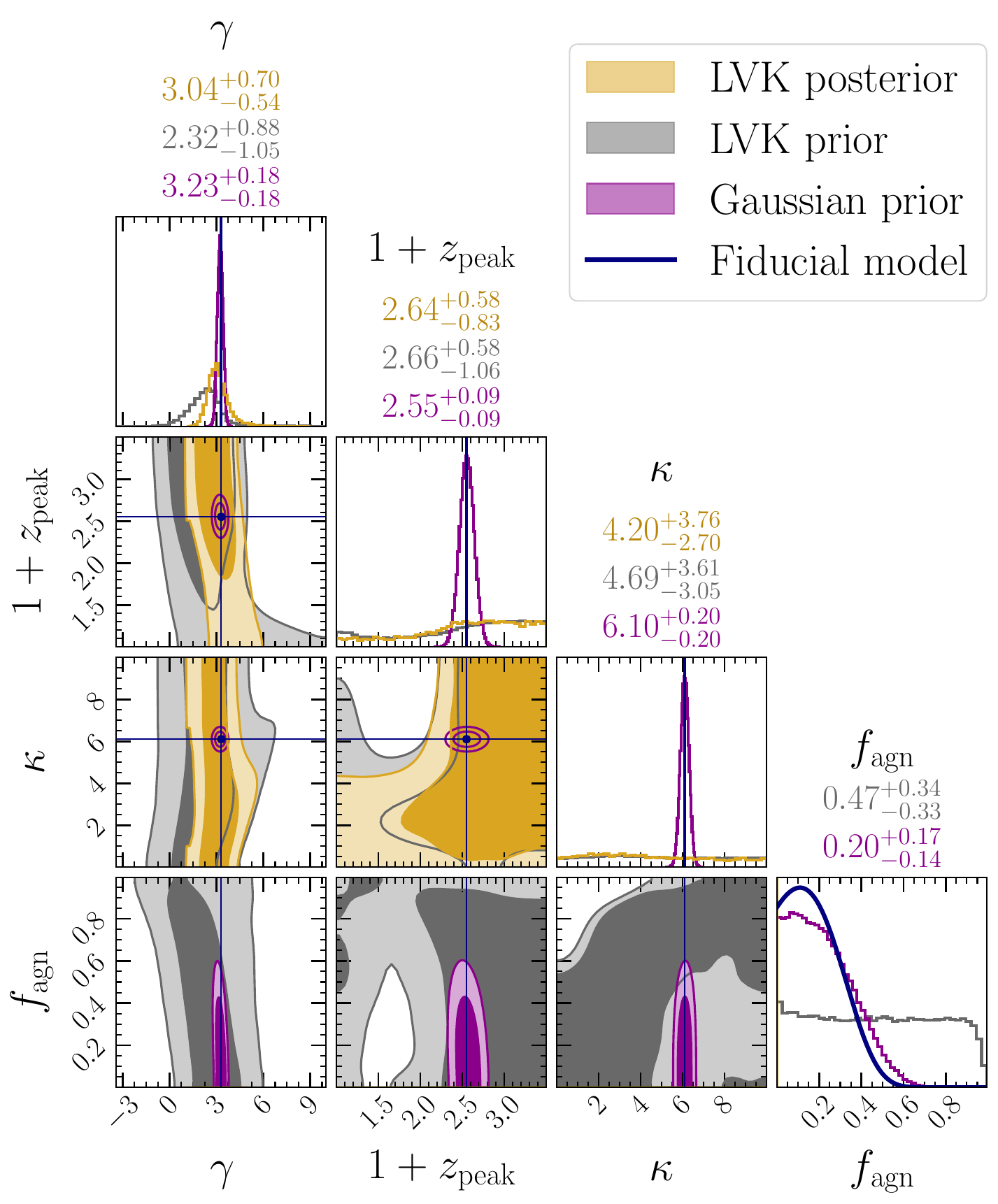}
    \caption{Joint posteriors on the fraction ($\fagn$) of gravitational wave (GW) events coming from unobscured AGN brighter than $\lbol{45.5}$, and the hyperparameters ($\gamma$, $z_{\rm peak}$, $\kappa$) of the merger rate evolution of the alternative-origin GW population. Joint posteriors are shown for three sets of hyperpriors: broad uniform priors used by LVK \citep{ligo:gwtc5pop} (grey), Gaussian priors from a fit to EM data \citep{Strolger:2020} (purple), and our fiducial model where $\fagn$ is the only free parameter (blue). Yellow contours show the hyperposteriors obtained by \citet{ligo:gwtc5pop} under their \texttt{Default BBH} model with a \texttt{Madau-Dickinson} merger rate evolution with redshift. Contours indicate the 68 percent and 95 percent credible levels.
    }
    \label{fig:corner45p5}
\end{figure}

\begin{figure}
    \centering
    \includegraphics[width=\linewidth]{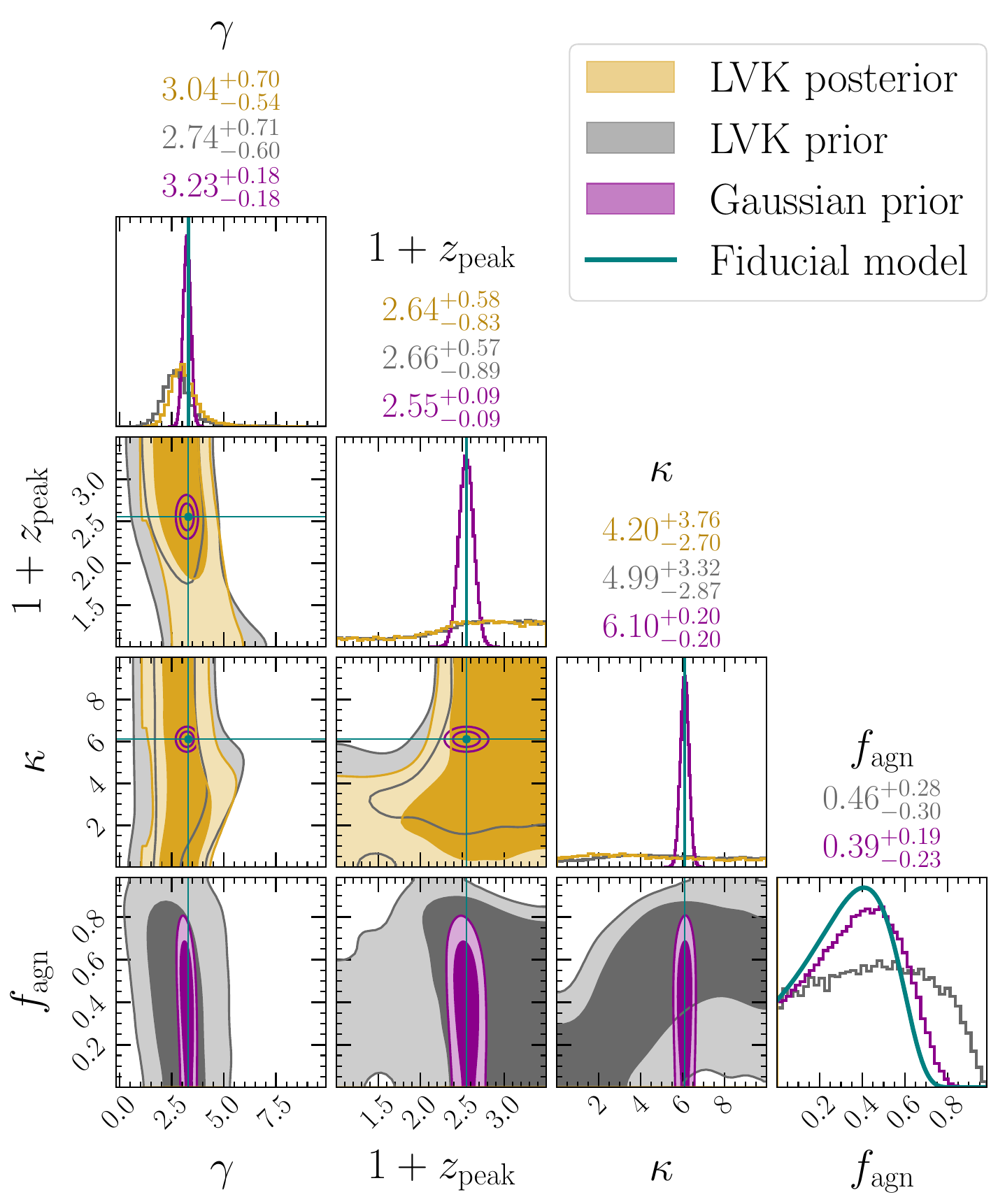}
    \caption{Joint posteriors on the fraction ($\fagn$) of gravitational wave (GW) events coming from unobscured AGN brighter than $\lbol{46.5}$, and the hyperparameters ($\gamma$, $z_{\rm peak}$, $\kappa$) of the merger rate evolution of the alternative-origin GW population. Joint posteriors are shown for three sets of hyperpriors: broad uniform priors used by LVK \citep{ligo:gwtc5pop} (grey), Gaussian priors from a fit to EM data \citep{Strolger:2020} (purple), and our fiducial model where $\fagn$ is the only free parameter (teal). Yellow contours show the hyperposteriors obtained by \citet{ligo:gwtc5pop} under their \texttt{Default BBH} model with a \texttt{Madau-Dickinson} merger rate evolution with redshift. Contours indicate the 68 percent and 95 percent credible levels.
    }
    \label{fig:corner46p5}
\end{figure}

Here, we provide supplementary figures and tables mentioned in the main text. Tab. \ref{tab:agn_probabilities} shows a ranking of all 256 GW events in our sample by their posterior probability of coming from an AGN, $p(\mathrm{AGN})$, as plotted in Fig. \ref{fig:p_AGN} in Sect. \ref{sect:results}. Figs. \ref{fig:corner45p5}-\ref{fig:corner46p5} show the joint hyperposteriors on $\fagn$ and the hyperparameters that govern the redshift evolution of the rate of alternative-origin GW events, assuming the subpopulations of AGN brighter than $\lbol{45.5}$ (Fig. \ref{fig:corner45p5}) and $\lbol{46.5}$ (Fig. \ref{fig:corner46p5}). These figures are similar to Fig. \ref{fig:corner44p5} in Sect. \ref{subsect:4Dpost}. Tab. \ref{tab:constraints} shows the 95 percent credible intervals on $\fagn$ for all hyperpriors and AGN subpopulations considered in this work.

\begin{table*}
\centering
\caption{Probability of each GW event being hosted by an AGN with $\lbol{44.5}$. The probability that at least one event has an AGN origin is 98.1 percent.}
\label{tab:agn_probabilities}
\begin{tabular}{c c @{\hspace{1cm}} c c @{\hspace{1cm}} c c @{\hspace{1cm}} c c}
\hline\hline
Event & $p({\rm AGN})$ & Event & $p({\rm AGN})$ & Event & $p({\rm AGN})$ & Event & $p({\rm AGN})$ \\
\hline
GW230704\_212616 & 0.250 & GW231102\_071736 & 0.185 & GW241225\_082815 & 0.151 & GW241009\_084816 & 0.130 \\
GW240824\_205609 & 0.238 & GW240531\_075248 & 0.184 & GW231014\_040532 & 0.151 & GW240627\_131622 & 0.130 \\
GW241201\_055758 & 0.235 & GW241007\_082943 & 0.184 & GW240420\_175625 & 0.150 & GW190408\_181802 & 0.130 \\
GW241229\_155844 & 0.235 & GW191127\_050227 & 0.183 & GW200129\_065458 & 0.150 & GW240921\_201835 & 0.130 \\
GW231119\_075248 & 0.234 & GW240515\_005301 & 0.182 & GW240716\_034900 & 0.150 & GW191103\_012549 & 0.129 \\
GW240621\_214041 & 0.233 & GW230708\_230935 & 0.181 & GW230706\_104333 & 0.149 & GW191129\_134029 & 0.129 \\
GW240618\_071627 & 0.233 & GW240930\_234614 & 0.181 & GW241006\_015333 & 0.149 & GW230630\_234532 & 0.129 \\
GW231230\_170116 & 0.232 & GW231029\_111508 & 0.181 & GW241130\_034908 & 0.148 & GW240520\_213616 & 0.129 \\
GW231005\_021030 & 0.232 & GW231008\_142521 & 0.180 & GW231118\_005626 & 0.148 & GW230927\_153832 & 0.129 \\
GW240612\_081540 & 0.230 & GW190731\_140936 & 0.180 & GW250118\_170523 & 0.148 & GW241110\_124123 & 0.128 \\
GW230630\_125806 & 0.229 & GW230608\_205047 & 0.179 & GW190915\_235702 & 0.148 & GW240920\_124024 & 0.128 \\
GW241116\_151753 & 0.229 & GW200209\_085452 & 0.179 & GW230924\_124453 & 0.147 & GW190630\_185205 & 0.128 \\
GW240107\_013215 & 0.227 & GW200219\_094415 & 0.179 & GW241129\_021832 & 0.147 & GW190708\_232457 & 0.128 \\
GW230922\_040658 & 0.226 & GW231005\_091549 & 0.179 & GW231104\_133418 & 0.147 & GW200224\_222234 & 0.128 \\
GW190805\_211137 & 0.226 & GW231113\_122623 & 0.178 & GW240703\_191355 & 0.146 & GW241130\_110422 & 0.127 \\
GW240601\_061200 & 0.223 & GW240908\_082628 & 0.177 & GW190517\_055101 & 0.146 & GW231123\_135430 & 0.127 \\
GW240527\_183429 & 0.222 & GW230708\_053705 & 0.177 & GW240530\_012417 & 0.146 & GW200225\_060421 & 0.126 \\
GW230806\_204041 & 0.219 & GW240615\_160735 & 0.177 & GW231110\_040320 & 0.146 & GW230605\_065343 & 0.126 \\
GW250118\_023225 & 0.217 & GW190620\_030421 & 0.176 & GW240513\_183302 & 0.145 & GW240428\_225440 & 0.126 \\
GW230928\_215827 & 0.216 & GW200128\_022011 & 0.176 & GW240825\_055146 & 0.145 & GW231224\_024321 & 0.126 \\
GW231001\_140220 & 0.214 & GW240705\_053215 & 0.175 & GW231206\_233901 & 0.144 & GW231223\_075055 & 0.125 \\
GW230819\_171910 & 0.213 & GW190413\_052954 & 0.175 & GW230624\_113103 & 0.143 & GW240915\_001357 & 0.125 \\
GW241125\_010116 & 0.213 & GW240531\_040326 & 0.174 & GW230911\_195324 & 0.143 & GW170809\_082821 & 0.125 \\
GW230803\_033412 & 0.212 & GW190602\_175927 & 0.174 & GW231226\_101520 & 0.142 & GW231118\_090602 & 0.124 \\
GW231221\_135041 & 0.211 & GW230805\_034249 & 0.173 & GW190503\_185404 & 0.142 & GW241111\_111552 & 0.124 \\
GW230825\_041334 & 0.211 & GW241109\_033317 & 0.172 & GW240601\_231004 & 0.142 & GW231223\_202619 & 0.124 \\
GW241009\_022835 & 0.211 & GW230609\_064958 & 0.171 & GW200208\_130117 & 0.141 & GW190728\_064510 & 0.124 \\
GW230831\_015414 & 0.210 & GW190519\_153544 & 0.169 & GW240622\_004008 & 0.141 & GW250118\_055802 & 0.123 \\
GW231004\_232346 & 0.209 & GW191222\_033537 & 0.169 & GW170814\_103043 & 0.140 & GW240615\_113620 & 0.123 \\
GW231127\_165300 & 0.206 & GW241230\_084504 & 0.169 & GW191215\_223052 & 0.140 & GW230904\_051013 & 0.122 \\
GW230709\_122727 & 0.204 & GW170729\_185629 & 0.168 & GW240507\_041632 & 0.140 & GW190707\_093326 & 0.122 \\
GW250116\_015318 & 0.204 & GW190803\_022701 & 0.168 & GW190828\_063405 & 0.140 & GW190720\_000836 & 0.121 \\
GW240908\_125134 & 0.204 & GW231206\_233134 & 0.168 & GW241113\_163507 & 0.139 & GW200316\_215756 & 0.121 \\
GW240621\_200935 & 0.202 & GW240514\_121713 & 0.167 & GW190910\_112807 & 0.139 & GW240930\_035959 & 0.119 \\
GW230824\_033047 & 0.202 & GW241230\_233618 & 0.167 & GW241002\_030559 & 0.139 & GW241114\_235258 & 0.119 \\
GW241124\_024914 & 0.202 & GW230704\_021211 & 0.167 & GW170104\_101158 & 0.138 & GW190725\_174728 & 0.119 \\
GW230814\_061920 & 0.201 & GW241009\_220455 & 0.167 & GW230811\_032116 & 0.138 & GW190930\_133541 & 0.118 \\
GW231118\_071402 & 0.199 & GW230927\_043729 & 0.167 & GW230729\_082317 & 0.138 & GW190512\_180714 & 0.117 \\
GW240501\_033534 & 0.198 & GW250109\_010541 & 0.166 & GW230723\_101834 & 0.138 & GW190521\_074359 & 0.117 \\
GW250104\_015122 & 0.198 & GW230606\_004305 & 0.166 & GW240916\_184352 & 0.138 & GW241225\_042553 & 0.117 \\
GW231129\_081745 & 0.198 & GW240519\_012815 & 0.164 & GW231020\_142947 & 0.138 & GW241109\_115924 & 0.115 \\
GW231223\_032836 & 0.197 & GW240630\_101703 & 0.163 & GW190513\_205428 & 0.137 & GW250114\_082203 & 0.115 \\
GW190706\_222641 & 0.196 & GW190527\_092055 & 0.163 & GW170823\_131358 & 0.137 & GW241231\_054133 & 0.115 \\
GW240426\_031451 & 0.195 & GW230920\_071124 & 0.162 & GW240512\_024139 & 0.137 & GW240915\_105151 & 0.113 \\
GW240525\_031210 & 0.195 & GW230914\_111401 & 0.161 & GW231018\_233037 & 0.136 & GW151226\_033853 & 0.112 \\
GW230930\_110730 & 0.195 & GW190421\_213856 & 0.161 & GW190828\_065509 & 0.135 & GW240910\_103535 & 0.112 \\
GW230707\_124047 & 0.194 & GW190701\_203306 & 0.161 & GW240830\_211120 & 0.135 & GW230627\_015337 & 0.111 \\
GW250108\_152221 & 0.194 & GW190727\_060333 & 0.160 & GW231113\_200417 & 0.135 & GW240919\_061559 & 0.111 \\
GW200216\_220804 & 0.193 & GW240902\_143306 & 0.159 & GW230731\_215307 & 0.134 & GW240621\_195059 & 0.110 \\
GW191230\_180458 & 0.193 & GW250109\_074552 & 0.159 & GW240109\_050431 & 0.134 & GW191204\_171526 & 0.109 \\
GW240907\_153833 & 0.193 & GW230702\_185453 & 0.159 & GW191109\_010717 & 0.134 & GW170608\_020116 & 0.109 \\
GW190521\_030229 & 0.193 & GW230628\_231200 & 0.159 & GW240629\_145256 & 0.134 & GW240925\_005809 & 0.109 \\
GW240924\_000316 & 0.192 & GW230712\_090405 & 0.158 & GW190925\_232845 & 0.133 & GW170818\_022509 & 0.108 \\
GW241210\_120900 & 0.190 & GW250119\_025138 & 0.158 & GW231114\_043211 & 0.133 & GW230814\_230901 & 0.108 \\
GW230820\_212515 & 0.190 & GW230726\_002940 & 0.155 & GW231231\_154016 & 0.133 & GW191216\_213338 & 0.107 \\
GW190929\_012149 & 0.190 & GW241114\_024711 & 0.155 & GW240920\_073424 & 0.133 & GW241011\_233834 & 0.106 \\
GW190413\_134308 & 0.189 & GW240104\_164932 & 0.155 & GW200311\_115853 & 0.132 & GW250119\_190238 & 0.105 \\
GW231028\_153006 & 0.189 & GW231108\_125142 & 0.153 & GW230919\_215712 & 0.132 & GW200202\_154313 & 0.105 \\
GW240923\_204006 & 0.189 & GW240526\_093944 & 0.152 & GW200112\_155838 & 0.132 & GW150914\_095045 & 0.103 \\
GW231213\_111417 & 0.188 & GW241210\_060606 & 0.152 & GW200302\_015811 & 0.132 & GW190924\_021846 & 0.103 \\
GW240505\_133552 & 0.187 & GW250101\_011205 & 0.152 & GW241127\_061008 & 0.131 & GW241102\_124058 & 0.103 \\
GW190719\_215514 & 0.187 & GW240511\_031507 & 0.152 & GW240922\_142106 & 0.131 & GW240413\_022019 & 0.102 \\
GW230601\_224134 & 0.186 & GW230922\_020344 & 0.151 & GW191105\_143521 & 0.131 & GW240527\_230910 & 0.097 \\
GW241102\_144729 & 0.186 & GW241101\_220523 & 0.151 & GW240414\_054515 & 0.130 & GW190412\_053044 & 0.092 \\
\hline
\end{tabular}
\end{table*}

\begin{table}\label{tab:constraints}
\centering
\caption{The 95 percent credible intervals on the fraction of GW events coming from AGN brighter than the bolometric luminosity $L_{\rm bol}^{\rm thresh}$, varying the priors on the hyperparameters governing the redshift evolution of the BBH merger rate from all alternative formation channels. The fiducial model and Gaussian priors are based on the fit by \citet{Strolger:2020}. The LVK prior is uniform on the indicated interval, adopted from \citet{ligo:gwtc5pop}. All posteriors are informed by both GWTC-5.0 and Quaia.}
\begin{tabular}{cccc}
\hline\hline
$L_{\rm bol}^{\rm thresh} [\mathrm{erg \, s^{-1}}]$ & LVK & Gaussian & Fiducial \\
\hline
$10^{44.5}$    & $[0.04, 0.99]$ & $[0.00, 0.58]$ &  $[0.00, 0.52]$ \\
$10^{45.5}$    & $[0.00, 0.92]$  & $[0.00, 0.48]$ &  $[0.00, 0.41]$ \\
$10^{46.5}$    & $[0.00, 0.86]$     &  $[0.00, 0.67]$  & $[0.00, 0.61]$  \\
\hline
\end{tabular}
\end{table}

\bsp	%
\label{lastpage}
\end{document}